\documentclass[sn-aps]{sn-jnl}

\usepackage{graphicx}
\usepackage{xcolor}
\usepackage{caption}
\usepackage{subcaption}
\usepackage{multirow}
\usepackage[numbers,sort&compress]{natbib}
\usepackage{threeparttable}
\usepackage{textcomp}

\title{Mass Testing and Characterization of 20-inch PMTs for JUNO}

\author[6,5]{\fnm{Angel} \sur{ Abusleme}}
\author[46]{\fnm{Thomas} \sur{ Adam}}
\author[67]{\fnm{Shakeel} \sur{ Ahmad}}
\author[67]{\fnm{Rizwan} \sur{ Ahmed}}
\author[56]{\fnm{Sebastiano} \sur{ Aiello}}
\author[67]{\fnm{Muhammad} \sur{ Akram}}
\author[67]{\fnm{Abid} \sur{ Aleem}}
\author[49]{\fnm{Tsagkarakis} \sur{ Alexandros}}
\author[30]{\fnm{Fengpeng} \sur{ An}}
\author[23]{\fnm{Qi} \sur{ An}}
% ...
\author*[]{(JUNO collaboration)}\email{juno\_pub\_comm@juno.ihep.ac.cn}

% ...
\affil[5]{Millennium Institute for SubAtomic Physics at the High-energy Frontier (SAPHIR), ANID, Chile}
\affil[6]{Pontificia Universidad Cat\'{o}lica de Chile, Santiago, Chile}
% ...
\affil[23]{University of Science and Technology of China, Hefei, China}
% ...
\affil[30]{East China University of Science and Technology, Shanghai, China}
% ...
\affil[46]{IPHC, Universit\'{e} de Strasbourg, CNRS/IN2P3, F-67037 Strasbourg, France}
% ...
\affil[49]{III. Physikalisches Institut B, RWTH Aachen University, Aachen, Germany}
% ...
\affil[56]{INFN Catania and Dipartimento di Fisica e Astronomia dell Universit\`{a} di Catania, Catania, Italy}
% ...
\affil[67]{Pakistan Institute of Nuclear Science and Technology, Islamabad, Pakistan}
% ...

\abstract{Main goal of the JUNO experiment is to determine the neutrino mass ordering using a 20\,kt liquid-scintillator detector. Its key feature is an excellent energy resolution of at least 3\,\% at 1\,MeV, for which its instruments need to meet a certain quality and thus have to be fully characterized. More than 20,000 20-inch PMTs have been received and assessed by JUNO after a detailed testing program which began in 2017 and elapsed for about four years. Based on this mass characterization and a set of specific requirements, a good quality of all accepted PMTs could be ascertained. This paper presents the performed testing procedure with the designed testing systems as well as the statistical characteristics of all 20-inch PMTs intended to be used in the JUNO experiment, covering more than fifteen performance parameters including the photocathode uniformity. This constitutes the largest sample of 20-inch PMTs ever produced and studied in detail to date, i.e. 15,000 of the newly developed 20-inch MCP-PMTs from Northern Night Vision Technology Co. (NNVT) and 5,000 of dynode PMTs from Hamamatsu Photonics K. K.(HPK).
}


\date{Received: date / Accepted: date}

\begin{document}
\maketitle
\flushbottom

\section{Introduction}
\label{sec:1:intro}

\begin{figure*}
\centering
\includegraphics[scale=0.5]{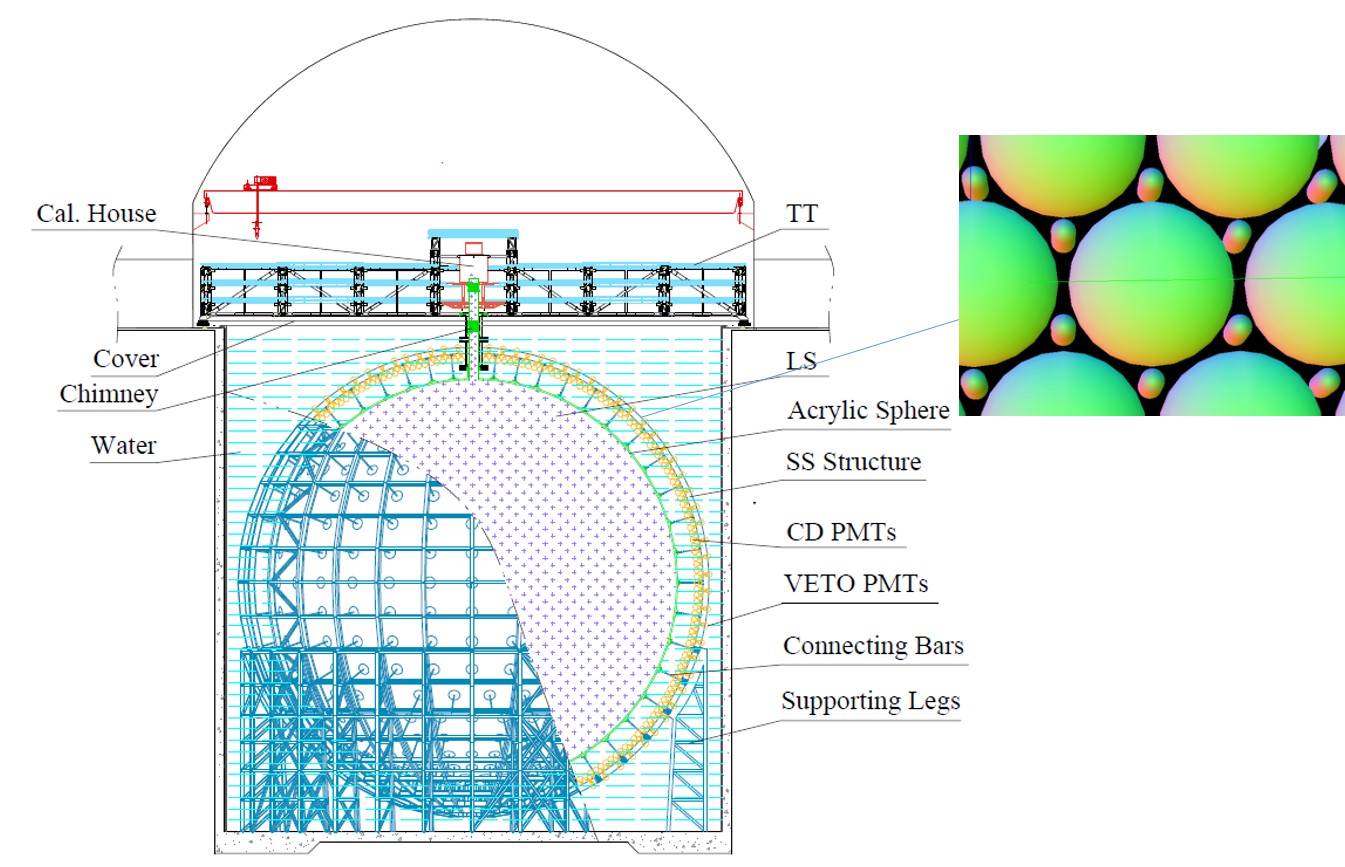}
\caption{Layout of the JUNO detector system\,\cite{JUNOdetector}. The zoomed in picture on the right is illustrating the proposed PMT arrangement of the JUNO central detector, with the 3-inch PMTs placed in the gaps between the 20-inch PMTs.}
\label{fig:juno:detector}
\end{figure*}

The Jiangmen Underground Neutrino Observatory (JUNO) experiment~\cite{JUNOCDR,JUNOphysics} is a new large-volume multi-purpose liquid-scintillator experiment currently under construction in southern China, located in a cavern with a 700\,m rock overburden. Its main goal is to determine the neutrino mass ordering from the neutrino oscillation spectrum of two close-by nuclear power plants at a distance of 53\,km each, with a sensitivity better than three standard deviations after six years of data taking~\cite{JUNOphysics,JUNOyufeng,JUNOliang}. The central detector (CD) of JUNO~\cite{JUNOCDlombardi,JUNOCDzhimin,JUNOCDyuekun,JUNOdetector} (Fig.\,\ref{fig:juno:detector}) consists of an acrylic sphere with a diameter of 35.4\,m and is filled with 20\,kt of LAB-based liquid scintillator (LS)~\cite{JUNO-LS-optimization}. A high transparency LS, high optical coverage (\textgreater 75\%) with high quantum efficiency photomultiplier tubes (PMTs), and low background/noise levels are needed to achieve an energy resolution of at least 3\,\% at 1\,MeV~\cite{JUNOdetector,JUNOCalibration}, which is essential for realizing the physics goals of JUNO. The high coverage of the CD is achieved by closely packing 17,612 high quantum efficiency 20-inch PMTs (Large PMTs or LPMTs) and 25,600 3-inch PMTs (Small PMTs or SPMTs) in the gaps between the LPMTs, placed at a distance of 19.8\,m to the detector's center. All of them are mounted on a stainless steel lattice shell (diameter of 40.1\,m) outside the acrylic sphere. The whole construction is further embedded into a cylindrical pure water pool with a diameter of 43.5\,m and depth of 44\,m (water filled to 43.5\,m), instrumented with another about 2,400 20-inch PMTs. The water pool will be used as an active Cherenkov veto against cosmic muons traversing the detector, but also acts as a shielding buffer from the surroundings.

In large volume LS detectors as JUNO, optimizing the energy resolution specifically concerns all parts related to the light collection and detection, including in particular the performance of the light-sensitive devices in the range from single photo-electrons (SPE) to hundreds of photo-electrons per channel. Aiming to achieve an unprecedented energy resolution, JUNO is designed with a set of 1\,GHz waveform sampling electronics to record the pulses of the 20-inch PMTs for a precise neutrino event detection and reconstruction~\cite{JUNOdetector,JUNOElectronics,JUNOEmbedded}.  
Playing the key role in this process, the PMTs' photon detection efficiency, dark count rate (DCR), and timing characteristics are most important to achieve the best possible detector performance~\cite{JUNOdetector}. At the same time, good performance in other parameters is needed for better event reconstruction, background reduction, and a successful long-term operation: this concerns the gain as a function of high voltage, charge resolution, peak-to-valley ratio, pulse shape features, correlated pulses (pre-/after-pulses), and, in particular, the photocathode response uniformity and Earth's magnetic field (EMF) sensitivity of the large volume PMTs. Detailed knowledge of the PMT characteristics moreover provides valuable input for precise detector simulations leading to a better understanding of the to-be-running experiment~\cite{JUNOdetector,JUNOSimulation,JUNO-Rec-QLiu,JUNO-Rec-Zhen,JUNO-Rec-Marcel}. All LPMTs used in JUNO have been measured in detail and checked against a list of preassigned requirements (see sec.\,\ref{sec:2:criteria}). These tests also include consistency checks between bare and waterproof potted PMTs\footnote{See sec.\,\ref{sec:2:facilities} as well as \cite{Qin_Neutrino,JUNOPMTinstr} for details about the potted PMTs.}.

Concerning the mechanical safety which is relevant for the large vacuum glass bulbs of PMTs working in pure water at a maximum depth of 44\,m, a careful inspection is performed on each PMT to identify  any mechanical/structural defects. Furthermore, following many valuable, previously performed studies on different PMT parameters, such as general studies on the large area PMT performance~\cite{largePMTlei,largePMTcomp,wavesamplingPMT,largePMTxia,JUNOPMTlinearity,JUNOPMTgain,PMTrelativeCE}, temperature effects~\cite{lowTPMT,lowTPMT2,HamManual,largePMTtemperature}, and the lifetime of the newly developed MCP-PMT by Northern Night Vision Technology Co. (NNVT)~\cite{lifetimeMCPPMT,lifetimeMCPPMT2,lifetimeMCPPMT3,argonneMCPPMT}, follow-up studies on the DCR as a function of temperature and the long-term stability of the MCP-PMTs are evaluated.

In this work, all details are presented about the measured characteristics of all accepted JUNO 20-inch PMTs, as well as brief introductions to the setup of the mass testing systems and the performed testing procedures. The selected 20-inch PMT types of JUNO will be described in sec.\,\ref{sec:1:pmt}. The testing facilities and performed procedures will be presented in sec.\,\ref{sec:1:setup}. The testing results of the bare PMTs (full 20,000 LPMT sample) will be shown in detail in sec.\,\ref{sec:1:results}. Expected features of waterproof potted PMTs and differences with respect to testing results from the acceptance tests of the bare PMTs will be discussed in sec.\,\ref{sec:1:exppotted}. Finally, a summary is given in sec.\,\ref{sec:1:summary}.
Meanwhile, it is valuable to stress that several other PMT characterization campaigns were performed in the past; we mention those of the IceCube~\cite{icecubePMT}, MiniBooNE~\cite{MiniBooNEPMT}, IMB~\cite{IMBPMT}, Borexino~\cite{BorexinoPMT,BorexinoPMT2}, Daya Bay~\cite{DayabayPMT}, Chooz~\cite{ChoozPMT}, SNO~\cite{SNOPMT}, Double Chooz~\cite{doubleChoozPMT,doubleChoozPMT2}, DAMA~\cite{DAMAPMT}, Auger~\cite{AugerPMT,AugerPMT2}, ANNIE~\cite{ANNIEPMT}, YBJ~\cite{YBJ8inch}, LHAASO~\cite{LHAASOPMT,LHAASOPMT2}, Super-K~\cite{sk-SUZUKI1993299}, Hyper-Kamiokande~\cite{HKPMT,HKPMT2}, and KM3Net~\cite{KM3NeTPMT,KM3NeTPMT2} experiments. For completeness, it should be noted here that the 3-inch PMTs used for JUNO are investigated as well in a separate campaign~\cite{JUNODCmiao2017,JUNO3inchPMT}.

\section{20-inch photomultiplier tubes for JUNO}
\label{sec:1:pmt}

\subsection{Selected 20-inch PMTs of JUNO}
\label{sec:2:selectedpmt}

\begin{figure*}
\centering
\includegraphics[scale=0.5]{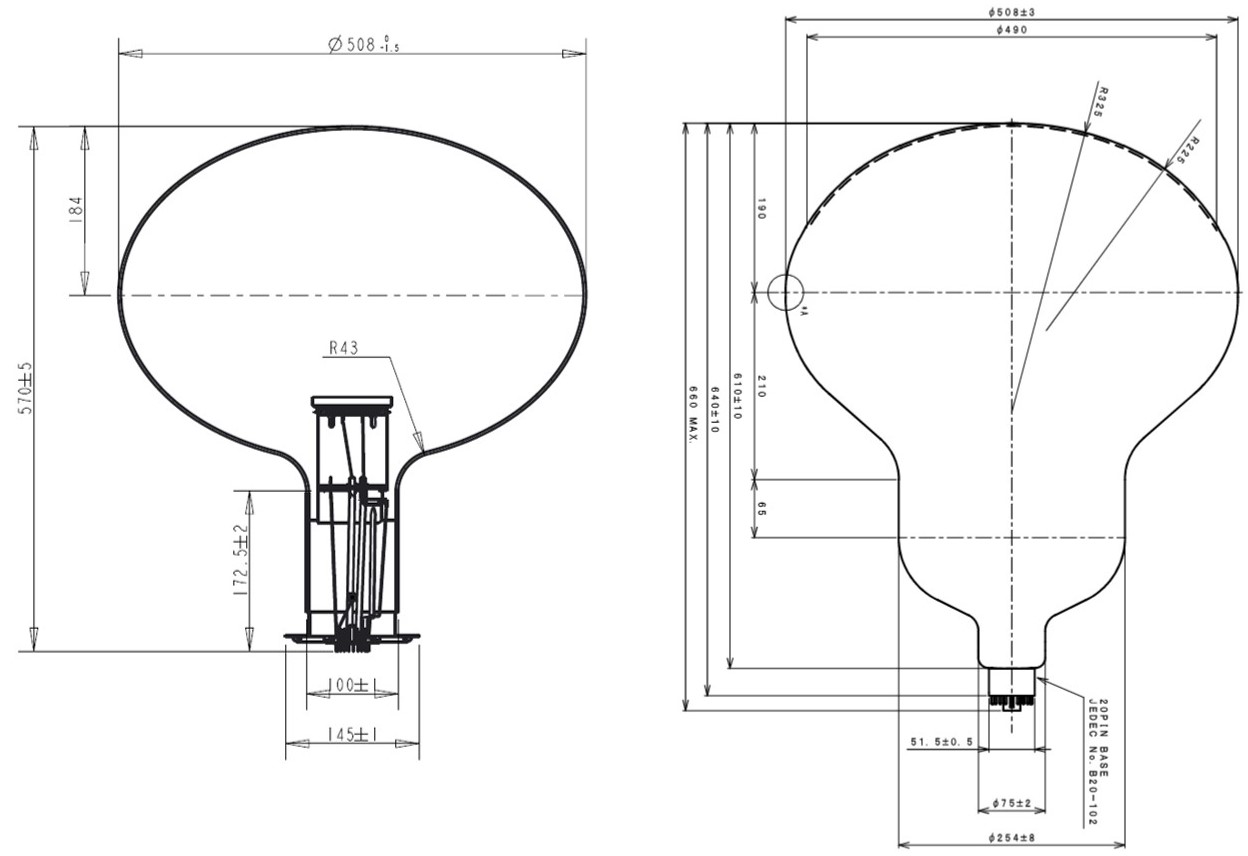}
\caption{Technical drawing of selected PMTs for JUNO: NNVT 20-inch MCP-PMT (type GDB-6201~\cite{NNVT-GDB6201-note}, left) and HPK 20-inch dynode PMT (type R12860~\cite{HPK-R12860}, right).}
\label{fig:PMT:dimension}
\end{figure*}

It is theoretically and practically significant to build a detector with maximum physics potential at minimal costs. To date, PMTs offer the best compromise between single-photon sensitivity and acceptance in noise and cost per unit area around room temperature, and large area PMTs are the first choice of photon sensors for large LS- or water-based neutrino experiments which require high photocathode coverage. Besides photon detection efficiency (PDE), other characteristics such as dark count rate (DCR), transit time spread (TTS), the radioactive background of the glass, peak-to-valley ratio (P/V), etc., will affect the photon detection and event reconstruction, and thus impact the physics measurements. Considering all related PMT parameters, costs and risks, evaluated with the physics goals of JUNO, a selection strategy for PMTs was proposed and applied, leading to a set of requirements for the performance of JUNO 20-inch PMTs~\cite{JUNOPMTliangjian}. With the bidding following the selection strategy, there are two types of 20-inch PMTs selected by JUNO: approx.~5,000 box and linear-focused dynode-PMTs (R12860-50 HQE or R12860) from Hamamatsu Photonics K. K. (HPK)~\cite{HPK-R12860} and approx.~15,000 MCP-PMTs (GDB-6201 or N6201) from North Night Vision Technology Co. (NNVT)~\cite{YWang_newMCP,LiuMCP,NNVT-GDB6201-improvement,NNVT-GDB6201-note}. The dimensions of the two types of PMTs are shown in Fig.\,\ref{fig:PMT:dimension}.

\subsection{Acceptance criteria}
\label{sec:2:criteria}

Based on joint considerations of the physics requirements discussed above and the PMT manufacturers, a list of combined performance criteria was defined for all ordered 20,000 20-inch PMTs of JUNO. Each of the received PMTs must be tested at least once to assure that all predefined requirements are satisfied. In Tab.\,\ref{tab:PMT:criteria} the required nominal values as well as lower or upper limits for part of the parameters are summarized. Each of the listed parameters will be discussed separately and in detail within this paper. The photon detection efficiency (PDE), as one critical parameter required by JUNO and discussed in detail in sec.\,\ref{sec:2:pde}, is not listed directly in Tab.\,\ref{tab:PMT:criteria}, but appears as a combined parameter of the quantum efficiency (QE), the collection efficiency (CE), and the effective area ratio (EAR)\footnote{The PDE is defined as \textit{PDE}\,($\lambda$) = \textit{QE}\,($\lambda$) $\times$ \textit{CE} $\times$ \textit{EAR}. The Effective Area Ratio (EAR), which is from the manufacture, is the relative effective area ratio, where the collection efficiency is greater than 95.6\% when the diameter of a PMT is 508\,mm.}. It is finally requested with an averaged value of $\geq 27\,\%$ for all the PMTs, and a minimum value of $\geq 24\,\%$ for a single PMT (in particular for use in the CD), both defined at a wavelength of 420\,nm to match the typical emission spectra of JUNO liquid scintillator \cite{JUNO-LS-optimization}. A catalog index of the content and all the checked parameters is listed in Tab.\,\ref{tab:checklist}. 

\begin{table}[!ht]
\centering
\caption{Main acceptance criteria for JUNO 20-inch PMTs.}
\label{tab:PMT:criteria}       
\begin{threeparttable}
\begin{tabular}{lll}
\hline\noalign{\smallskip}
   & HPK R12860-50 & NNVT GDB-6201 \\
\noalign{\smallskip}\hline\noalign{\smallskip}
Parameter & Average (Limit) & Average (Limit) \\
\noalign{\smallskip}\hline\noalign{\smallskip}
\hline
QE & $30.3\,\%$ ($\geq 27\,\%$) & $28.5\,\%$ ($\geq 26.5\,\%$) \\
CE & $95.6\,\%$ &  $98\,\%$ ($\geq 96\,\%$) \\
Effective area ratio & $96\,\%$ ($93\,\%$) &  $97\,\%$ ($\geq 96\,\%$) \\
Gain & $10^7$ & $10^7$ \\
HV (for a $10^7$ gain) & 2000\,V ($\leq 2500$\,V) & 2500\,V ($\leq 2800$\,V) \\
QE uniformity & $5\,\%$ ($\leq 15\,\%$\, inside $70^{\circ}$)  & $8\,\%$ ($\leq 10\,\%$) \\
   & $20\,\%$ ($\leq 30\,\%$\, inside $80^{\circ}$) &  \\
TTS (FWHM) & 2.7\,ns ($\leq 3.5$\,ns) & 12\,ns ($\leq 15$\,ns) \\
P/V ratio & 3 ($\geq 2.5$) & 3.5 \\
Pre-pulse ratio  & & \\
  (80\,ns window, & $0.8\,\%$ ($\leq 1\,\%$)  & $0.5\,\%$ ($\leq 1\,\%$) \\
  main pulse $\sim$160\,p.e.) &  &   \\
After-pulse ratio  & & \\
  (0.5$\sim$20\,\textmu s window, & $10\,\%$ ($\leq 15\,\%$)  & $10\,\%$ ($\leq 15\,\%$)  \\
  main pulse $\sim$160\,p.e.) &  & \\
Dark count rate & 10\,kHz ($\leq 50$\,kHz) & $\leq 50$\,kHz (if $24\,\% \leq$ PDE $< 27\,\%$) \\
  (0.25\,p.e., 22$^{\circ}$C)&             &  $\leq 60$\,kHz (if $27\,\% \leq$ PDE $< 28\,\%$) \\
  &             &  $\leq 80$\,kHz (if $28\,\% \leq$ PDE $< 29\,\%$) \\
  &             &  $\leq 100$\,kHz (if $29\,\% \leq$ PDE) \\
Glass radioactivity & $^{238}$U: $< 400$\,ppb & $^{238}$U: $< 75$\,ppb  \\
  & $^{232}$Th: $< 400$\,ppb & $^{232}$Th: $< 75$\,ppb \\
  & $^{40}$K: $< 40$\,ppb & $^{40}$K: $< 30$\,ppb \\
Pressure tolerance & $\geq 0.8$\,MPa & $> 1$\,MPa \\
Dimension tolerance\tnote{1} & $508\,(\pm 3$\,mm) (diameter) & $508\,(\pm 3$\,mm) (diameter) \\
  & $ <10$\,mm (height)  &  $< 10$\,mm (height)  \\
Lifetime\tnote{2} & $\geq 20$\,years & $\geq 25$\,~years \\
\noalign{\smallskip}\hline
\end{tabular}
\begin{tablenotes}
\item[1] The tolerance is $\pm$3\,mm with the lower limit required by the optical coverage and the upper limit constrained by the detector installation. It is larger than the specification of the NNVT PMT shown in Fig.\,\ref{fig:PMT:dimension}.
\item[2] Gain decrease less than 50\% in 20 years with the same HV.
\end{tablenotes}
\end{threeparttable}
\end{table}

\begin{table}[!htb]
\centering
\caption{Catalog index of content and checked parameters.}
\label{tab:checklist}       
\resizebox{\linewidth}{!}{
\begin{tabular}{lll}
\hline 
 Category & Measurement/Test & Section \\
\hline \hline
    \multirow{2}{*}{Testing system} & Facilities  & \ref{sec:2:facilities} \\
    & Procedure & \ref{sec:2:datalogure}  \\ 
\hline
  \begin{tabular}[c]{@{}l@{}}
     Mechanical Safety \\  \& Physical Characteristics
  \end{tabular}  & Visual Inspection, Diameter, Weight & \ref{sec:3:visual}\\
\hline
  \multirow{10}{*}{\begin{tabular}[c]{@{}l@{}}
     Performance   \\ for precise detection   \\ and reconstruction 
  \end{tabular}} & Gain/Operating Voltage & \ref{sec:2:gainhv} \\
   & SPE Amplitude & \ref{sec:3:amplitude} \\
   & SPE Pulse Shape & \ref{sec:3:risefall} \\
   & SPE Charge Resolution & \ref{sec:3:resolution} \\
     & SPE Peak-to-Valley Ratio (P/V) & \ref{sec:3:pv} \\
    & Photon Detection Efficiency(PDE)  & \ref{sec:2:pde} \\
    & Dark Count Rate (DCR) & \ref{sec:2:dcr}  \\
     & Transit Time Spread (TTS) & \ref{sec:2:tts} \\
     & Correlated Pulses & \ref{sec:2:afterpulse} \\
     & Anode Non-linearity & \ref{sec:2:nonlinaerity} \\
     & Earth Magnetic Field (EMF) Sensitivity & \ref{sec:2:emf} \\
     & Photocathode Response Uniformity & \ref{sec:2:uniform} \\
\hline
   Long-term Stability & DCR, Gain, Light Intensity & \ref{sec:2:aging} \\
\hline
\multirow{2}{*}{Potted PMT}  & Gain & \ref{sec:2:potted:HV}\\
    & DCR  &  \ref{sec:2:potted:dcr} \\
\hline
\end{tabular}
}
\end{table}

\section{Testing setup and procedure}
\label{sec:1:setup}

\subsection{Testing facilities}
\label{sec:2:facilities}

To concentrate all actions related to PMT testing, a warehouse with around 4500\,$\mathrm{m}^2$ was rented since 2017 at Zhongshan Pan-Asia Electric Co., Ltd., Guangdong Province, China, which is about 150\,km away from the JUNO experimental site. In this warehouse, named \emph{Zhongshan Pan-Asia 20-inch PMT testing and potting station}, all relevant steps of preparing the 20,000 20-inch PMTs for JUNO were processed, including receiving (database management for recording and statistics), acceptance tests, storage, and waterproof potting. The warehouse runs in a controlled environment with a temperature of $25^{\circ}$C (in a range of $\pm 3^{\circ}$C) and relative humidity of $50$\,\% (in a range of 30\%-70\%), and is maintained by a local team which is also in charge of the daily PMT testing routine and all storage operations. At the same time, most of the 20-inch PMT testing shifts focusing on data acquisition and analysis are shared by the whole JUNO collaboration.

Based on the specific criteria of the selected 20-inch PMTs, a set of acceptance testing facilities were developed to realize a process of semi-automatic PMT testing covering most of the parameters listed in Tab.\,\ref{tab:PMT:criteria}\footnote{Glass radioactivity~\cite{JUNOPMTradioactivity} and pressure tolerances are investigated in separate tests, since those tests may require to destroy the PMTs as part of the measurement process.}. Main units are a commercial-container-based multi-channel testing system developed and installed to perform full characterizations for each of the 20,000 20-inch PMTs, as well as a scanning station system designed and built aiming to sample about $5\,\%$ of all the PMTs to check i.e. their photocathode uniformity. All PMT tests are performed under a shielding against Earth's magnetic field (EMF) with a remaining strength of less than $10\,\%$ of EMF (which is about 5\,\textmu T).

All PMTs are tested with JUNO-optimized high voltage (HV) dividers using a positive HV~\cite{JUNOPMTinstr,JUNOPMTsignalover,JUNOPMTsignalopt,JUNOPMTflasher}. The initial tests of the full sample of PMTs used a pluggable HV divider with an integrated HV-signal de-coupler on board. Later, a large sub-sample of the PMTs were tested again with the final JUNO version of the HV divider directly soldered to the PMT pins, where the entire PMT and HV divider were encapsulated with a waterproof housing~\cite{JUNOPMTinstr,JUNOCDPMT}. In the latter case the PMT pulses are picked up from a stand-alone HV-signal decoupler through a 2\,m (central detector PMTs) or 4\,m (water pool PMTs) cable\footnote{Where another 2\,m extension cable is used for the connection between potted PMT and testing electronics}. Although these versions differ in their mechanical setup, they feature the same HV-divider ratio and direct current (DC). However, the first few thousand bare NNVT PMTs were measured by another HV-divider\footnote{Based on the PMT testing results and feedback from the manufacturer, the design of the HV dividers for NNVT PMTs was improved over the testing campaign and updated to its final version after testing the first few thousand bare PMTs.} which only affects the rise- and fall-time of the pulses of these PMTs.

Two containers and two scanning stations are built and used for the bare PMT testing. Another two containers of slightly modified interior design are set up, one dedicated to long-term stability tests of PMTs, and another to the testing of waterproof potted PMTs together with JUNO (final front-end) 1F3 electronics~\cite{JUNOdetector}.

\subsubsection{PMT mass testing: container systems}
\label{sec:3:container}

The main setup of the individual (acceptance) tests of all 20,000 20-inch JUNO PMTs is the so-called \textit{container system}. This system has been described extensively in a stand-alone paper~\cite{JUNOPMTcontainer}, where all details about mechanical setup, data taking electronics, measurement process and accuracies can be found. To enhance understanding of the later presented results from the PMT mass testing, the setup will be briefly described here as well.

\begin{figure}[!ht]
\centering
\includegraphics[width=1.0\textwidth]{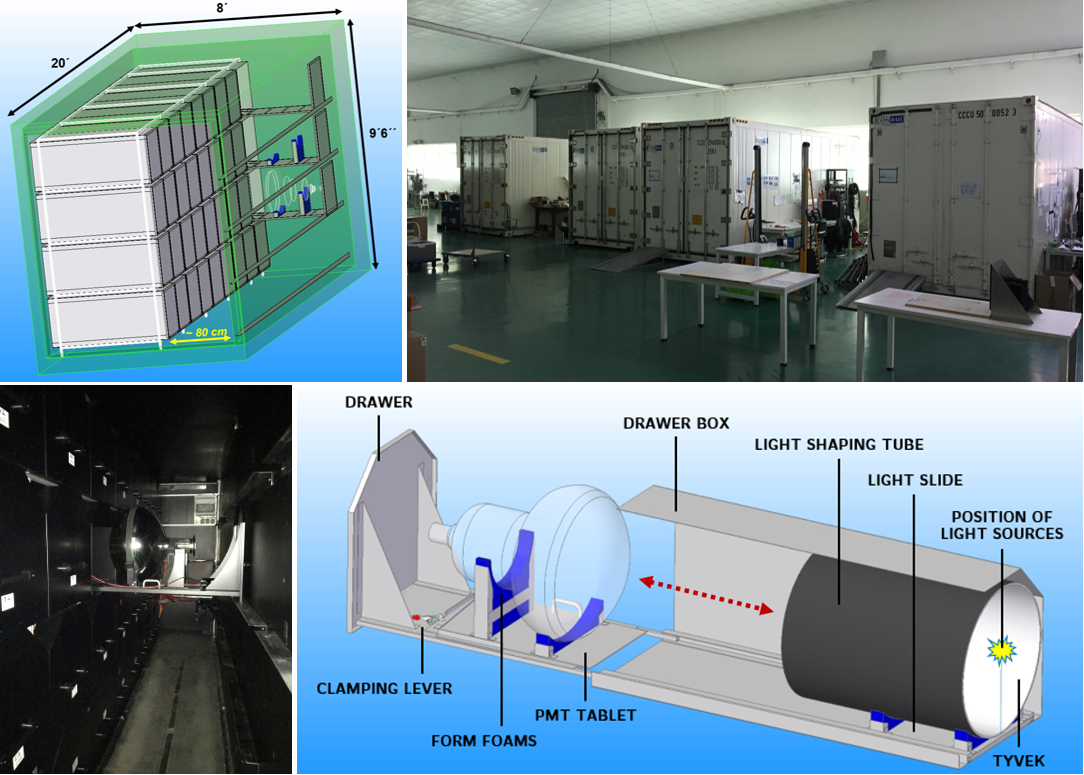}
\caption{Pictures of the PMT container system, located at \textit{Zhongshan Pan-Asia 20-inch PMT testing and potting station}, China. Top row: Conceptual and actual picture of the four testing containers; the sketch drawing shows the shelf system mounted inside. Bottom left: Interior of one of the testing containers, with a PMT loaded onto a drawer. Bottom right: Schematic view of the inside of a drawer box.}
\label{fig:PMT:container}
\end{figure}

The container system consists of four 20-feet high-cube reefer containers, which act as darkrooms and are able to host 36 (containers $\#$A and $\#$B) or 32 (containers $\#$C and $\#$D) PMTs in optically separated measurement channels (shelf system with drawer boxes), see Fig.\,\ref{fig:PMT:container}. Each container is passively shielded against the EMF by a multi-layer silicon-iron shielding and equipped with a high power HVAC (heating, ventilation, and air conditioning) unit, granting environmental control over the container's interior. All drawer boxes are further equipped with temperature sensors, which allow to measure the relationship between dark count rate and temperature of the surroundings. Containers $\#$A and $\#$B are equipped with commercial data taking electronics (main unit is a CAEN V1742 switched-capacitor digitizer~\cite{CAEN-FADC}), controlled and supervised by a custom-made data acquisition software (DAQ) based on LabView. This DAQ performs a sequence of individual measurements to enable a full PMT characterization within 24 hours, covering absolute PDE, DCR, gain, TTS\footnote{Due to mechanical issues at container $\#$A, precise values for the TTS by laser could be measured only in container $\#$B and only for a representative sub-sample of all PMTs, see sec.\,\ref{sec:2:tts} for more details.}, P/V, charge resolution, pulse characteristics such as rise-/fall-time, pre-pulse ratio and signal-to-noise ratio. These containers are intended to perform the acceptance tests of the full LPMT sample as well as functionality tests of a large sub-sample of potted PMTs. Container $\#$C is equipped with a similar set of commercial electronics and intended to act as long-term stability test setup, while container $\#$D is modified to host the JUNO 1F3 underwater boxes~\cite{JUNOdetector}, allowing additional functionality and performance tests of waterproof potted PMTs together with the final JUNO electronics.

Each drawer in all four containers is equipped with a LED device from company \textit{HVSys}~\cite{HVSYS-LED}, together with a collimator, attenuator, diffuser, and a cylindrical reflector to illuminate the full photocathode uniformly for SPE- and multi-p.e.-related measurements at optimized light intensities~\cite{PMTintensity} and uniformity~\cite{JUNOPMTcontainer} as shown at the bottom of Fig.\,\ref{fig:PMT:container}. These devices are pulsed LEDs (at 420\,nm) featuring a feedback loop on a small micro-controller, which stabilizes the light intensity over time to a calibrated level (based on a dedicated calibration campaign prior to the testing campaign, see sec.\,\ref{sec:2:gain} and \ref{sec:2:pde} for details). Each of the containers $\#$A and $\#$B are further equipped with a picosecond laser flashing system based on a PiLas 420X picosecond laser from Advanced Laser Systems (A.L.S.)~\cite{container-laser}. This device can produce short light pulses with a wavelength of 420\,nm and a width of $\sim 80$\,ps, which are distributed by a fiber splitter over all channels and used for precise TTS measurements at SPE level\footnote{The laser light does not pass the diffuser, however the opening angle of the emitted light from the fibers still covers most of the PMT's photocathode. TTS measurements using the LED system are performed as well (after ADC jitter correction of container LED measurement), but only for a relative check, see also sec.\,\ref{sec:2:tts} and \ref{sec:3:uniform:tts}.}. In container $\#$C specifically, an additional set of LEDs was introduced into the drawer boxes pulsing with different frequencies and light intensities in order to simulate an accelerated aging of the PMTs due to higher stress.

The performance and stability of each container (except for container $\#$C) is monitored by a small number of 20-inch PMTs (``reference PMTs'' or ``monitoring PMTs'', consisting of up to three HPK and two NNVT PMTs per container), which are included in every measurement run with one staying at a fixed drawer and the others being circulated over all channels. The containers achieve a stable performance over time as well as a sufficient accuracy of e.g. $\leq 1\,\%$ for the absolute PDE and $\leq 1$\,ns for the timing parameters such as the TTS (only container $\#$B)~\cite{JUNOPMTcontainer} -- this will be discussed also in secs.\ref{sec:2:pde} and \ref{sec:2:tts}. Hence, the system can provide reliable and comparable results for all available 20-inch PMTs characterized in this PMT testing campaign. More details about this system can be found also in~\cite{JUNOPMTcontainer,TietzschDiss,Tietzsch_Neutrino}.

\subsubsection{PMT sample testing: scanning stations}
\label{sec:3:station}

The second setup for the individual testing of a large sub-sample of 20-inch JUNO PMTs is the so-called \textit{scanning station system}. Also this system has been described already in other publications~\cite{Tietzsch_Neutrino,JUNOPMTstation}, where details about mechanical setup, electronics, and measurement process can be found. Similar to before, this setup shall be briefly described, to enhance understanding of the later presented results.

\begin{figure}[!ht]
  \centering\includegraphics[width=1.0\textwidth]{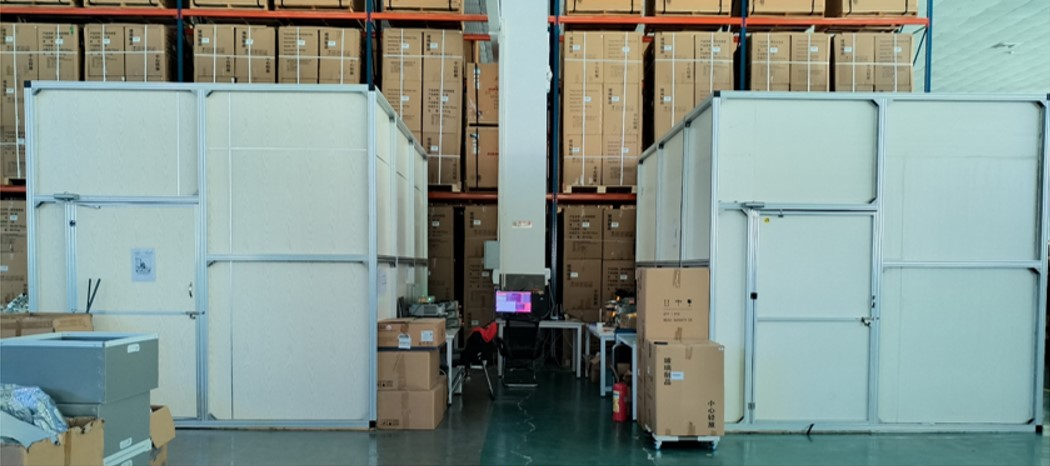}
  \centering\includegraphics[width=1.0\textwidth]{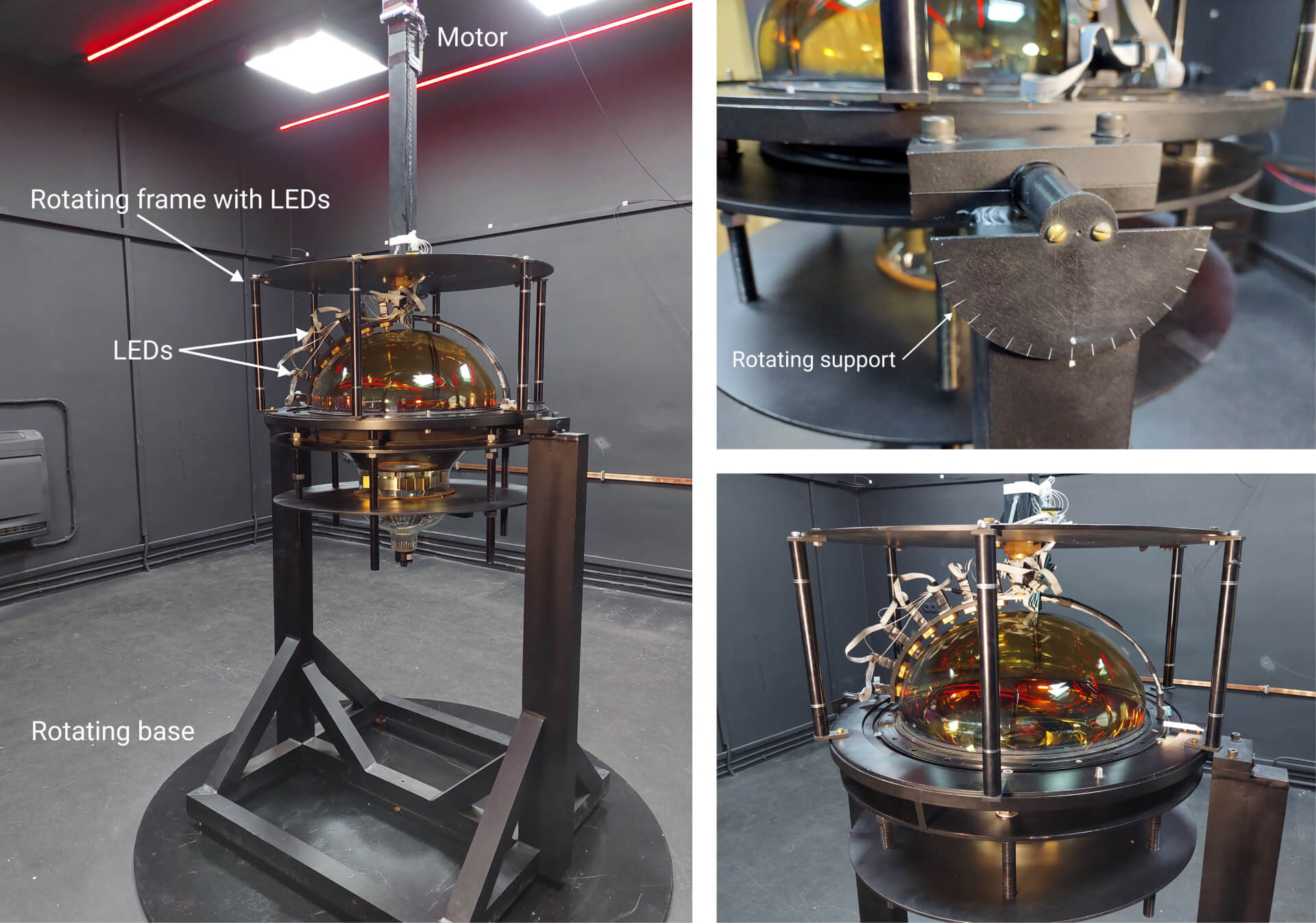}
\caption{Pictures of the scanning station system. Top: the darkrooms used for the two scanning stations; bottom: mechanical setup and rotating support of a single scanning station.}
\label{fig:PMT:station}       
\end{figure}

The scanning station system constitutes a complementary system, which is able to investigate possible inhomogeneities of characteristics along the PMT's photocathode surface (in complementary to the container system, which is not sensitive to local inhomogeneities). This can be achieved using a rotatable arch with seven stabilized pulsed LEDs mounted at different zenith angles (\#1 (pole), ..., \#7 (equator)) covering all 360$^{\circ}$ azimuthal angles in 15$^{\circ}$ steps and thus the full photocathode surface. The light sources used by the scanning stations are LEDs also from \textit{HVSys} company (similar to the ones in the container systems) but only equipped with collimators and attenuators (detected light intensity is 1\,$\sim 1.5$\,p.e. per flash). Two of these scanning stations are set up 
for mass sampling and operated in separate darkrooms (Fig.\,\ref{fig:PMT:station}). Due to an active magnetic compensation of the scanning stations provided by Helmholtz coils installed within the walls, floor, and ceiling of the housing darkrooms of each station, they also allow for the testing of a PMT's magnetic field sensitivity in a range from -50\,\textmu T to +50\,\textmu T as well as a PMT characterization in a completely EMF-free environment\footnote{The residual magnetic field applied for the general testing is $\lesssim 1\,$\textmu T.}. The scanning station uses a DRS4-based ADC~\cite{DRS4-web} with an additional $\times$\,10 amplifier to acquire full PMT waveforms. The scanning procedure and data analysis are fully automated in a four to six hours cycle per PMT and station. The absolute light intensities of all LEDs in both scanning stations are frequently calibrated and monitored using a small calibration PMT (head-on-type linear-focused $1\frac{1}{8}$-inch R1355 PMT with a PDE of $23.9\pm1.0\,\%$ at 420\,nm). The calibrated LED intensity only shows a spread of less than 2\% relatively over the whole running period and thus proves the high stability in light output of the LEDs over time.

The scanning stations allow the characterization of individual PMTs in all relevant parameters by scanning the PMT's photocathode (i.e.~local gain and average number of detected photo-electrons), and enable a deeper understanding of their performance based on the PMT's uniformity (see also sec.\,\ref{sec:2:uniform}). This may help to further understand the PMTs' performance and identify potential problems not detectable by only using the containers as testing facilities. The scanning stations are moreover used to define the weighted photon detection efficiency (PDE) of the whole 20-inch photocathode surface, which is one of the key parameters to be discussed in sec.\,\ref{sec:2:pde}. Since the characterization process is more complex, and is complementary to the container system each scanning station can take only one PMT to be tested, these features will be obtained only for a sub-sample $\sim$\,5\% of all 20,000 PMTs (originally aimed to cover about 5\,\% randomly selected, added by cross-checks between containers and scanning stations).

\subsection{Receiving, cataloging and testing procedures}
\label{sec:2:datalogure}

\begin{figure}[!ht]
	\centering\includegraphics[width=0.75\textwidth]{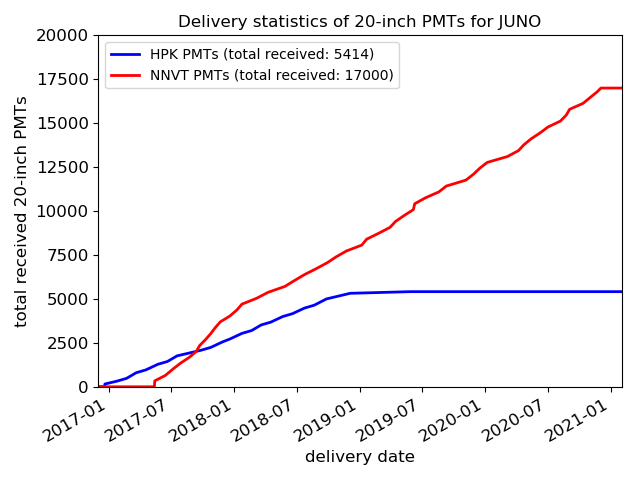}
	\caption{Statistics of PMTs received at \textit{Zhongshan Pan-Asia 20-inch PMT testing and potting station} and later tested within the testing systems. In total 22,414 PMTs were delivered.}
	\label{fig:PMT:TransportStatistics}       
\end{figure}

\begin{figure}[!ht]
  \centering\includegraphics[width=0.75\textwidth]{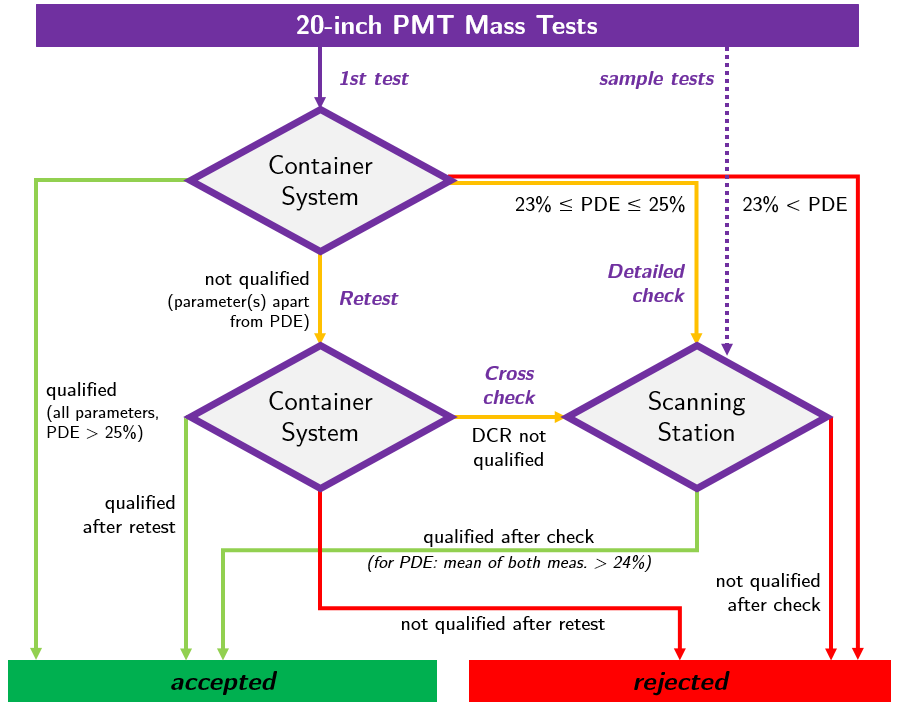}
\caption{Measurement and classification procedure performed by the two testing systems. Details of the full PMT characterization process performed by the container system can be also found in~\cite{JUNOPMTcontainer}.}
\label{fig:PMT:procedure2}       
\end{figure}

Starting from early 2017, the first batches of JUNO 20-inch PMTs arrived at \textit{Zhongshan Pan-Asia 20-inch PMT testing and potting station}, and continued to do so over a period of about four years (usually in batches of several hundred PMTs) (see Fig.\,\ref{fig:PMT:TransportStatistics}). After their arrival, a checklist was applied step by step to realize a detailed characterization and full documentation of all PMTs. 
This includes the whole process of checking-in, multiple testings (including visual inspections as well as performance tests with the containers and scanning stations), waterproof potting, and storage until further processing. Barcodes representing the PMT ID are used as a keyword to identify all individual records. The management of all PMTs is handled using a database with a web page link~\cite{JUNOrecieving} following the scheme in Fig.\,\ref{fig:PMT:procedure2}, which includes the vendor data of each PMT, inspection results, measurement records of both container and scanning station systems, measurement results management, waterproof potting status, potted PMT testing results, and storage location in individual charts, everything tagged by the individual unique PMT ID. Generally, it takes about three to six months to complete the whole process following the PMT receiving and storage. This cautiously executed procedure then leads to a final classification (accepted to use in JUNO or rejected). Since PDE and DCR are critical parameters, they will be cross-checked between containers and scanning stations if the initially measured value is around the boundary of the requirements as listed in Tab.\,\ref{tab:PMT:criteria} and indicated in Fig.\,\ref{fig:PMT:procedure2}).

\begin{figure}[!ht]
  	\begin{subfigure}[c]{0.49\textwidth}
		\centering
		\includegraphics[width=\linewidth]{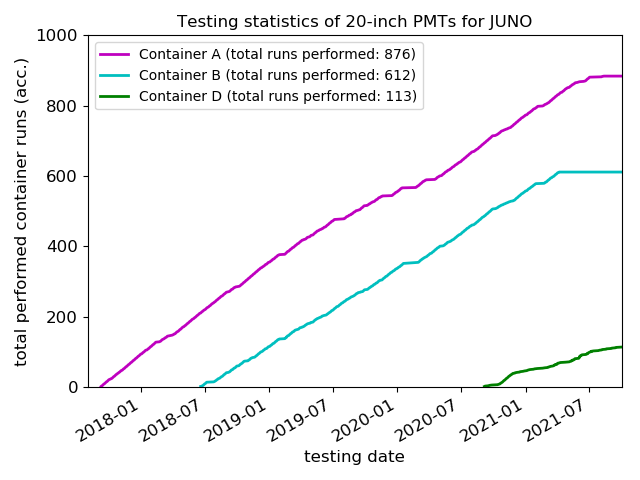}
		\caption{PMT testing statistics (containers)}
		\label{fig:PMT:TestingStatistics.a}
	\end{subfigure}\hfill
	\begin{subfigure}[c]{0.49\textwidth}
		\centering
		\includegraphics[width=\linewidth]{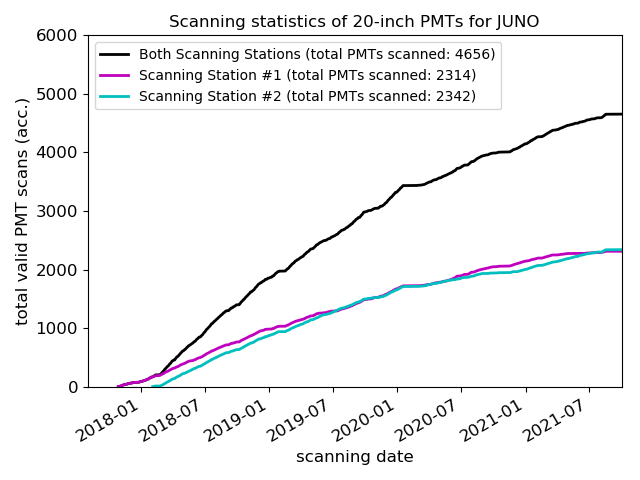}
		\caption{PMT testing statistics (scanning stations)}
		\label{fig:PMT:TestingStatistics.b}
	\end{subfigure}	
	\caption{Time evolution of the PMT testing. Left: more than 52,000 individual PMT characterizations were performed in about 1,600 runs using containers $\#$A, $\#$B and $\#$D (some PMTs have been tested multiple times including calibration runs, acceptance tests, re-tests, and continuous measurement campaigns). 
	Right: 4,656 valid characterizations from more than 5,200 scans (including special testings, and calibration) were performed using the scanning stations $\#$1 and $\#$2.}
	\label{fig:PMT:TestingStatistics}       
\end{figure}

During the whole PMT testing campaign, 22,414 PMTs have been received and stored in the warehouse, more than 52,000 individual PMT characterizations with the containers
and more than 5,200 PMT scans with the scanning stations have been completed. Some PMTs have been tested multiple times including the systems' calibrations, bare PMT testing, tests of the monitoring PMTs, potted PMT testing (using commercial electronics and JUNO 1F3 electronics) and special testing campaigns (see Fig.\,\ref{fig:PMT:TestingStatistics}). In total, 20,065 PMTs have finally passed all testing criteria and then are selected to operate in the JUNO experiment.

\section{PMT testing and characterization: bare PMT results}
\label{sec:1:results}

With the designed testing systems, each of the received PMTs has been tested at least once for the required parameters.
Their performance will be discussed in detail in the following sections, leading to the selection of a sample of individually qualified PMTs for JUNO.
During the daily testing, the reference PMTs (compare sec.\,\ref{sec:3:container}) are loaded into the containers in each run together with the untested PMTs to monitor the comparability of the measured parameters (and i.e. the stability of the container performance) over time.

\subsection{Visual inspection}
\label{sec:3:visual}

\begin{figure}[!ht]
	\begin{subfigure}[c]{0.495\textwidth}
		\centering
		\includegraphics[width=\linewidth]{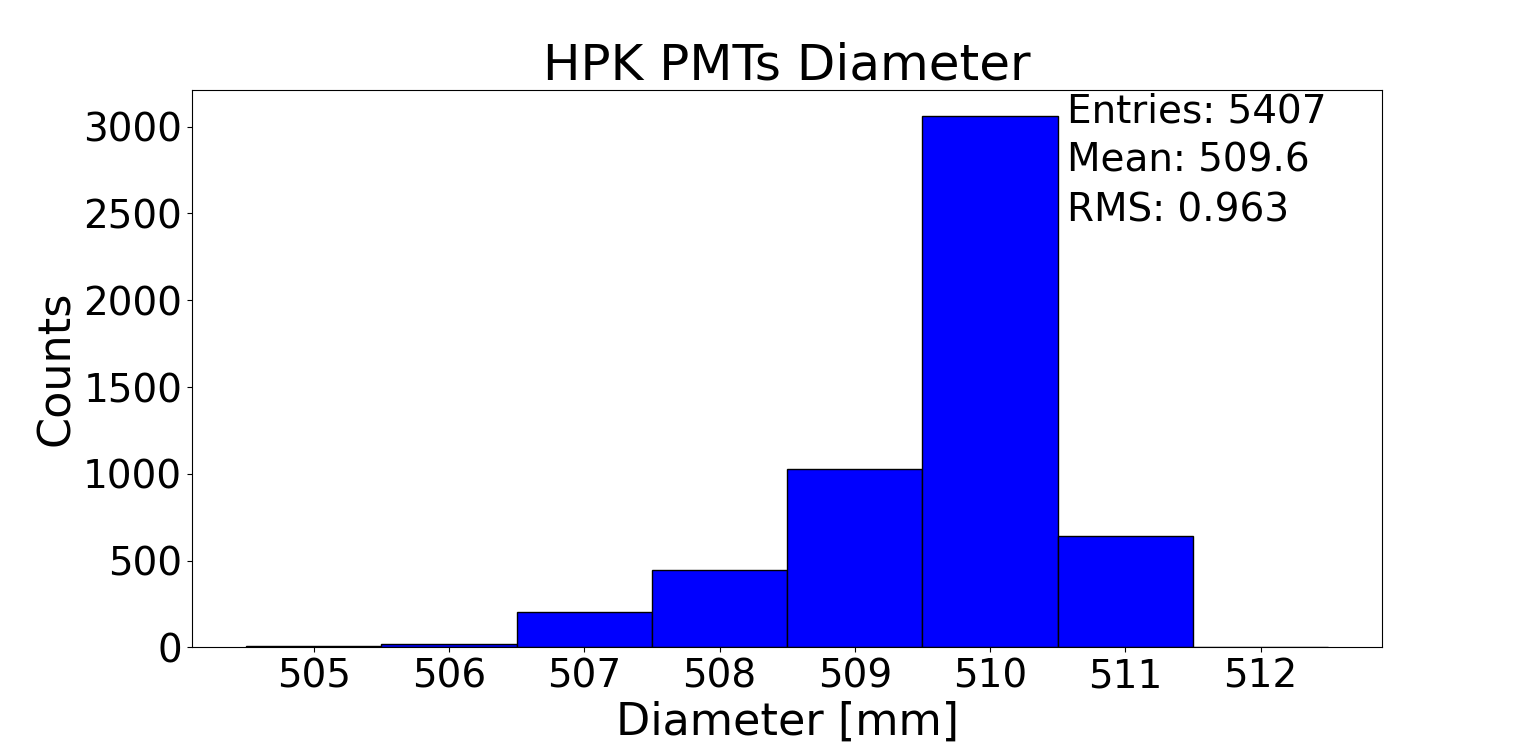}
	\end{subfigure}\hfill
	\begin{subfigure}[c]{0.495\textwidth}
		\centering
		\includegraphics[width=\linewidth]{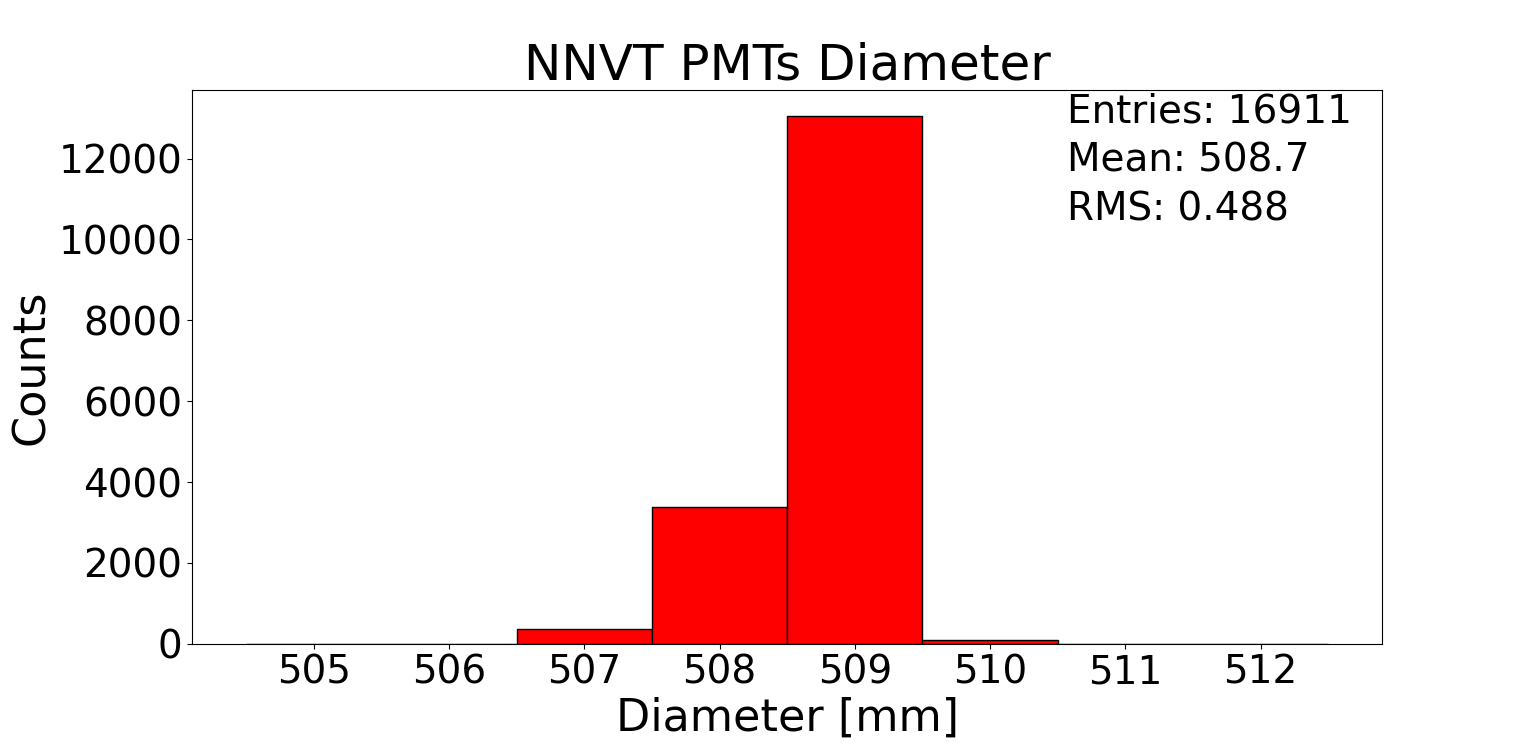}
	\end{subfigure}	
	\caption{Measured diameter in mm of all checked 20-inch PMTs. Some PMTs were rejected before the measurement after visual inspection, therefore the total number of tested PMTs is slightly reduced. Left: HPK; Right: NNVT.}
	\label{fig:PMT:diameter}       
\end{figure}

When a 20-inch PMT is working in up to 44\,m water depth, there is an implosion risk for its evacuated glass bulb~\cite{SKPMTaccident,sk-PhysRevD.78.032002,sk-PhysRevD.81.092004}. To protect the PMTs from a possible implosion chain triggered by any accident or defective PMT, all PMTs will be capsuled using a protection cover design featuring an acrylic shell on its upper half and a stainless steel shell around its bottom half~\cite{JUNOPMTwang,JUNOdetector}. The protection cover was optimized considering its structural strength, a good transparency, dimension limitations from PMT photocathode coverage and installation clearance, a compatibility with pure water, and a low radioactivity~\cite{JUNOPMTwang}. The smaller diameter of PMT will affect the physics coverage and the fix structure of the installation, while the larger diameter is possible to introduce some conflicts among PMTs. In order to satisfy the needs for the protection covers and assure an installation tolerance on the $\sim 3$\,mm level, a precise knowledge of the dimensions of all individual PMTs is required. Therefore, the dimensions (dameters)s) of the PMTs' glass bulbs were measured for all individual PMTs. The typical diameter of HPK dynode PMTs was found to be 510\,mm in a range of 505 to 512\,mm; the typical diameter of NNVT MCP-PMTs was found to be 509\,mm in a range of 506 to 511\,mm  (Fig.\,\ref{fig:PMT:diameter}). Only four HPK PMTs are out of the tolerance range, which has negligible impact on both physics and installation.

\begin{figure}[!ht]
	\begin{subfigure}[c]{0.495\textwidth}
		\centering
		\includegraphics[width=\linewidth]{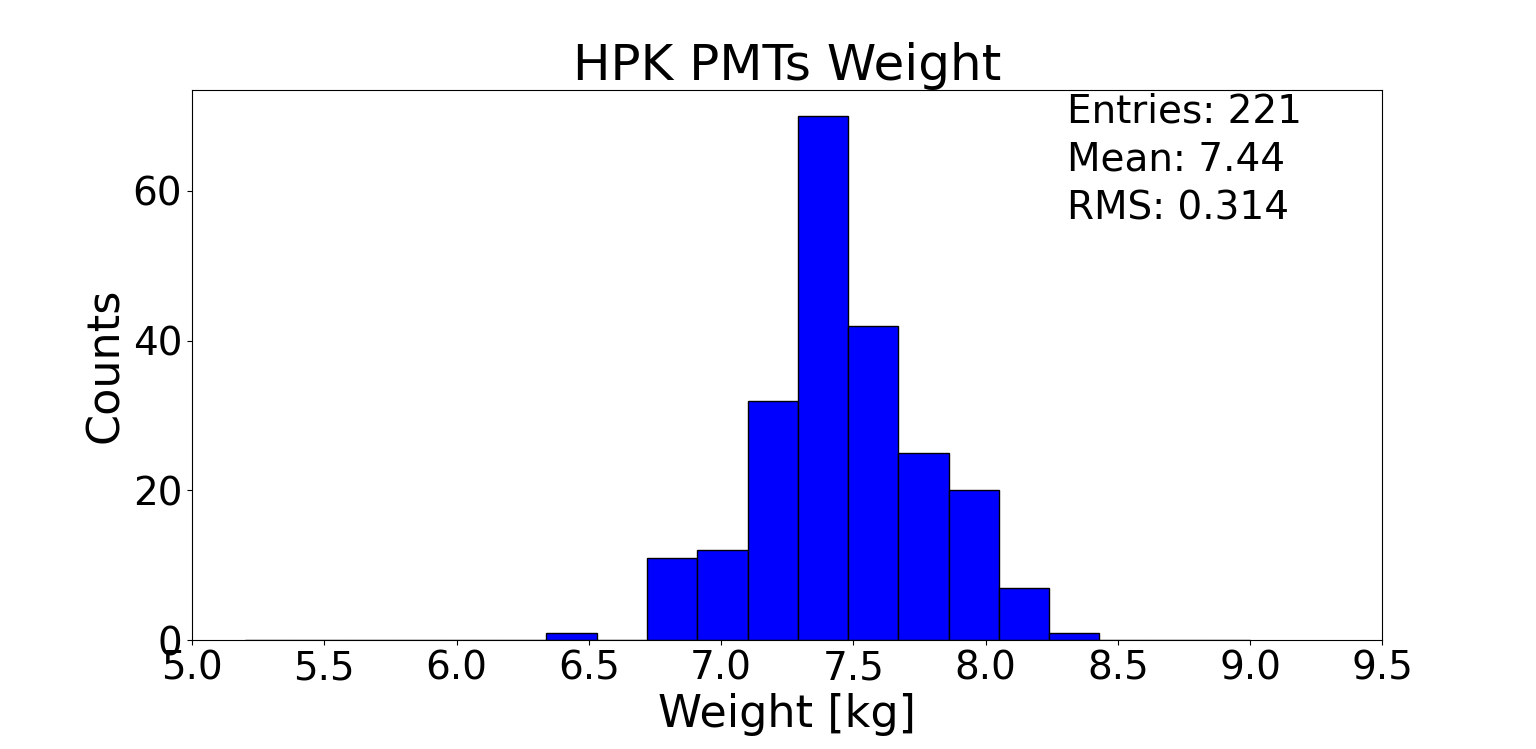}
	\end{subfigure}\hfill
	\begin{subfigure}[c]{0.495\textwidth}
		\centering
		\includegraphics[width=\linewidth]{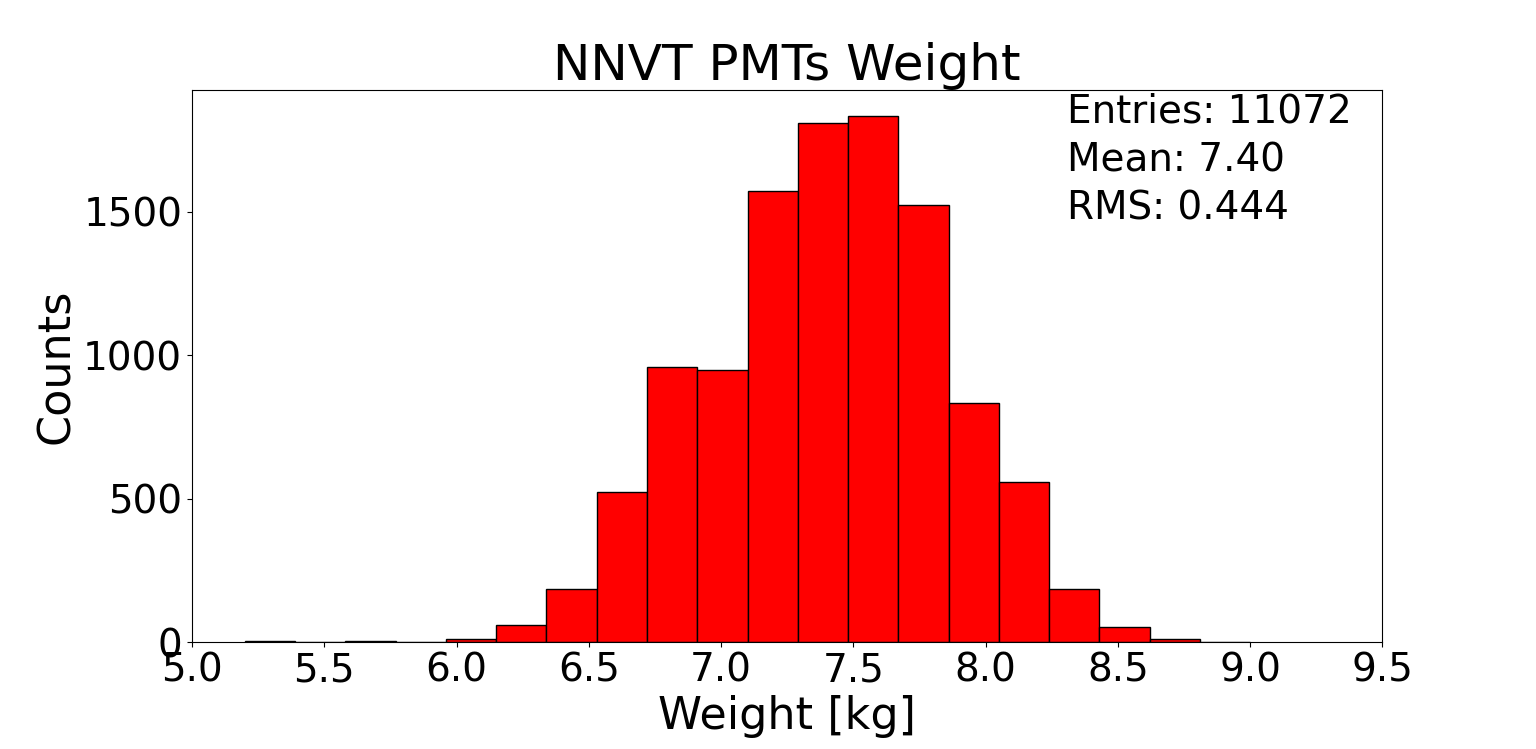}
	\end{subfigure}	
	\caption{Measured weight in kg of all checked 20-inch PMTs. Weighing was performed only
with a sub-sample of all PMTs. Left: HPK; right: NNVT.}
	\label{fig:PMT:weight}       
\end{figure}

The glass thickness\footnote{Only a small sample of the PMTs was measured directly for the thickness of its glass bulb using an ultrasonic thickness gauge to find possible weak points.} is assumed to be proportional to the PMT weight statistically, thus measuring the PMTs' weight helps to evaluate possible weaknesses, where the lighter PMTs can be cross-checked further in more detail\footnote{Some of the glass bulbs from NNVT featuring a very low weight are further tested under different water pressures, resulting in all PMTs satisfying the structural strength requirements for JUNO.}. According to the measurement on a sub-sample of all PMTs, the typical weight of an HPK dynode PMT is about 7.4\,kg in a range of 6.5 to 8.4\,kg; the typical weight of an NNVT MCP-PMT is about 7.5\,kg in a range of 5.2 to 9.1\,kg (see Fig.~\ref{fig:PMT:weight}).

A high-quality glass bulb is essential to avoid any dangerous defect related to its structural strength, especially for JUNO since all PMTs are closely packed. Even though both vendors have improved their glass bulb's strength via an optimized geometry as well as the identification of any small bubbles, cracks, weaknesses on the glass thickness, etc., a detailed visual check on every PMT is still required. Thus, all individual PMTs were visually checked for mechanical related issues (including cracks, bumps, scratches), bubbles and open bubbles, broken holes (glass notches) on the outside glass surface, defects at the sealing or KOVAR section\footnote{The KOVAR section of the NNVT MCP-PMTs includes the KOVAR plate for sealing the vacuum glass bulb and the output pins which feed the signals into the glass bulb, as well as the glass transition section,
see also the end part of the PMT neck with output pins on the left of Fig.\,\ref{fig:PMT:dimension}.} (including a gap between the KOVAR plate and the glass, dislocation of the KOVAR plate, broken glass or glass shedding at the transitional section of NNVT PMTs, and issues related to the socket), glass devitrification, gas leakage, and all other possible defects. Furthermore, internal impurities such as metal contaminants in the photocathode or glass were checked since they might affect the electronic features of the PMTs. The results of the visual inspection are listed in Tab.~\ref{tab:PMT:visual-rej} including several PMT defects. 


\begin{table}[!ht]
	\centering
	\caption{Main defects of all 20-inch PMTs rejected after the visual inspection.}
	\label{tab:PMT:visual-rej}
	\begin{tabular}{lcc}
		\hline\noalign{\smallskip}
		Defect & HPK & NNVT \\
	       & \# (\% to total) & \# (\% to total)\\
		\noalign{\smallskip}\hline\noalign{\smallskip}
		Impurity inside & 5 (0.09\%) & 304 (1.78\%) \\
		Mechanical related & 6 (0.11\%) & 86 (0.50\%)\\
		(Open) bubbles & 353 (6.52\%) & 38 (0.22\%)\\
		Broken glass hole & 0 (0.00\%) & 29 (0.17\%)\\
		Seal (KOVAR) section & 1 (0.02\%) & 61 (0.36\%) \\
		Gas leakage & 6 (0.11\%) & 9 (0.05\%)\\
		Glass devitrification & 0 (0.00\%) & 87 (0.51\%) \\
		others & 2 (0.04\%) & 29 (0.17\%)\\
		\noalign{\smallskip}\hline\noalign{\smallskip}
		Total & 373 (6.9\,\%) & 643 (3.8\,\%) \\
		\noalign{\smallskip}\hline
	\end{tabular}
\end{table}

The visual inspection is a primary step of the PMTs' acceptance testing procedure. All PMTs with good tags are transferred to the following performance checks. In total, 5,041 PMTs of 5,414 delivered HPK PMTs, and 16,357 PMTs from 17,000 delivered NNVT PMTs were selected as they satisfied the requirements of the visual inspection for JUNO.

\subsection{Working high voltage and gain}
\label{sec:2:gainhv}

\subsubsection{Charge spectrum calculation}
\label{sec:2:gain}

\begin{figure}[!ht]
	\begin{subfigure}[c]{0.49\textwidth}
		\centering
		\includegraphics[width=\linewidth]{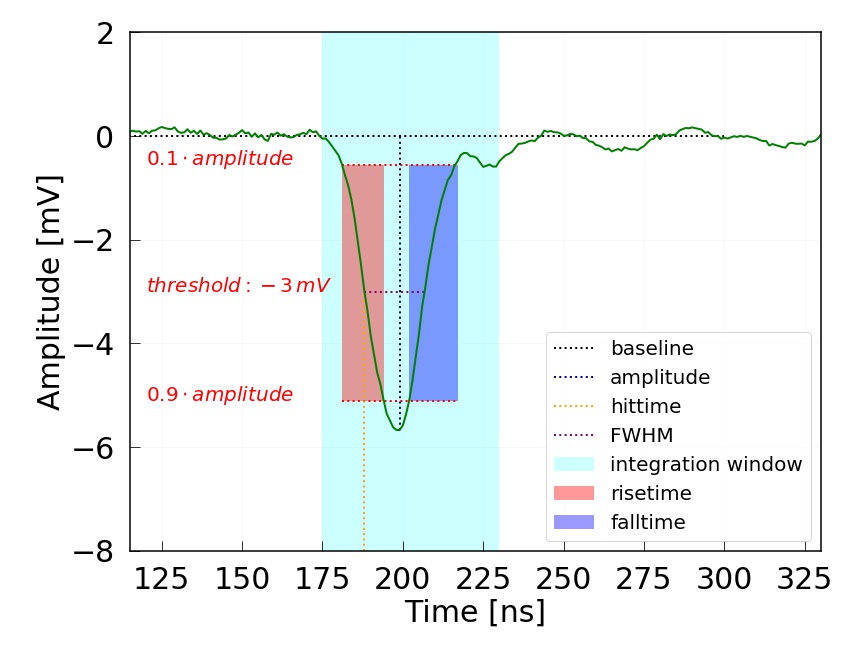}
		\caption{Example of a waveform of a single PMT as recorded by the ADC.}
		\label{fig:gain:wave}
	\end{subfigure}\hfill
	\begin{subfigure}[c]{0.49\textwidth}
		\centering
		\includegraphics[width=\linewidth]{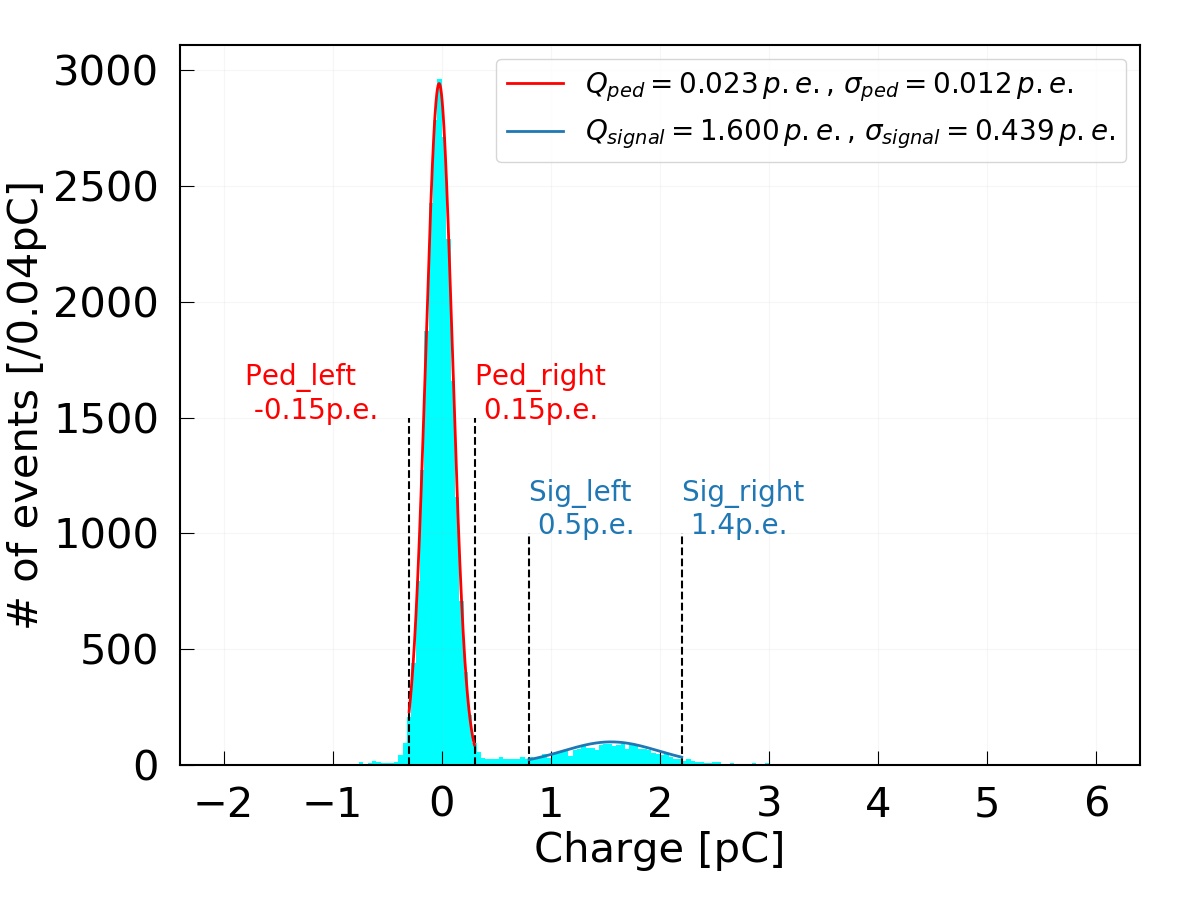}
		\caption{Typical charge distribution of a single PMT for \textmu$\sim0.1$\,p.e.}
		\label{fig:gain:charge}
	\end{subfigure}	
   \caption{Typical plots of an averaged waveform from a single PMT (left, also showing the typical definitions used for the following analysis) and the corresponding integrated charge spectrum from these waveforms (right). The SPE peak is located around 1.602\,pC in the case of a gain of $1\times10^7$.}
	\label{fig:PMT:waveformCharge}       
\end{figure}

Using the containers, all PMT waveforms are acquired by a 1\,GS/s (samples per second) Analog-Digital Converter (ADC, type CAEN V1742) in 512\,ns long window, synchronized to the auto-stabilized low intensity LED light pulses. To calculate the charges of the PMT pulses, each waveform is integrated in a time window [-20,+55]\,ns relative to the peak location of the signal (see Fig.\,\ref{fig:gain:wave}), considering the external trigger timing, an input impedance of 50\,$\Omega$ of the electronics~\cite{waveAnalysisHaiqiong}. An example waveform and charge spectrum in pC is shown in Fig.\,\ref{fig:PMT:waveformCharge}. The single photo-electron (SPE) spectrum can be de-convoluted with Poisson statistics as in Eq.\,\ref{Equ:Poisson}, where \textmu~is the mean p.e.~number received by the first dynode or MCP, and $P(n)$ is the probability to have $n$ p.e.s. When the value of \textmu~is around 0.1, the probability $P(1)$ of 1\,p.e.~events is $0.090$ (and $P(2) \sim 0.005$ for 2\,p.e.~events etc.), thus single p.e. events will dominate the signal distribution in this case. The resulting distribution or SPE response spectrum (SPR) can be used in the following analyses of e.g.~the gain (using a comparable and constant light intensity of \textmu~$\sim0.1$ for all LEDs) or the PDE (using individual light intensities and a calibration function for a set DAC of each LED, which is calibrated using a small sample of 20-inch reference PMTs\footnote{These calibration PMTs are not the same PMTs as the ones introduced as monitoring PMTs in sec.\,\ref{sec:3:container}.} with known PDE).

\begin{equation}
P(n)=\frac{e^{-\mu}\mu^n}{n!}
\label{Equ:Poisson}
\end{equation}

In the resulting charge spectrum (see Fig.\,\ref{fig:gain:charge}), the pedestal represents all trigger/signal events with no light; it will be fitted in the range [-0.15,+0.15]\,p.e. ($Q_\textit{ped}$ and $\sigma_\textit{ped}$). The second peak right of the pedestal peak corresponds to single p.e.~events, located around 1.602\,pC for a gain of $1\times10^{7}$ (as assumed in this example). This SPE peak will be fitted in the range [0.5,1.4]\,p.e.\footnote{The range is selected to enable a unique definition for all the PMTs considering also the long tail effect of MCP-PMTs~\cite{JUNOPMTgain}.} ($Q_{sig}$ and $\sigma_{sig}$). All the fitting mentioned here is done with a Gaussian function.

\subsubsection{Gain and HV determination}
\label{sec:2:hv}

To calculate the PMT gain $G$ at a set voltage, the information from the charge spectrum as shown before in Fig.\,\ref{fig:gain:charge} can be used as shown in Eq.\,\ref{Equ:gain}:

\begin{equation}
G = \frac{Q_\textit{sig}-Q_\textit{ped}}{e}
\label{Equ:gain}
\end{equation}

\begin{figure}[!ht]
	\begin{subfigure}[c]{0.47\textwidth}
		\centering
		\includegraphics[width=\linewidth]{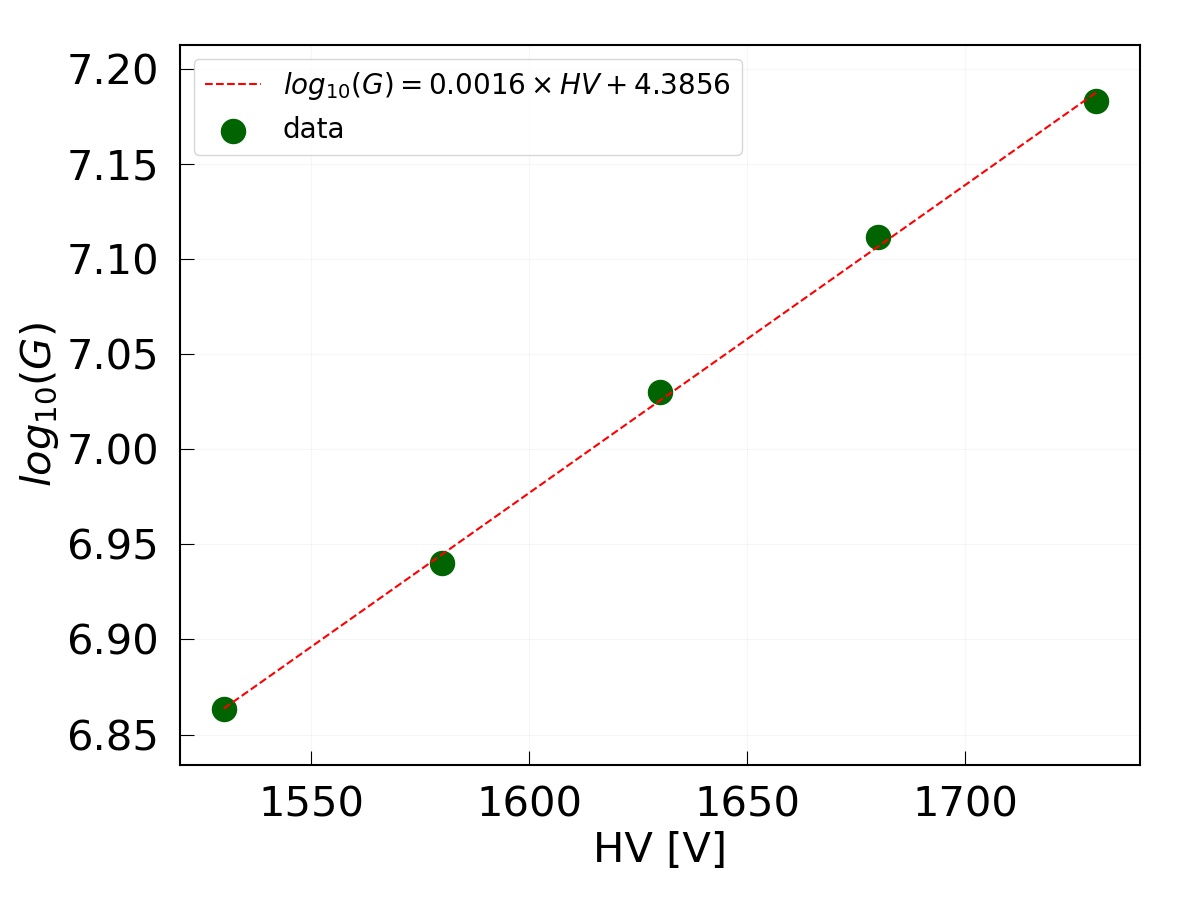}
		\caption{Example plot and fitting of the HV vs. measured gain, with corresponding fit results for the data.}
		\label{fig:pmt:hvmodel}
	\end{subfigure}\hfill
	\begin{subfigure}[c]{0.46\textwidth}
		\centering
		\includegraphics[width=\linewidth]{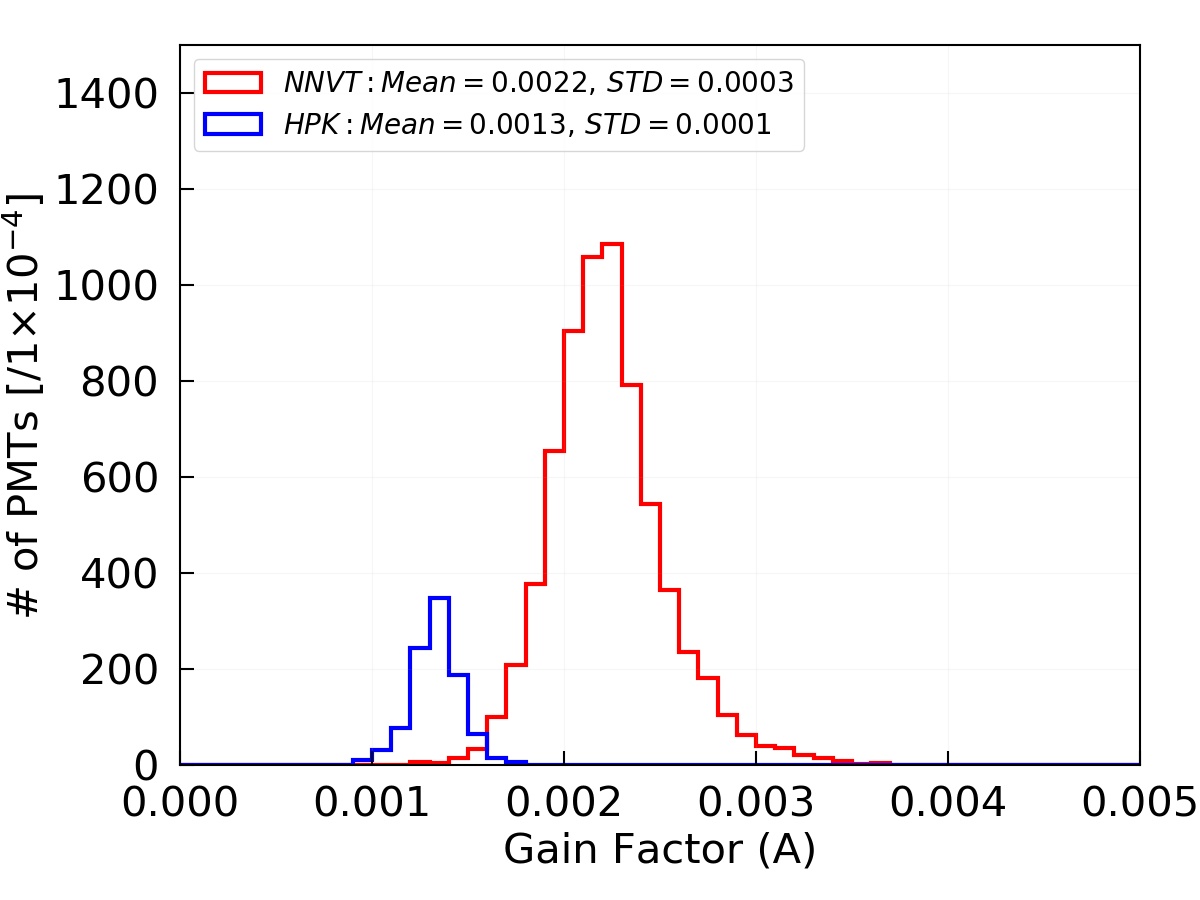}
		\caption{Gain factor $A$ for the gain of the qualified PMTs. Red: NNVT; blue: HPK.}
		\label{fig:gainfactor}
	\end{subfigure}	
   \caption{Example plot for a HV vs. gain fit (left, for a single PMT) and the distribution of the fitted gain factor $A$ (as introduced in Eq.\,\ref{Equ:gainhvFiting}) of all qualified PMTs (right).}
	\label{fig:PMT:HVgain}       
\end{figure}

The voltage to be applied for a gain of $1 \times 10^7$ is determined from a series of measurements with different high voltages (HV), performed between -150\,V to +150\,V (executed in 50\,V steps) relative to the HV value proposed by the manufacturer. 
The optimal working HV for a $1\times10^{7}$ gain is then identified using a fit between applied HV and measured gain, as shown in Fig.\,\ref{fig:pmt:hvmodel}. The used fitting function is defined in Eq.\,\ref{Equ:gainhvFiting} with $G$ the measured gain, which is consistent with the applied fitting function used for dynode PMTs in the selected range~\cite{HamManual}. Additional information about the HV determination can be found in~\cite{JUNOPMTcontainer}. 
A discussion about different methods for the gain determination and their effects can be found also in separate papers~\cite{PMTgainmodel1994,PMTgainmodel2017,JUNOPMTgain}.

\begin{equation}
\log_{10}\,(G) = A~\times~\textit{HV}+B
\label{Equ:gainhvFiting}
\end{equation}

The fitted correlation factor $A$ between HV and gain $G$ of all the measured PMTs is shown in Fig.\,\ref{fig:gainfactor}. The mean value of $A$ is $0.0013$ for HPK and $0.0022$ for NNVT, which means the working gain will be doubled when increasing the HV by additional $\sim$\,230\,V for HPK PMTs and $\sim$\,140\,V for NNVT PMTs around this working gain range.

With having tested the monitoring PMTs hundreds of times during the whole campaign, the standard deviation of the determined working HV repeatability was found to be about 1\,\% (15$\sim$18\,V) for both HPK and NNVT PMTs on average, and the standard deviation of the determined $1\times 10^7$ gain is $2$\,\% for HPK dynode PMTs and $3$\,\% for NNVT MCP-PMTs, which indicates the uncertainty in determining the working HV and gain with the container system.

\begin{figure}[!ht]
	\begin{subfigure}[c]{0.49\textwidth}
		\centering
		\includegraphics[width=\linewidth]{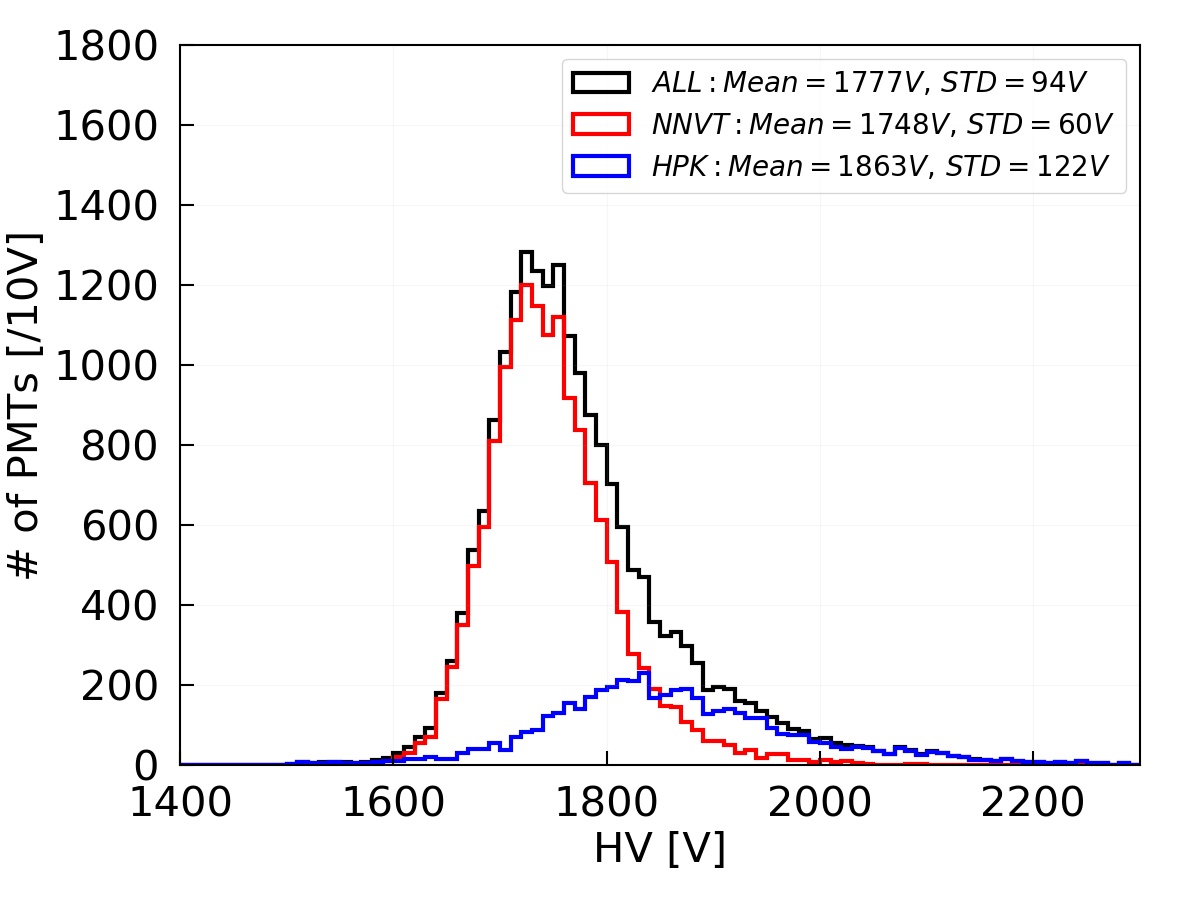}
		\caption{Determined HV of all qualified bare PMTs for a gain of $1\times10^{7}$.}
		\label{fig:hv}
	\end{subfigure}
	\begin{subfigure}[c]{0.49\textwidth}
		\centering
		\includegraphics[width=\linewidth]{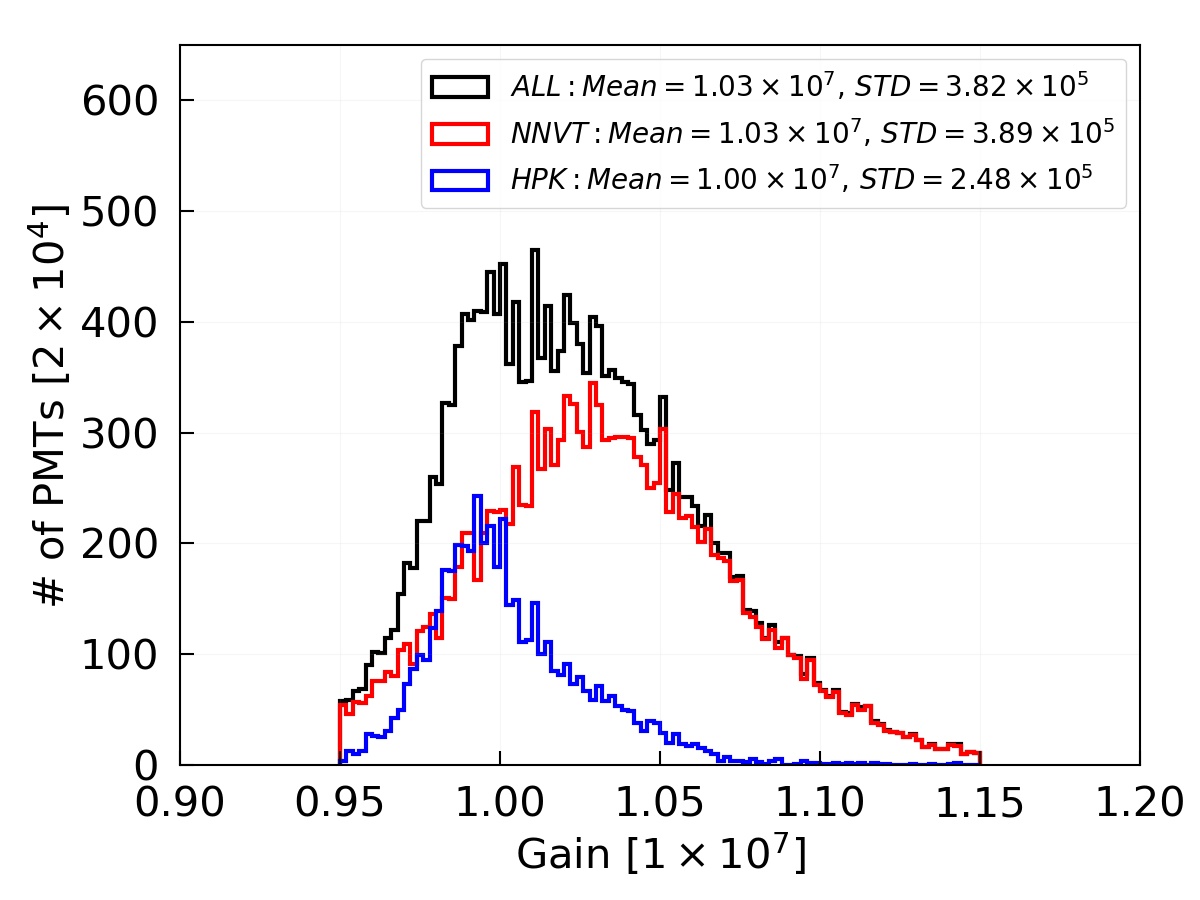}
		\caption{Measured gain distribution of all qualified PMTs during the PMT characterization.}
		\label{fig:gain}
	\end{subfigure}	
   \caption{Determined HV and gain of all qualified PMTs. Measured gains not contained in $[0.95,1.15] \times 10^7$ lead to a retest of the corresponding PMT. Black: all PMTs; red: NNVT; blue: HPK.}
	\label{fig:PMT:HVgain:distribution}       
\end{figure}

\begin{figure}[htb]
	\begin{subfigure}[c]{0.495\textwidth}
		\centering
		\includegraphics[width=\linewidth]{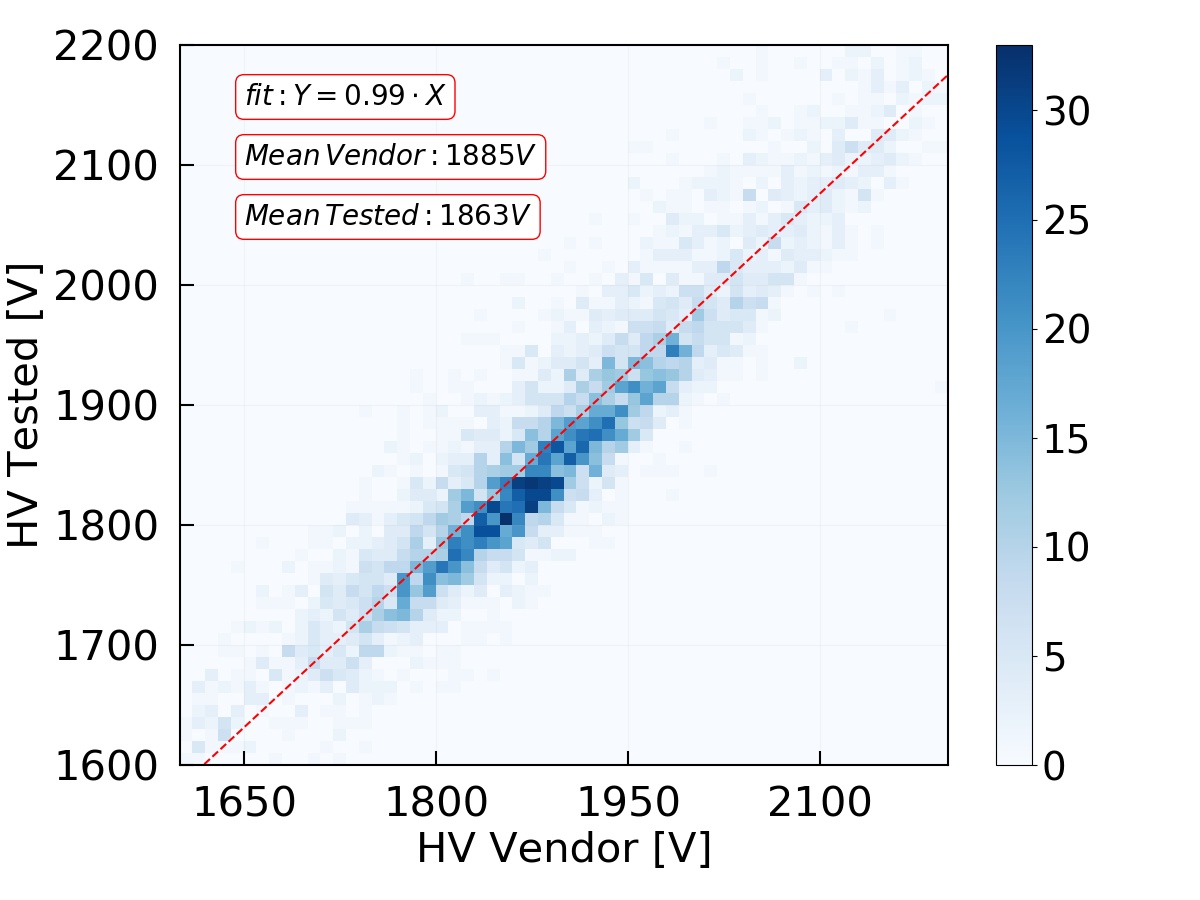}
		\label{fig:hvvendorvstested2}
	\end{subfigure}	
	\begin{subfigure}[c]{0.495\textwidth}
		\centering
		\includegraphics[width=\linewidth]{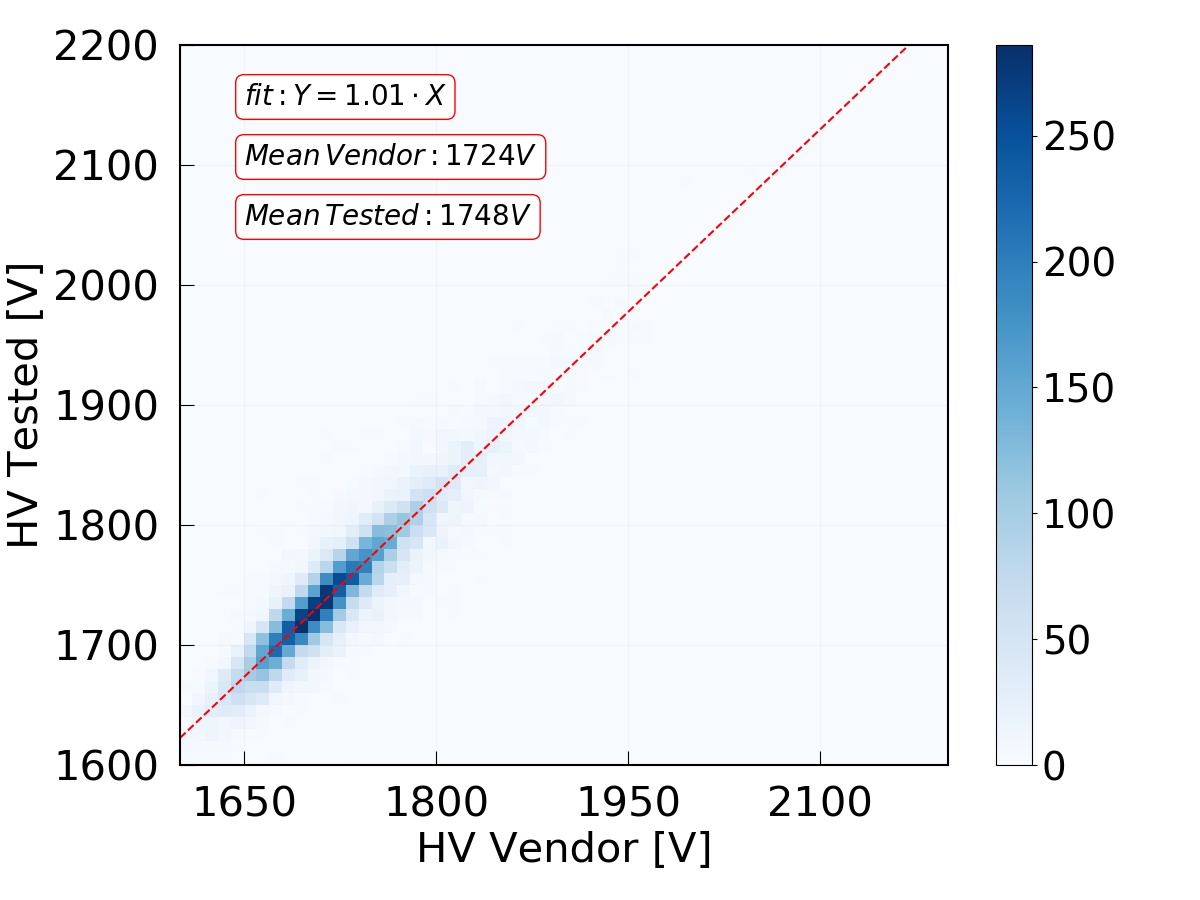}
		\label{fig:hvvendorvstested1}
	\end{subfigure}
	\caption{Comparison between the working HV for a gain of $1\times10^{7}$ of all qualified bare PMTs, measured by JUNO and the suggested HV values from the manufacturers. Left: HPK; right: NNVT.}
	\label{fig:hvvendorvstested}       
\end{figure}

The determined working HV of all qualified bare PMTs producing a gain of $1 \times 10^7$ is shown in Fig.\,\ref{fig:hv}; the determined mean HV is about $1863$\,V for HPK, and $1748$\,V for NNVT PMTs. The determined HV for a $1\times10^{7}$ gain is a slightly biased with respect to the values suggested by the manufacturers of both PMT types (the working HV is $20$\,V higher for HPK PMTs and $20$\,V lower for NNVT PMTs on average compared to the supply voltages suggested by the manufacturers), even though they show a very good correlation (the fitted slope is 0.99 for HPK PMTs and 1.01 for NNVT PMTs), as illustrated in Fig.\,\ref{fig:hvvendorvstested}. This difference could arise from different testing methods, cable attenuations, used HV dividers and electronics. However, all PMTs are reaching the requirements for the working HV at a $1\times10^{7}$ gain, i.e. the NNVT PMTs show a much lower working HV value than the allowed maximum value as fixed in Tab.\,\ref{tab:PMT:criteria}.

The final gain applied in the container system during the PMT characterization is cross-checked once more as a part of the regular analysis. The results show that the set gains do not exactly match a value of $1\times10^{7}$ in all cases, but follow a distribution with a width of 2-4\,\% (which is consistent to the estimation about the gain determination uncertainty from the monitoring PMTs) as well as an offset of $3\,\%$ for the NNVT MCP-PMTs towards a higher gain, as presented in Fig.\,\ref{fig:gain}. This is mainly introduced by the used HV-gain model in the data analysis, which is more applicable for the dynode PMTs than for the MCP-PMTs.

\subsection{Single photo-electron features}
\label{sec:2:spe}

With the determined HV for a $1\times10^7$ gain, approximately 20,000 waveforms are acquired for each PMT in the SPE mode (with \textmu$\simeq0.1$\,p.e.) using an external trigger synchronizing the LED pulses to the data acquisition. The recorded waveforms are analyzed for their characteristic features in order to extract the SPE spectrum and a number of important parameters characterizing the PMT's performance. These parameters and the resulting distribution for the full data set will be presented in following sections.

\subsubsection{Signal-to-noise ratio (S/N)}
\label{sec:3:sn}

\begin{figure}[!htb]
	\begin{subfigure}[c]{0.495\textwidth}
		\centering
		\includegraphics[width=\linewidth]{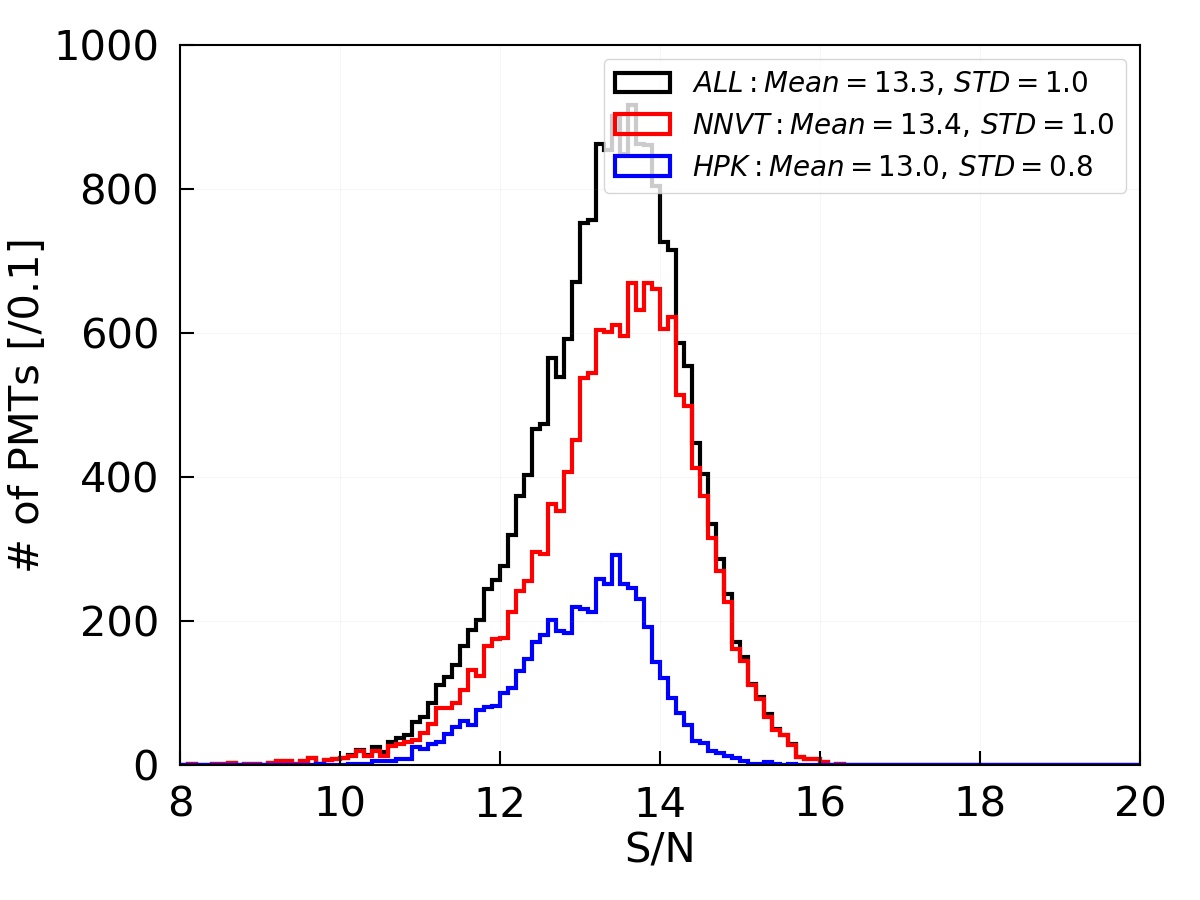}
        \caption{S/N ratio of the container systems.}
		\label{fig:sn}
	\end{subfigure}	
	\begin{subfigure}[c]{0.495\textwidth}
		\centering
		\includegraphics[width=\linewidth]{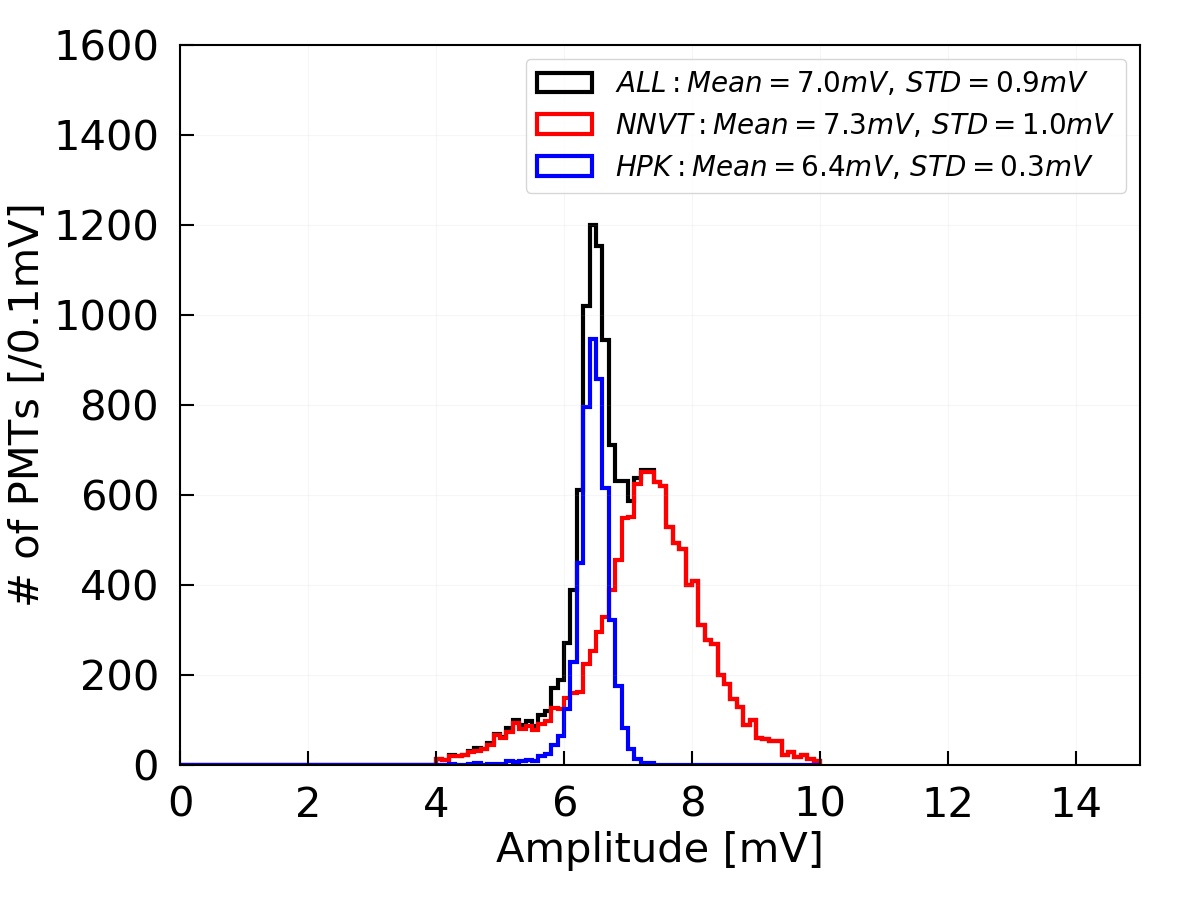}
		\caption{SPE amplitude of container systems.}
		\label{fig:amplitude}
	\end{subfigure}
	\caption{The S/N ratio of all the measurements with qualified PMTs (left). The SPE amplitude of all qualified PMTs (right). Black: all PMTs; red: NNVT; blue: HPK. Note: the amplitude is only selected within [4,10]\,mV.}
	\label{fig:PMT:snAmp}       
\end{figure}

The signal-to-noise (S/N) ratio in the container measurements is defined as a control parameter to ensure a valid measurement of the charge resolution and peak-to-valley ratio, as well as for a reliable threshold setting in the following PDE and DCR measurements. The S/N ratio is defined in Eq.\,\ref{equ:sn}, where $\sigma_{\textit{ped}}$, $Q_{\textit{sig}}$ and $Q_{\textit{ped}}$ all are calculated as described in sec.\,\ref{sec:2:gain}. The S/N ratio is required to be larger than 10 in all individual SPE measurements, corresponding to a noise level smaller than 0.1~p.e. At the container systems, the mean S/N ratio for both HPK and NNVT PMTs is around around 13, as depicted in Fig.\,\ref{fig:sn}. \\
For comparison, the scanning station has reached a S/N ratio of $20$ in its measurements due to the use of an additional amplifier (see sec.~\ref{sec:3:station}). This also allows high quality cross-checks of the PDE and DCR measurement with the scanning station.

\begin{equation}
S/N=\frac{Q_{\textit{sig}}-Q_{\textit{ped}}}{\sigma_{\textit{ped}}}
\label{equ:sn}
\end{equation}

\subsubsection{Amplitude of single photo-electron}
\label{sec:3:amplitude}

In case of a $1\times10^7$ working gain, the pulse amplitude of SPE events is picked out from the recorded waveforms as described in Fig.\,\ref{fig:gain:wave}, and corresponds to the minimum value of the negative SPE pulse form. Typical values for each PMT are shown in Fig.\,\ref{fig:amplitude}; HPK dynode PMTs have a slightly smaller SPE amplitude of about 6.5\,mV compared to about 7.5\,mV for the NNVT MCP-PMT, whereas HPK PMTs have a narrow distribution (with a standard deviation (STD) of $0.4$\,mV) compared to NNVT PMTs (STD of $2.5$\,mV). The different mean values of SPE amplitudes of the two PMT types are mainly influenced by the different pulse shapes related to the PMTs itself and the designs of the HV dividers.

\subsubsection{Rise-time, fall-time and full width at half maximum of SPE pulses}
\label{sec:3:risefall}

Rise-time, fall-time and full width at half maximum (FWHM) of a typical pulse can be extracted from the full sample of recorded waveforms of each PMT, following the definition illustrated in Fig.\,\ref{fig:gain:wave}. As commonly known, the rise-time and fall-time are more related to the PMT itself, i.e. using the JUNO optimized version of the HV divider for the NNVT tubes, the rise-time was observed to be slower than in the case of using an early design version of the HV divider (which was used also for early testings).
All qualified PMTs were tested in the containers with different
pluggable JUNO optimized HV dividers~\cite{JUNOPMTsignalover,JUNOPMTsignalopt,JUNOPMTflasher}. Their SPE pulses show typical values for rise-time, fall-time, and FWHM of $6.9$\,ns, $10.2$\,ns and $11.6$\,ns for HPK PMTs, and $4.9$\,ns, $17.3$\,ns, and $7.9$\,ns for NNVT PMTs, respectively.

\subsubsection{Charge resolution}
\label{sec:3:resolution}

\begin{figure}[!htb]
	\begin{subfigure}[c]{0.495\textwidth}
		\centering
		\includegraphics[width=\linewidth]{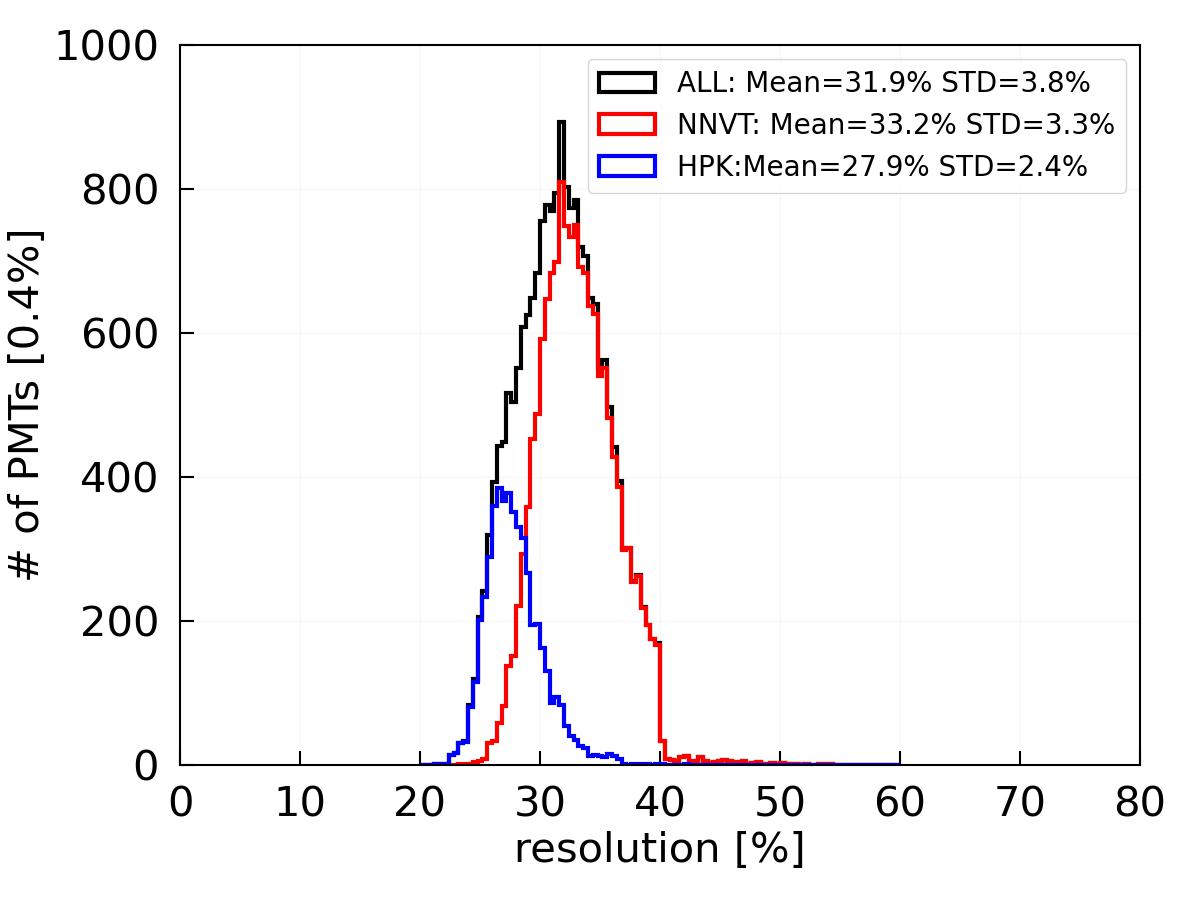}
		\caption{SPE charge resolution.}
		\label{fig:resolution}
	\end{subfigure}\hfill
	\begin{subfigure}[c]{0.495\textwidth}
		\centering
		\includegraphics[width=\linewidth]{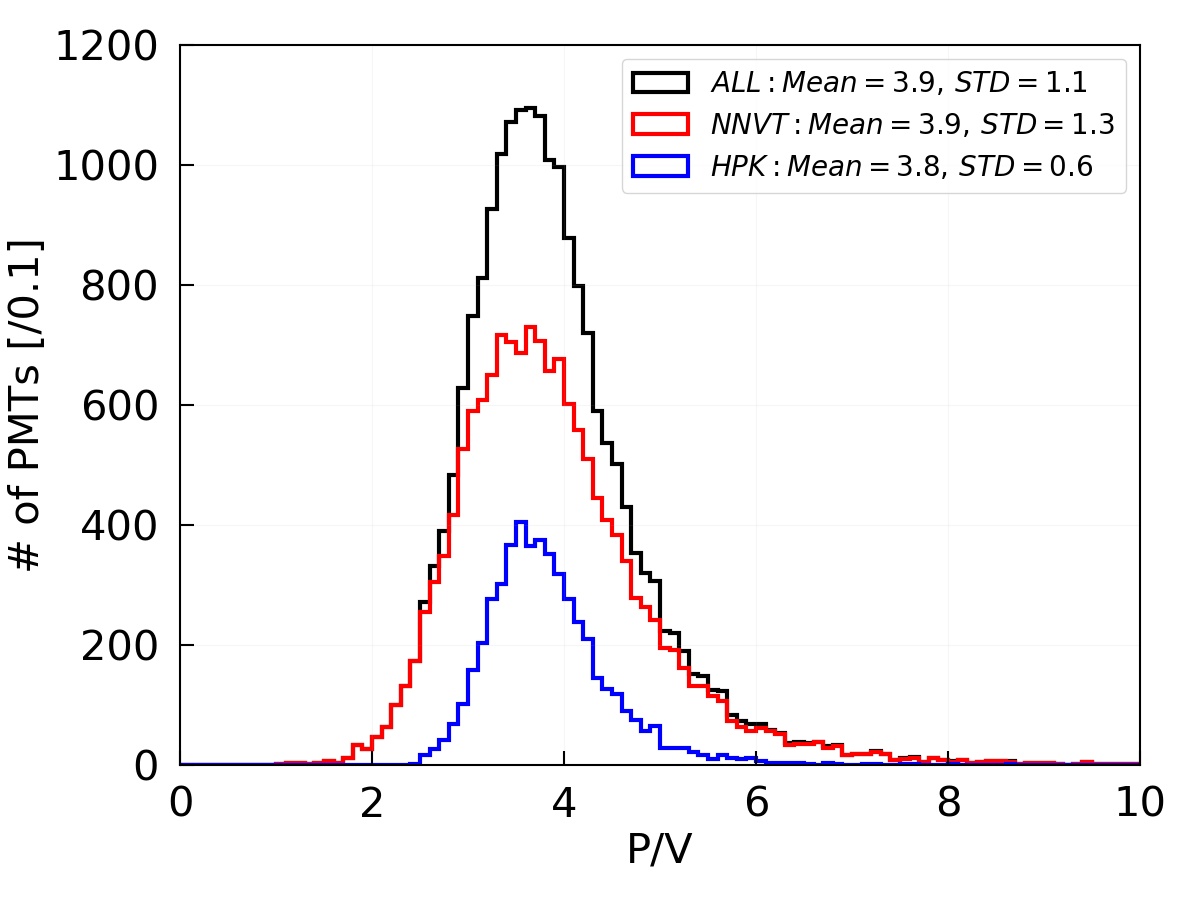}
	    \caption{Measured P/V.}
		\label{fig:pv}
	\end{subfigure}
	\caption{Left: SPE charge resolution of all qualified PMTs; right: measured P/V of all qualified PMTs. Black: all PMTs; red: NNVT; blue: HPK. Please note that the requirement of the charge resolution is relaxed for a few NNVT tubes for the JUNO veto detector, which is higher than 40\%.}
	\label{fig:pv:res}       
\end{figure}

The charge resolution of the PMTs' SPE response is calculated using Eq.\,\ref{equ:resolution}, where the $\sigma_{\textit{sig}}$ originates from the $\sigma$ of the Gaussian fitting of the PMT's SPE charge spectrum as introduced in sec.\,\ref{sec:2:gain}. The distribution of the SPE charge resolution for the full PMT sample as measured by the container system is shown in Fig.\,\ref{fig:resolution}, where the mean values are around 27.9\,$\%$ for the HPK PMTs and around 33.2\,$\%$ for the NNVT PMTs. However, the characteristic PMT charge response does not necessarily follow a Gaussian distribution~\cite{AugerPMT}, especially in case of the NNVT PMTs which normally show a long tail in their charge spectrum~\cite{JUNOPMTgain}. Nevertheless, the use of a Gaussian fit is justified since the parameter is used only as a relative control for charge response spread. The SPE charge resolution is required to be less than 40\% for all individual PMTs of the JUNO CD, while it releases for a few NNVT tubes for the JUNO VETO detector.

\begin{equation}
\textit{Res.}\,(\%)=\frac{\sigma_{\textit{sig}}}{Q_{\textit{sig}}-Q_{\textit{ped}}}\times100
\label{equ:resolution}
\end{equation}

\subsubsection{Peak-to-valley ratio (P/V)}
\label{sec:3:pv}

The peak-to-valley ratio (P/V) is a parameter to estimate the PMT's ability to distinguish between noise and photo-electron signals which can be calculated using Eq.\,\ref{equ:pv} and is based on the measured SPE charge spectrum (see Fig.\,\ref{fig:gain:charge}), where the valley $N_v$ is defined as the level of the local minimum between pedestal and SPE peak and is fitted by a parabolic function, while the peak $N_p$ describes the level of the SPE peak picked out from their Gaussian fit. The measurement results from the container systems of all qualified PMTs are shown on the right of Fig.\,\ref{fig:pv}, where its typical values are around 3.8 for HPK PMTs and 3.9 for NNVT PMTs. The mean of the P/V of both PMT types satisfies the defined acceptance criteria (larger than 3 of HPK PMT and 3.5 of NNVT PMT, respectively), while some of individual PMTs show much smaller values ($< 2.5$), especially the NNVT PMTs.

\begin{equation}
P/V=\frac{N_p}{N_v}
\label{equ:pv}
\end{equation}

\subsubsection{Excess noise factor (ENF) and gain excess normal distribution factor (gENF)}
\label{sec:2:genf}

\begin{figure}[htb]
	\begin{subfigure}[c]{0.495\textwidth}
		\centering
		\includegraphics[width=\linewidth]{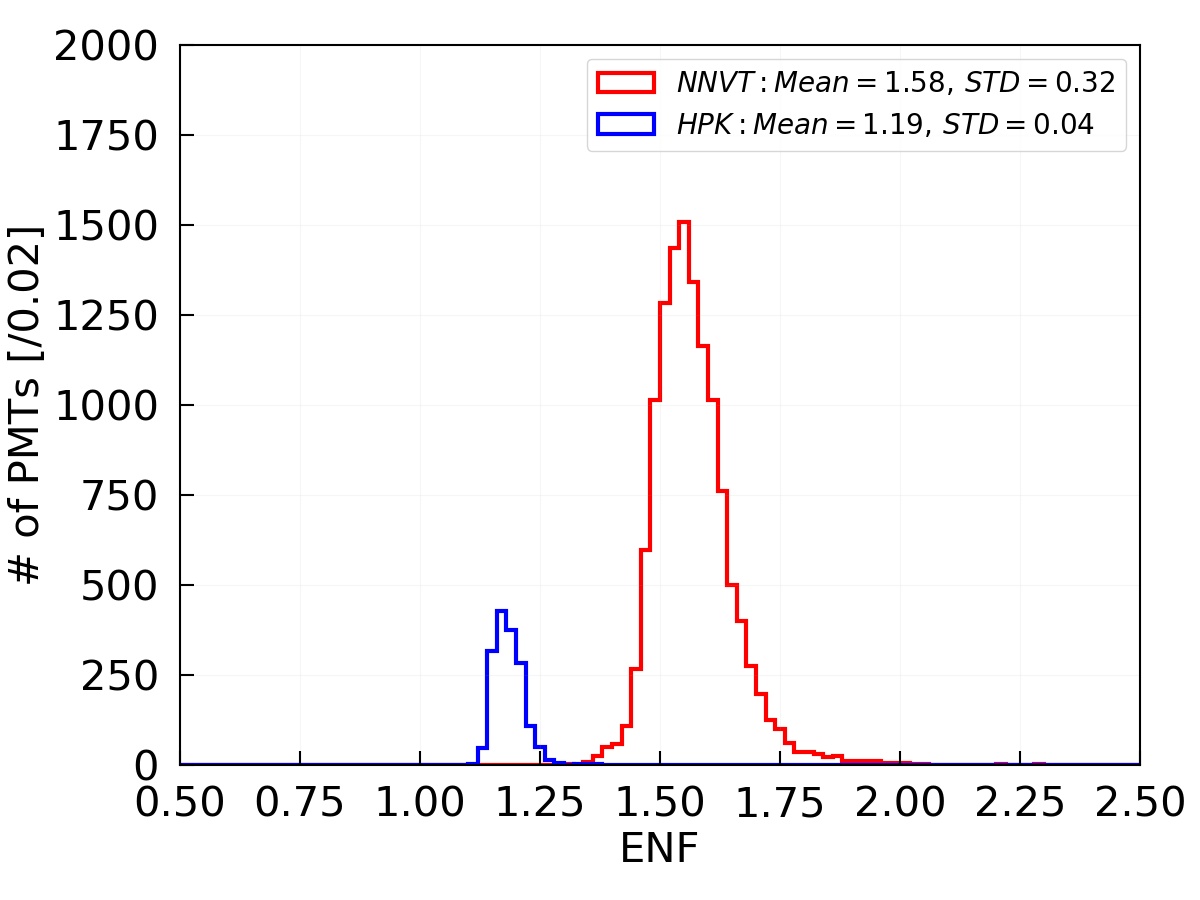}
		\caption{ENF distribution}
		\label{fig:enf}
	\end{subfigure}\hfill
	\begin{subfigure}[c]{0.495\textwidth}
		\centering
		\includegraphics[width=\linewidth]{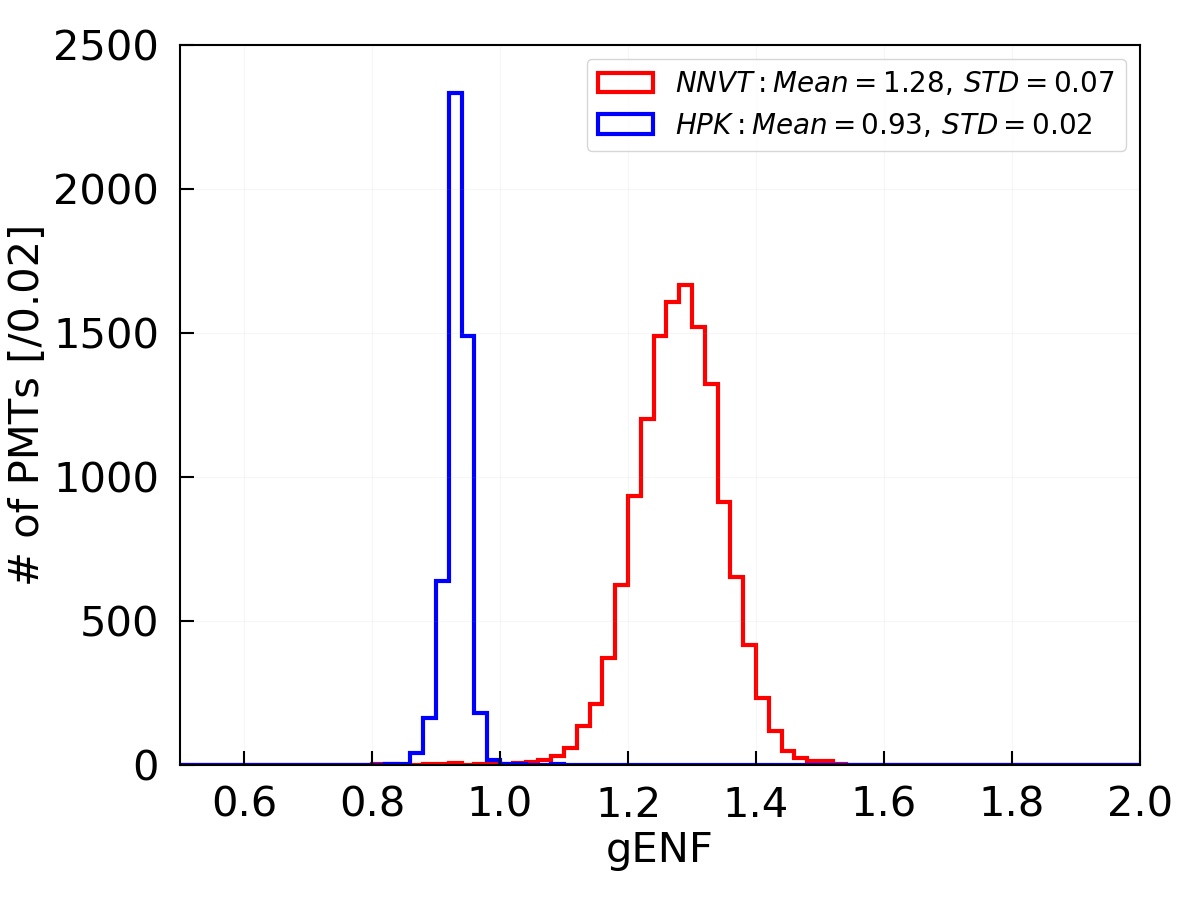}
	    \caption{gENF distribution}
		\label{fig:PMT:gENF}
	\end{subfigure}
	\caption{Left: ENF distribution of all qualified bare PMTs. Right: gENF distribution of all qualified bare PMTs. Red: NNVT; blue: HPK.}
	\label{fig:enf:genf}       
\end{figure}

A previous study~\cite{AugerPMT} also mentioned an excess noise factor (ENF), defined in Eq.\,\ref{equ:enf} in order to effectively represent the measured charge spread. The larger the ENF under a similar light intensity, the broader the distribution of output signals will be for the same input\footnote{If the input is following a Poisson distribution, and the smearing is following a Gaussian distribution (in an ideal case), it is that $\textit{ENF} = 1 + \textit{Resolution}_{\textit{ideal}}^{2}$.}. $N_{\textit{pe}}$, $\sigma_{\textit{spec}}$ and $M_{\textit{spec}}$ in Eq.\,\ref{equ:enf} are the measured number of p.e.s (mean number of p.e. from a measurement as \textmu~in Eq.\,\ref{Equ:Poisson}), the standard deviation and the mean of the whole charge spectrum distribution including the pedestal\footnote{The pedestal's mean value is shifted to zero.}, respectively. The ENF distribution of all qualified PMTs is shown in Fig.\,\ref{fig:enf}, where the typical values are 1.17 for HPK PMTs and 1.55 for NNVT PMTs. The distribution indicates that the variation of output signals of NNVT PMTs for the same input is much larger than in case of HPK PMTs.

\begin{equation}
\textit{ENF}=N_{\textit{pe}}~\times~\left(\frac{\sigma_{\textit{spec}}}{M_{\textit{spec}}}\right)^2
\label{equ:enf}
\end{equation}

The gain excess normal distribution factor (gENF), as defined in Eq.\,\ref{equ:gENF} and discussed in~\cite{PMTgainmodel2017,JUNOPMTgain}, is further characterizing the mismatch between the determined charge in p.e.~at a gain $G$ (determined from the SPE peak based on a normal distribution as in Eq.\,\ref{Equ:gain}) and the expected charge based on the Poisson distribution (as in Eq.\,\ref{Equ:Poisson}), where $Q_{\textit{pC}}$, $Q_{\textit{pe}}^{\textit{G}}$ and $Q_{\textit{pe}}^{\textit{Poisson}}$ are the measured charge\footnote{With $Q_{\textit{pe}}^{\textit{Poisson}} = Q_{\textit{pe}}^{G_{\textit{avg}}}$, compare \cite{JUNOPMTgain}.} of the PMT pulses in pC, the charge in p.e.~expected for a gain $G$ and the individually expected charge of a PMT pulse in p.e.~based on Poisson statistics. $Q_{\textit{pe}}^{\textit{Poisson}}$ is only defined with a pulsed light source (such as the used LED) taken under a synchronized external trigger. Using the gENF factor, the expected charge in p.e.~of a PMT measurement can be calculated as defined Eq.\,\ref{equ:gENFcharge} for a spectrum on average or for a single signal.

\begin{equation}
\textit{gENF}=\frac{Q_{\textit{pe}}^{G}}{Q_{\textit{pe}}^{\textit{Poisson}}}
    =\frac{Q_{\textit{pC}}/G}{Q_{\textit{pe}}^{\textit{Poisson}}}
\label{equ:gENF}
\end{equation}

\begin{equation}
Q_{\textit{pe}}^{\textit{exp}}=\frac{1}{\textit{gENF}}~\times~Q_{\textit{pe}}^{G}
\label{equ:gENFcharge}
\end{equation}

With the container systems, the gENF factor was measured and calculated for all qualified PMTs as shown in Fig.\,\ref{fig:PMT:gENF}. One can find that the gENF factor of the HPK dynode PMTs is smaller than 1 ($0.93$), while that of NNVT MCP-PMTs is bigger than 1 ($1.28$). The expected charge in p.e.~is little larger than the calculated charge at a gain $G$ for HPK dynode PMTs, while it is smaller for NNVT MCP-PMTs. This effect is caused mainly by the imperfect amplification process and correlated noise of the dynode PMT on one hand, and by the long tail of the charge spectrum which is typical for MCP-PMTs on the other hand (see the SPE amplitude spectra in Sec.\,\ref{sec:3:dcr:threshold}); those effects are discussed in more details also in~\cite{JUNOPMTgain}.

\subsection{Photon detection efficiency (PDE)}
\label{sec:2:pde}

Instead of setting a requirement on the QE of the PMTs, JUNO uses the photon detection efficiency (PDE) to measure the sensitivity of the PMTs. Here, the PDE of the 20-inch PMTs is defined as the photon detection efficiency for light impinging vertically on the PMT surface averaged over the whole photocathode area with a surface area weight. This can be measured directly by the scanning station by calculating a surface weighted average of all its measurement points across the PMT area\footnote{The average PDE here is calculated as $\textit{PDE} (\lambda) = \sum_{\textit{spot}\_\textit{i}} \textit{PDE}_{\textit{spot}\_\textit{i}} (\lambda, \vec{r}) \times \textit{Weight}_{\textit{spot}\_\textit{i}}$.} considering the geometry differences of HPK and NNVT PMTs, where a reference PMT (HPK R1355 as mentioned in sec.\,\ref{sec:3:station}) is used for normalization. Following the QE measurements for samples of both types of 20-inch PMTs in~\cite{largePMTlei,largePMTxia,waveAnalysisHaiqiong,MCPPMTTTSsen,MCPPMT2016,MCPPMT2018}, a special comparison study on the relative collection efficiency among PMTs with different collection structure~\cite{PMTrelativeCE} has concluded that the reference PMT's QE can be dealt as its PDE, assuming a $100\,\%$ collection efficiency in case of a 5\,mm light spot impinging on the PMT center. Furthermore, the measured PDE of the container systems is normalized drawer by drawer to the scanning station using a set of 20-inch PMTs that have been measured in both the container and scanning station systems. This set of 20-inch PMTs with known PDE was also used to individually calibrate the light intensities in the drawers to enable a reliable PDE measurement in the first place.

\subsubsection{PDE measurement}
\label{sec:3:pde}

The PDE measurements of the containers and scanning stations are realized by employing the photon counting method on the charge spectra obtained with the stabilized pulsed LEDs operated at low intensity, where the effects of threshold and dark count rate are checked and considered as uncertainty.
Considering the observed noise level and charge resolution, a threshold of around 0.25\,p.e.~was chosen to separate the pedestal from photon induced events. The light intensity used in both container and scanning station system is a low multi-photon level (with $1-2$\,p.e.s observed by the PMTs) to minimize the uncertainty from system and statistics as suggested in~\cite{PMTintensity}. Based on the repeatability of the daily measurements of the monitoring PMTs (see sec.\,\ref{sec:3:container}), a measurement uncertainty of better than 1\,\% of the absolute PDE ($3$\,\% relative PDE)\footnote{Considering the later discussed light source aging in sec.\ref{sec:3:aging:PDEcorrection}, the acceptable uncertainty is enlarged to $1.5$\,\% abs. ($4-5$\,\% relative) for the directly measured PDE values 
for both containers before applying any aging corrections.} is achieved for both containers~\cite{JUNOPMTcontainer} as well as for both scanning stations (see sec.\,\ref{sec:3:station}).

\begin{figure}[!ht]
	\begin{subfigure}[c]{0.495\textwidth}
		\centering
		\includegraphics[width=\linewidth]{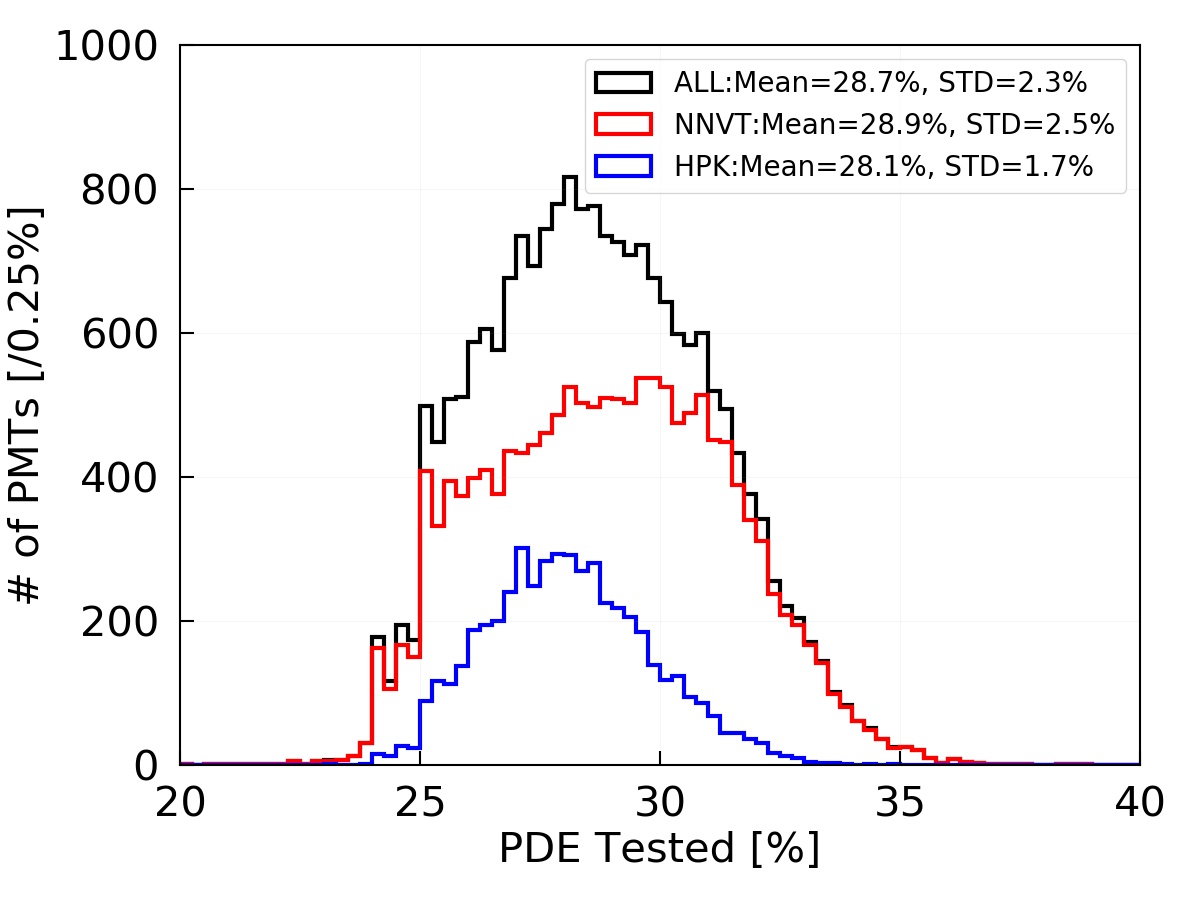}
		\caption{Measured PDE of all qualified PMTs}
		\label{fig:PDEtested:1}
	\end{subfigure}\hfill
	\begin{subfigure}[c]{0.495\textwidth}
		\centering
		\includegraphics[width=\linewidth]{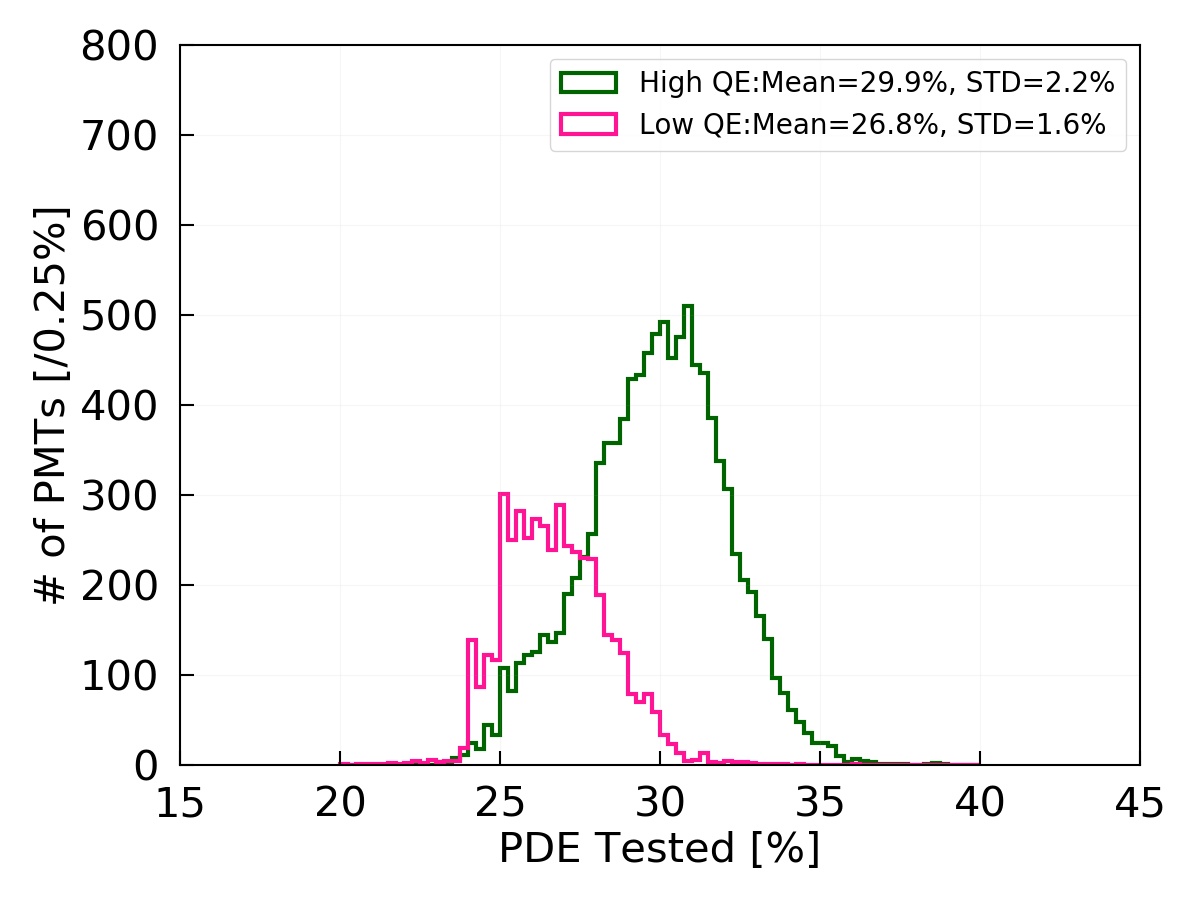}
		\caption{Measured PDE of NNVT PMTs only}
		\label{fig:PDEtested:2}
	\end{subfigure}	
	\caption{Measured PDE distribution of all qualified PMTs (left), and of all qualified NNVT PMTs (right). Black: all PMTs; red: NNVT; blue: HPK; Right: dark green: High\_QE; deeppink: Low\_QE.}
	\label{fig:PDEtested}       
\end{figure}

The PDE distribution of all qualified PMTs measured by the container system is shown in Fig.\,\ref{fig:PDEtested:1}, where the mean PDE is $28.1\,\%$ for HPK PMTs, and $28.9\,\%$ for NNVT PMTs. The sample of qualified NNVT PMTs can be classified as 4,609 NNVT low-QE PMTs (early version of the MCP-PMTs and identified by production dates) with a mean PDE of $26.8\,\%$, and 10,456 NNVT high-QE PMTs (latest version of the MCP-PMTs with higher QE, again identified by production dates) with a mean PDE of $29.9\,\%$ (Fig.\,\ref{fig:PDEtested:2}). Following the proposed selection to distribute the qualified PMTs between JUNO CD and veto detector, the mean PDE for the 17,612 PMTs selected for the CD is $29.1\,\%$ (including all HPK PMTs\footnote{Due to their clearly better timing performance (see sec.\ref{sec:2:tts}), all selected HPK PMTs will be used for the CD instrumentation.}), while the mean PDE for the 2,400 PMTs selected to be used in the veto detector is $25.6\,\%$.

\subsubsection{Light system aging and PDE correction}
\label{sec:3:aging:PDEcorrection}

\begin{figure}[!ht]
	\begin{subfigure}[c]{0.495\textwidth}
		\centering
		\includegraphics[width=\linewidth]{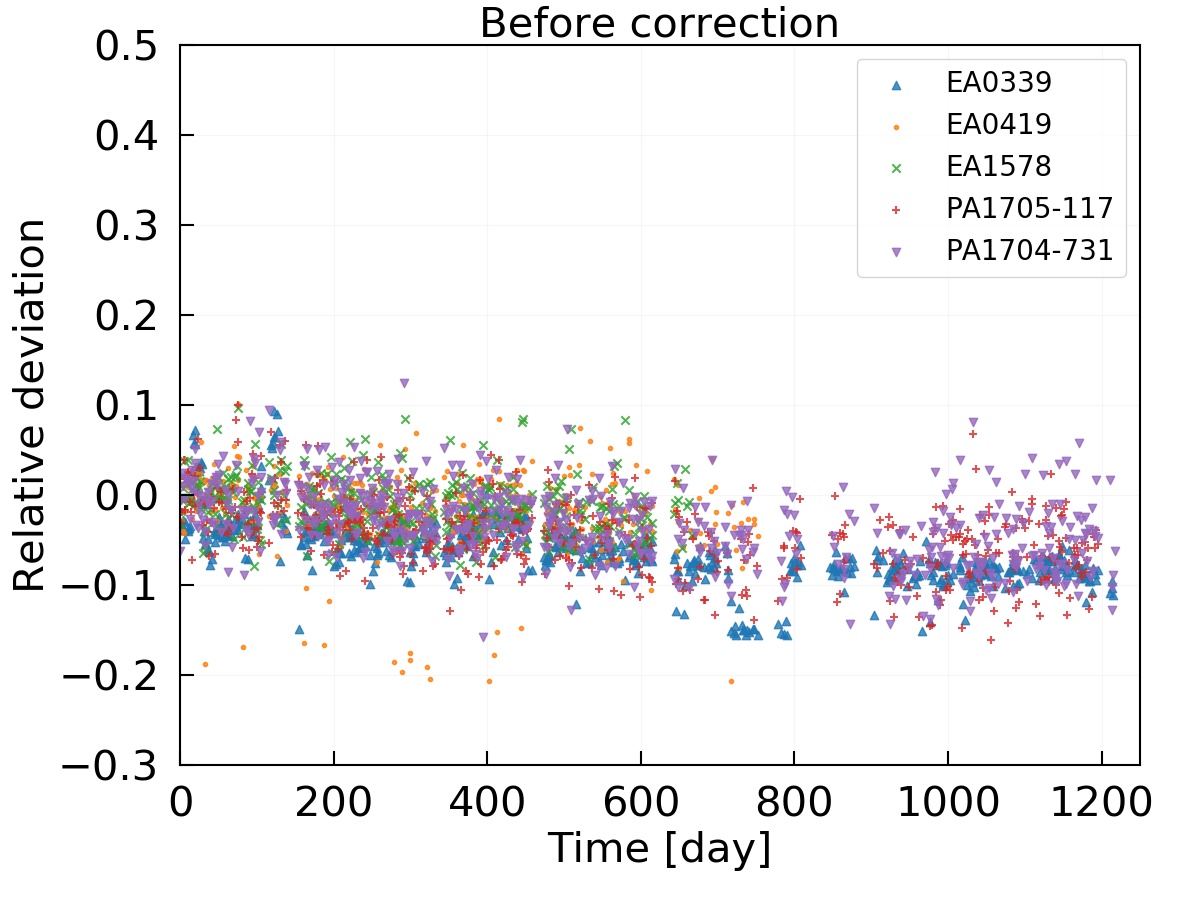}
		\caption{Container \#A before correction}
		\label{fig:PDEAtested:a}
	\end{subfigure}\hfill
	\begin{subfigure}[c]{0.495\textwidth}
		\centering
		\includegraphics[width=\linewidth]{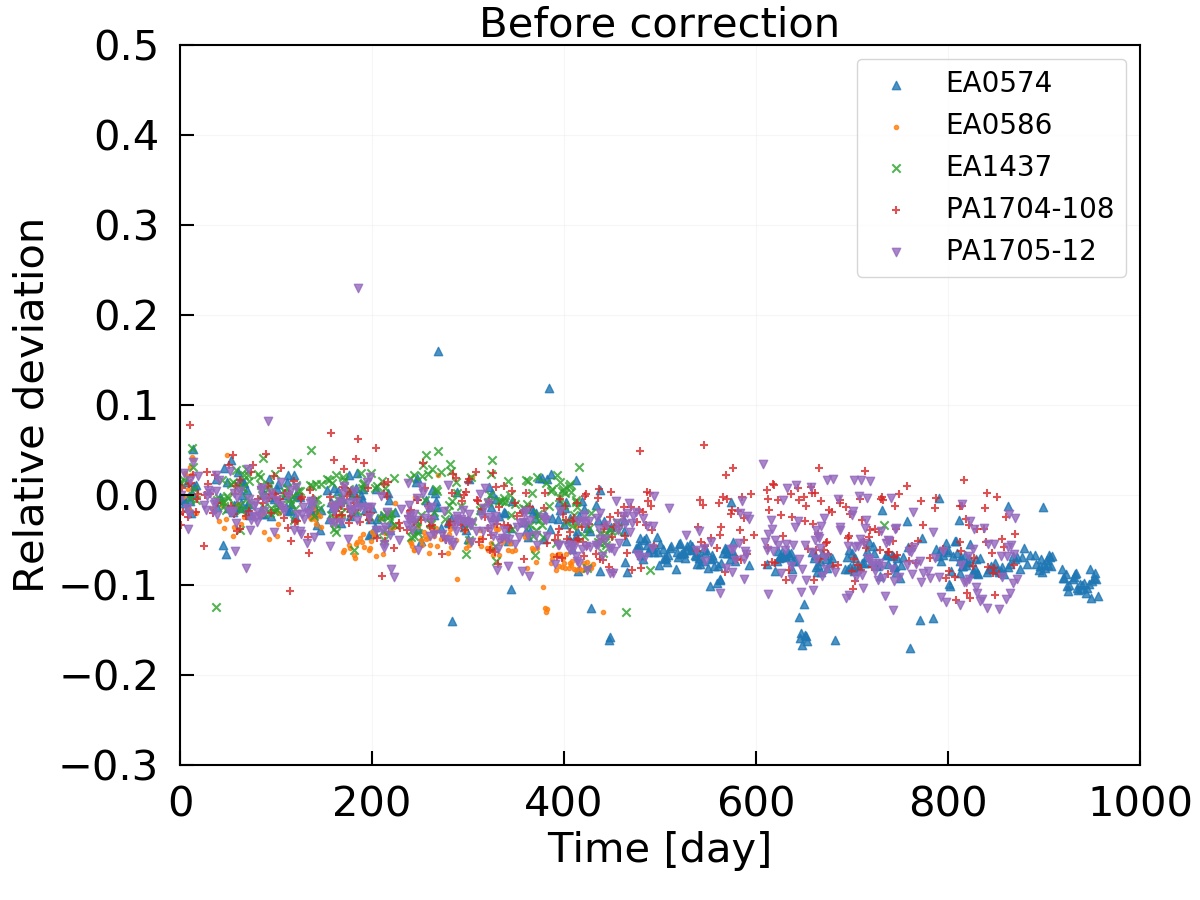}
		\caption{Container \#B before correction}
		\label{fig:PDEBtested:b}
	\end{subfigure}	\hfill
	\begin{subfigure}[c]{0.495\textwidth}
		\centering
		\includegraphics[width=\linewidth]{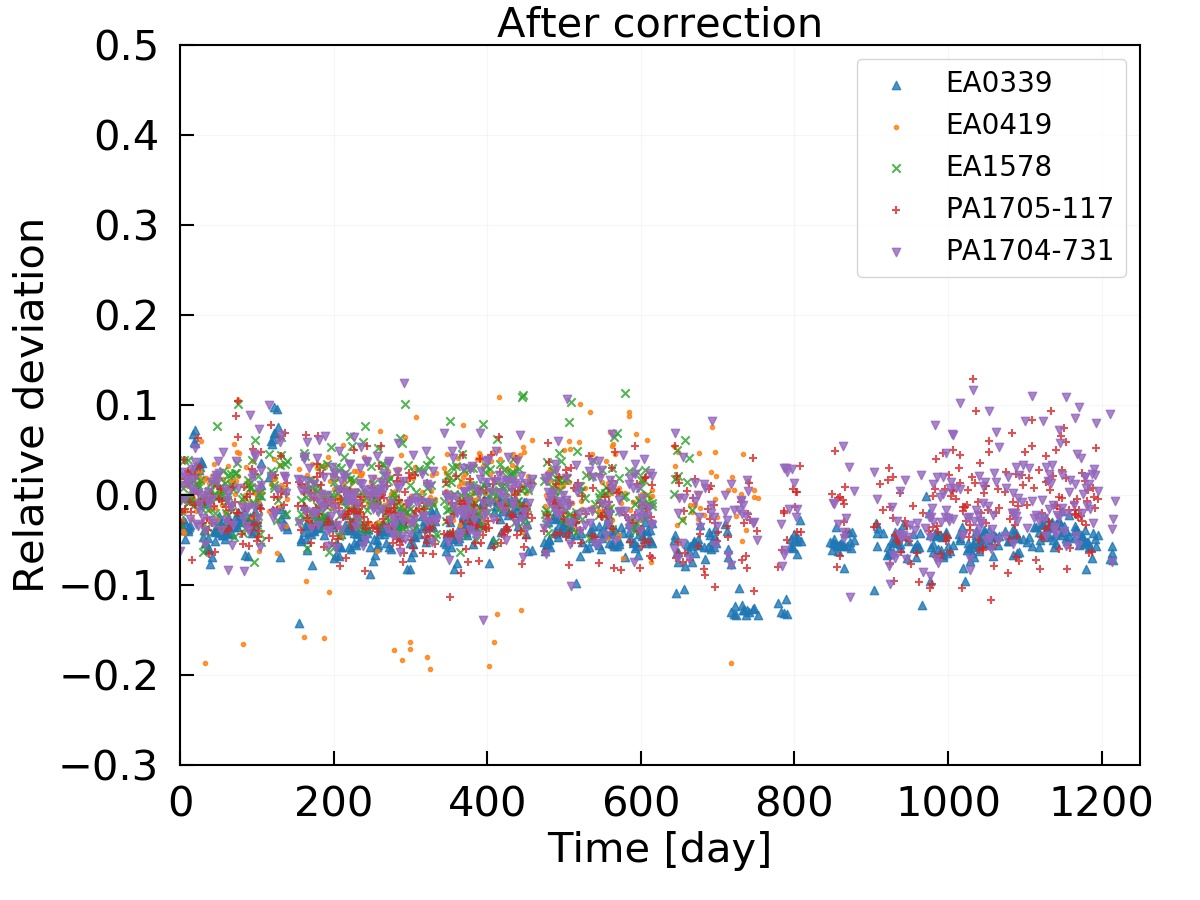}
		\caption{Container \#A after correction}
		\label{fig:PDEAtested:c}
	\end{subfigure}\hfill
	\begin{subfigure}[c]{0.495\textwidth}
		\centering
		\includegraphics[width=\linewidth]{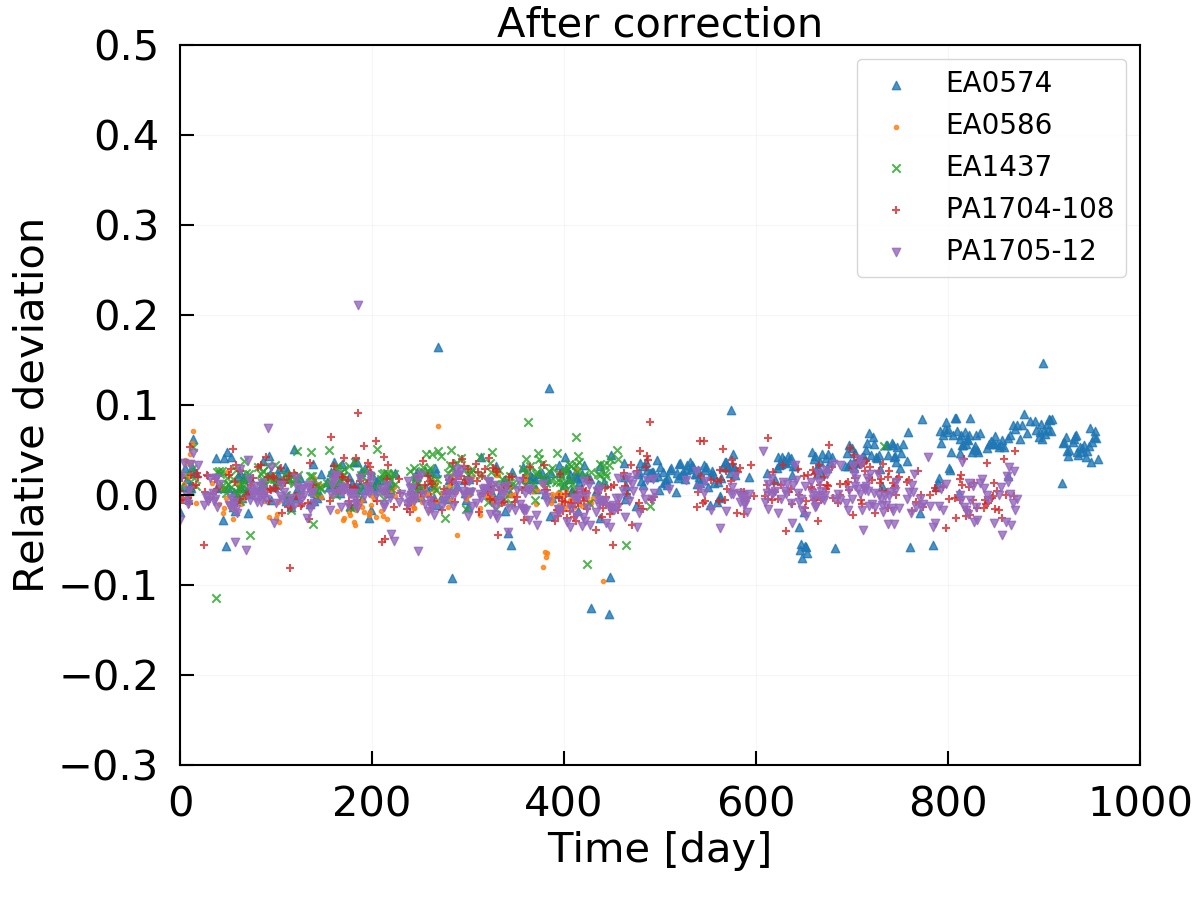}
		\caption{Container \#B after correction}
		\label{fig:PDEBtested:d}
	\end{subfigure}	
\caption{Relative variation of the measured PDE of the monitoring PMTs (HPK PMTs tagged by "EA", NNVT PMTs tagged by "PA") at the container system, before and after a correction based on a recalibration at the end of the regular testing period. Indicated variation is relative to the beginning of the testing period, which is used to compare among the PMTs. Top: directly measured PDE (with top left: container $\#$A; top right: container $\#$B). Bottom: PDE after aging correction (with bottom left: container $\#$A; bottom right: container $\#$B).}
\label{fig:PMT:containerPDEaging}       
\end{figure}

Within the $\sim$four years of operation of the two container systems, a decrease in the measured PDE of the monitoring PMTs was found for both of container $\#$A and $\#$B as shown on top of Fig.\,\ref{fig:PMT:containerPDEaging}. Both HPK and NNVT monitoring PMTs show a similar trend, which however required a long period of observation to be confirmed within the stated PDE uncertainty. The results were checked in detail based on the results of the monitoring PMTs as well as based on additional cross-checks with additional PMTs on the consistency among drawers, between the two containers, between the containers and scanning stations, as well as between the scanning stations themselves. It is concludes that the decrease is mainly caused by the LED light system\footnote{This includes LED, collimator, attenuator (neutral density filter), PTFE diffuser plate and reflector (light shaping tube in the drawer boxes).} itself used in the containers rather than by direct PMT aging. The effect directly depends on the container drawer as well. The scanning stations do not observe such a decreasing effect in the same period, according to the calibration and monitoring by the reference PMTs.

Based on a dedicated recalibration of all individual container channels at the end of the regular testing campaign and the observations of the monitoring PMTs from all individual drawers, a correction factor for the measured PDE was specified and applied for each drawer box of both containers, using a linear assumption for the PDE decrease relative to the testing date. The correction scale on average of all drawers was found to be $6.5 \times 10^{-5}$/day for container $\#$A and $\sim 7.6 \times 10^{-5}$/day for container $\#$B and was applied for both HPK and NNVT PMTs.
Finally, the corrected PDE results show a stable behaviour of the monitoring PMTs in both container systems, as well as consistent results with the scanning stations and between both containers.


The largest residual factor after the aging correction of all drawers is considered as the final bias uncertainty of the container system, which is about 2.3\% relatively for the tubes tested after 1,000 days with an aging factor of $2.3\times10^{-5}$/day, translating to an absolute value of $0.7$\,\% for tubes with 30\% PDE. After applying the aging correction of each drawer to the measured PDE of the container systems, the distributions of the corrected PDE results from the measurements of the monitoring PMTs (as shown at the bottom of Fig.\,\ref{fig:PMT:containerPDEaging}) are checked again as an uncertainty control. The variations are now within the aimed uncertainty range of $\lesssim 1\,\%$ absolute PDE, in particular when taking the spread of the corrected PDE distribution of each monitoring PMT into account. The largest variation of container \#B is provided by the fixed drawer monitoring PMT that is not used for daily measurements, which also includes a possible variation of the PMT itself.

\begin{figure}[!ht]
	\begin{subfigure}[c]{0.495\textwidth}
		\centering
		\includegraphics[width=\linewidth]{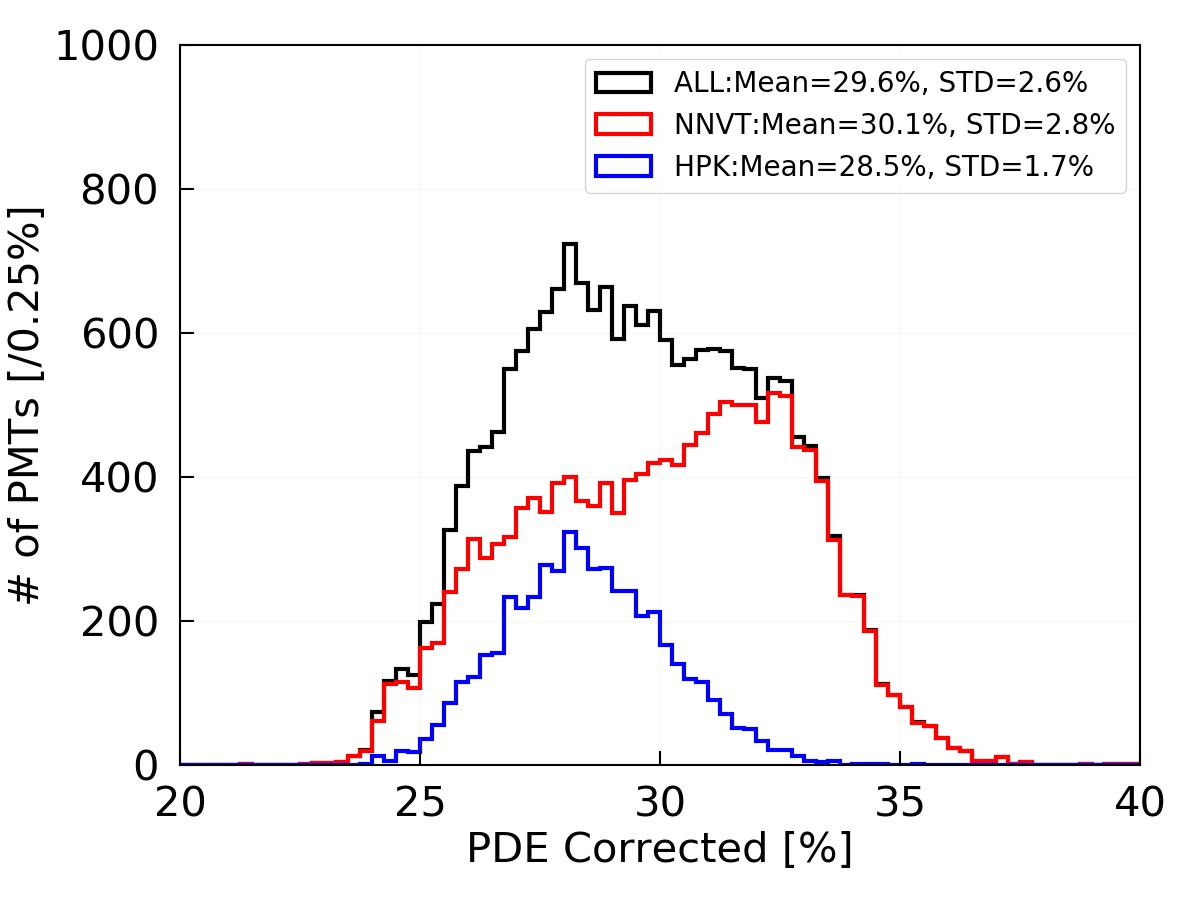}
		\caption{Corrected PDE of all qualified PMTs.}
		\label{fig:PDEcorrected1}
	\end{subfigure}\hfill
	\begin{subfigure}[c]{0.495\textwidth}
		\centering
		\includegraphics[width=\linewidth]{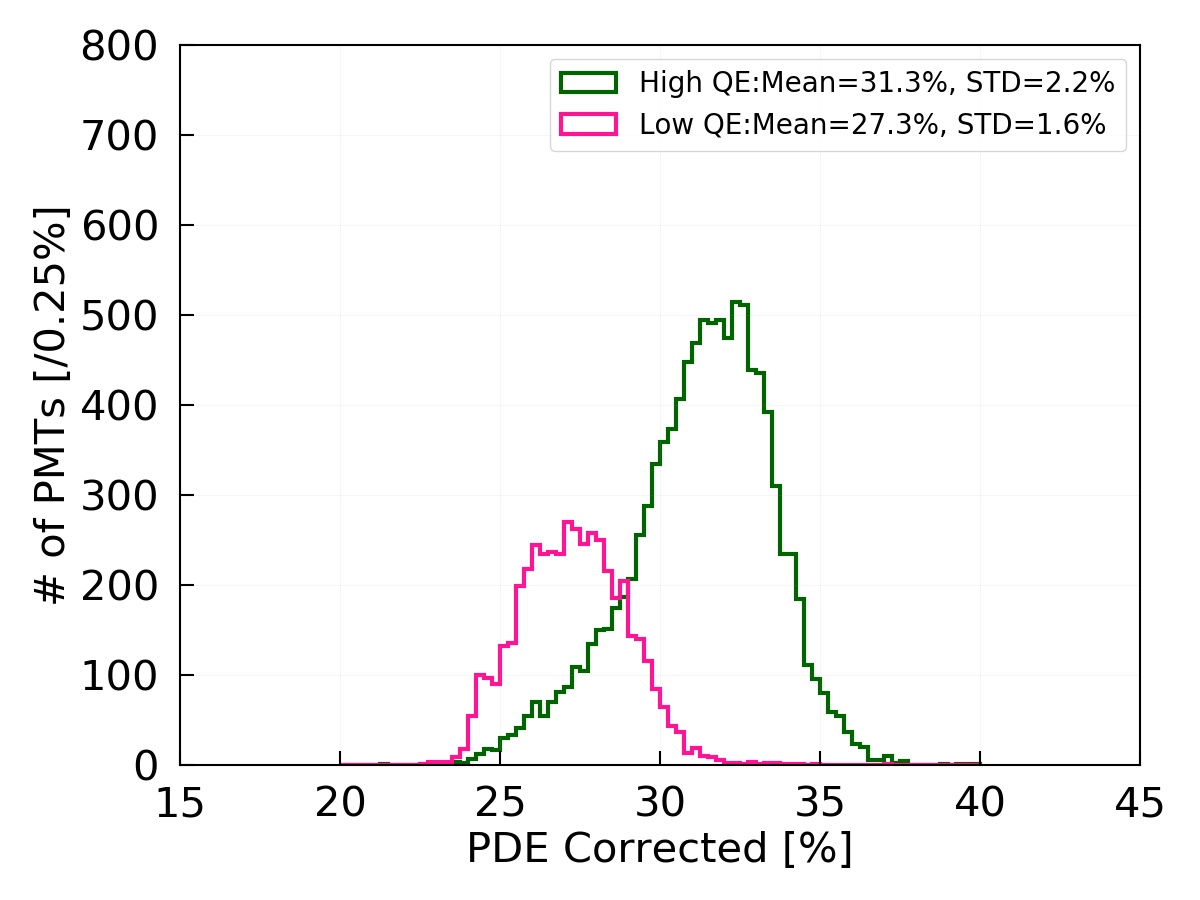}
		\caption{Corrected PDE of NNVT PMTs only.}
		\label{fig:PDECorrected2}
	\end{subfigure}	
	\caption{Corrected PDE distribution of all qualified PMTs (left), and of all NNVT PMTs (right). Black: all PMTs; red: NNVT; blue: HPK; dark green: high-QE MCP-PMTs; deep pink: low-QE MCP-PMTs.}
	\label{fig:PDECorrected}       
\end{figure}

The distributions of the corrected PDE results of all qualified PMTs are updated and depicted in Fig.\,\ref{fig:PDEcorrected1}, where the averaged values now increased to $28.5\,\%$ for the HPK PMTs and $30.1\,\%$ for NNVT PMTs. The average PDE of the qualified NNVT low-QE PMTs was updated to $27.3\,\%$, and $31.3\,\%$ for the qualified NNVT high-QE PMTs respectively\footnote{Since the difference between the corrected and original PDE is related to the testing date, the HPK PMTs are affected less as they were tested at an early stage, while the NNVT high-QE PMTs were tested later and thus are affected more.}, as shown in Fig.\,\ref{fig:PDECorrected2}. Furthermore, the corrected results of the proposed PMTs for JUNO CD and veto detector are also updated, with the averaged PDE increase to $30.1\,\%$ in the CD and to $26.1\,\%$ in the veto detector.

\subsection{Dark count rate (DCR)}
\label{sec:2:dcr}

Using the container system, the dark count rate (DCR) of each individual PMT was measured under a $1\times10^7$ gain. Data acquisition was performed using a CAEN V830AC scaler together with a CAEN V895B leading-edge discriminator, as described in~\cite{JUNOPMTcontainer}. The measured DCR value can be affected by the applied signal threshold, the time the PMT initially stayed in darkness in order to stabilize its dark rate prior to the measurement (``cooling time''), the photocathode's temperature, and the applied gain. The effect of each factor was checked in detail and will be discussed in the following.

\subsubsection{Amplitude threshold}
\label{sec:3:dcr:threshold}

\begin{figure}[!ht]
	\begin{subfigure}[c]{0.495\textwidth}
		\centering
		\includegraphics[width=\linewidth]{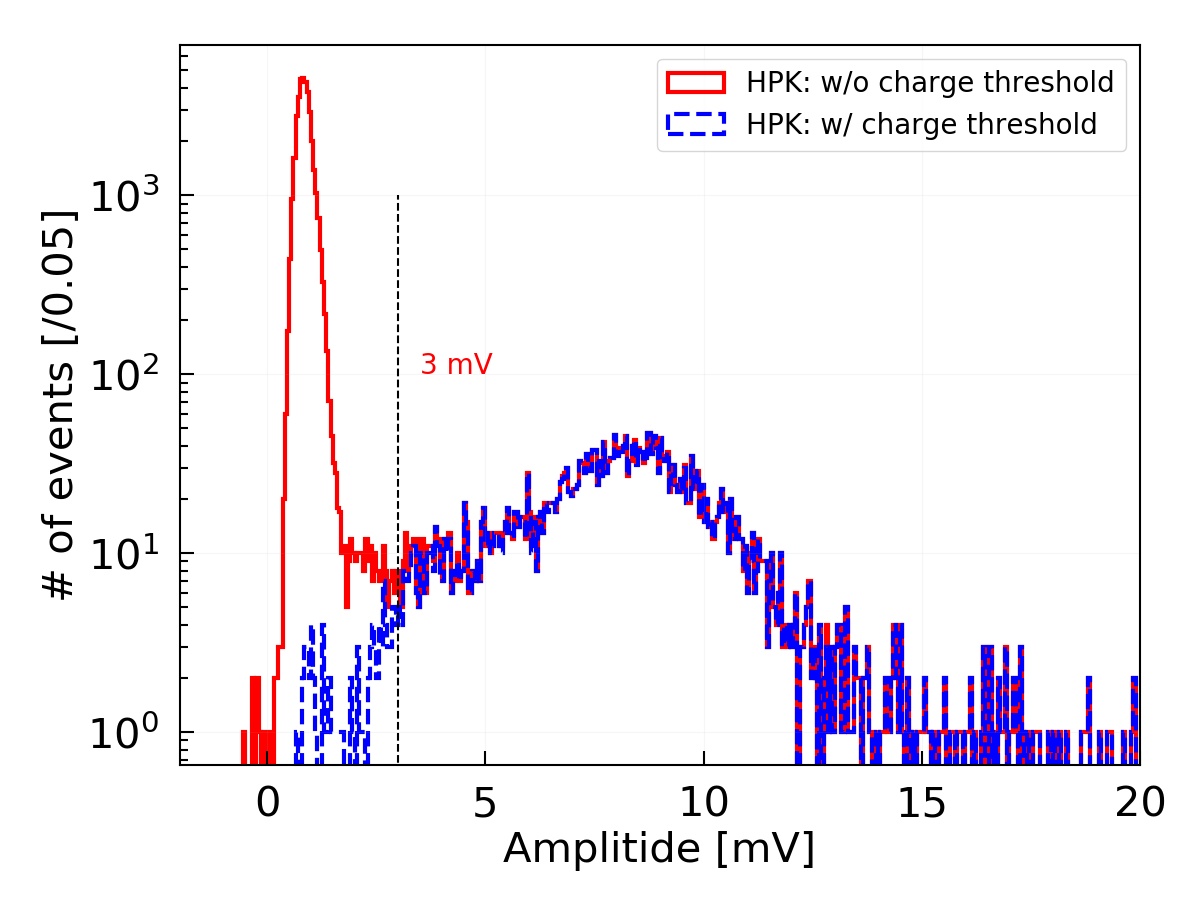}
		\label{fig:dcrthreshold:2}
	\end{subfigure}
	\begin{subfigure}[c]{0.495\textwidth}
		\centering
		\includegraphics[width=\linewidth]{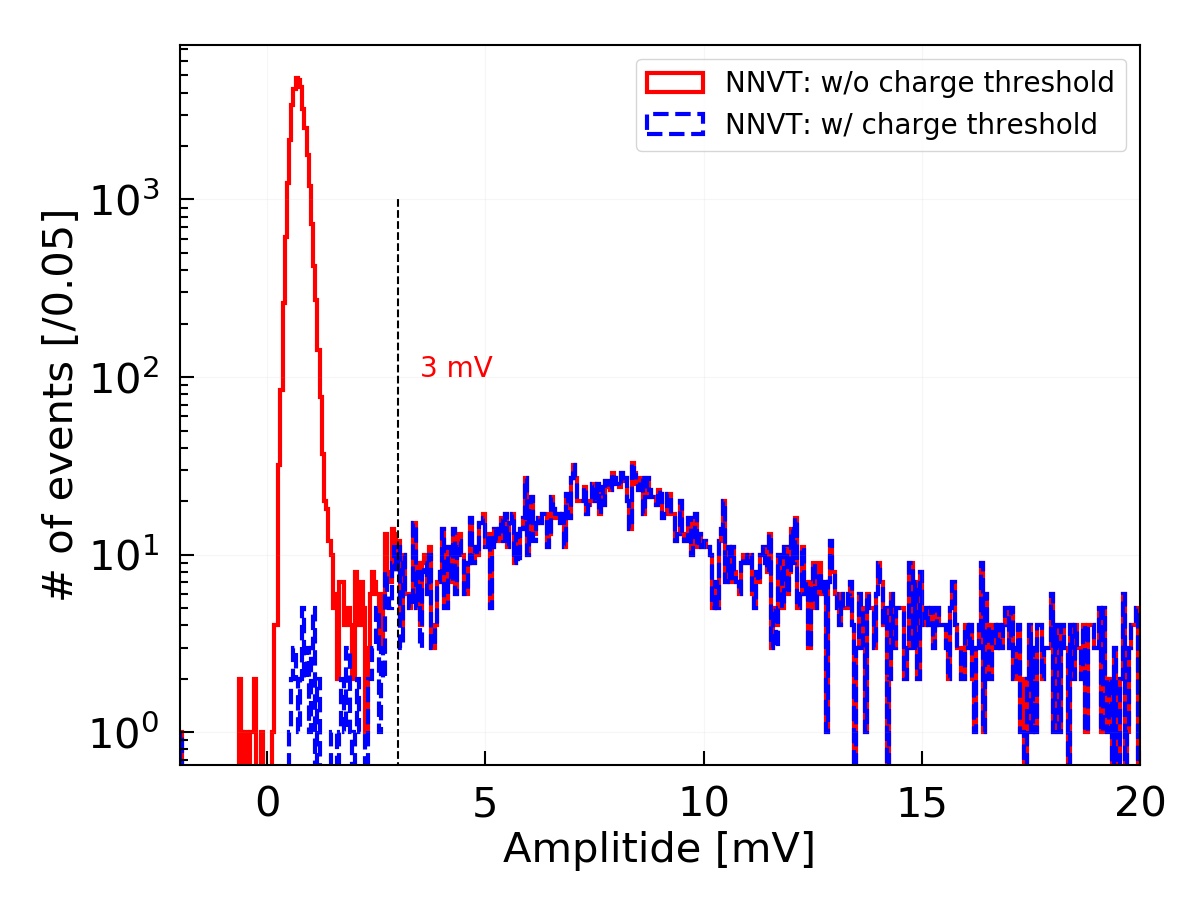}
		\label{fig:dcrthreshold:1}
	\end{subfigure}
	\caption{Effect of the set amplitude threshold on the DCR measurement of a single, typical PMT. The cut distributions (in blue) represent all pulses with a charge above the threshold of 0.25\,p.e., matching well to a pulse amplitude threshold of 3\,mV. Left: HPK pulse amplitudes; right: NNVT pulse amplitudes.}
	\label{fig:dcrthreshold}       
\end{figure}

A quarter p.e.~threshold was selected to consider the noise level of the system, the SPE charge resolution of the PMTs, as well as possible future JUNO conditions. The target threshold in charge corresponds to a signal threshold of about 3\,mV for the pulse amplitude of both HPK and NNVT PMTs.
The optimal threshold is not located precisely at the same level for both PMT types due to pulse shape differences between HPK and NNVT PMTs -- moreover, it is expected to be specific for each PMT rather than only refer to the mean amplitude value for an SPE pulse at a gain of $1\times10^7$. Since there are operational constraints set by the electronics used in the test\footnote{The used CAEN V895 discriminators provide a minimum step size for the threshold setting of only 1\,mV, see also~\cite{CAEN-disc}.}, a threshold of 3\,mV for the pulse amplitude was selected for the DCR measurements and was found to be a good compromise between measurement accuracy and comparability on one side, and mentioned constraints and present noise level on the other side, as illustrated in Fig.\,\ref{fig:dcrthreshold}.

As indicated in Fig.\,\ref{fig:PMT:procedure2}, PMTs failing the DCR acceptance criteria in the containers have been reevaluated in another container run or finally using the scanning station to exclude any other possible uncertainties on threshold or noise level. Since the scanning stations are using a factor $\times10$ amplifier prior to the counting electronics (see sec.\,\ref{sec:3:station} again), they can apply i.e.~an adequate DCR threshold with much higher precision than possible in the container systems.

\subsubsection{Cooling time}
\label{sec:3:dcr:cooltime}

\begin{figure}[!ht]
	\centering
	\includegraphics[width=0.5\linewidth]{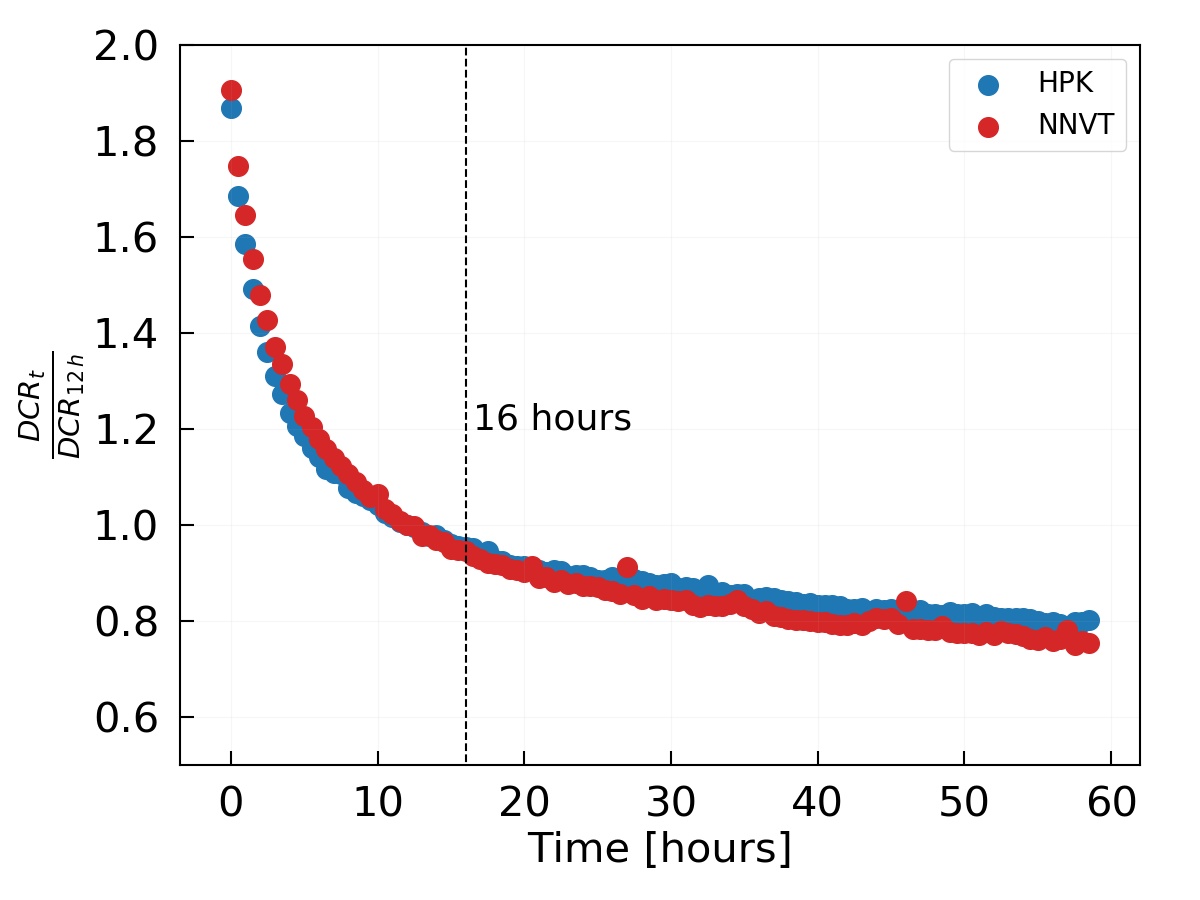}
	\caption{Measured DCR vs. cooling time for a typical HPK and NNVT PMT. The measured DCR values are normalized to the value taken after 12\,h (time after loading the PMT into the container). The indicated time of $\sim 16$\,h marks the usual time when the final DCR measurement is performed as part of the regular PMT characterization process.}
	\label{fig:PMT:dcrcoolingdown}       
\end{figure}

Large area PMTs such as the used 20-inch PMTs of JUNO generally need several hours in a dark environment to stabilize their dark rate after applying the HV. This is valid i.e.~after the PMT was exposed to ambient light. As discussed in~\cite{JUNOPMTcontainer}, all PMTs tested in the container system are resting for at least 12 hours in darkness before the start of the PMT characterization to ensure an effective (and significant) DCR measurement\footnote{This cooling time also considers operational constraints for an optimized testing cycle (max. of 24 hours including the reloading of the containers).}. During this time, the DCR will be frequently monitored as exemplarily presented in Fig.\,\ref{fig:PMT:dcrcoolingdown}. The concluding DCR measurement is usually performed after about 16 hours in darkness, as explained in~\cite{JUNOPMTcontainer}\footnote{Some container runs, i.e. the ones performed over the weekends, are performed with extended cooling times, in order to investigate the stabilization behavior of PMTs after a longer time in darkness, as well as to use the best possible conditions for a significant DCR measurement. With the monitoring data, it was confirmed that the bare HPK PMTs usually show a stable performance after $\sim$12 hours cooling time while the bare NNVT PMTs usually show a stable performance after
$\sim$25 hours of cooling time.}.
The gain has been set to a level of $1\times10^7$ following the procedure described in sec.\,\ref{sec:2:gain} prior to this measurement.

\subsubsection{Temperature}
\label{sec:3:dcr:temperature}

\begin{figure}[!ht]
	\begin{subfigure}[c]{0.495\textwidth}
		\centering
		\includegraphics[width=\linewidth]{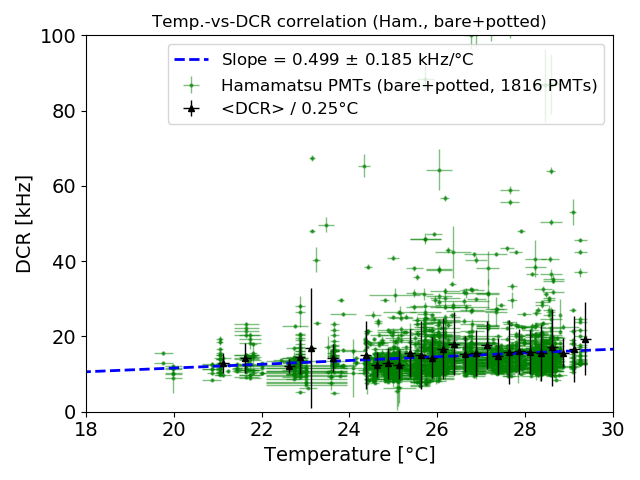}
		\caption{HPK PMTs (bare and potted)}
		\label{fig:PMT:DCRtrend.a}
	\end{subfigure}\hfill
	\begin{subfigure}[c]{0.495\textwidth}
		\centering
		\includegraphics[width=\linewidth]{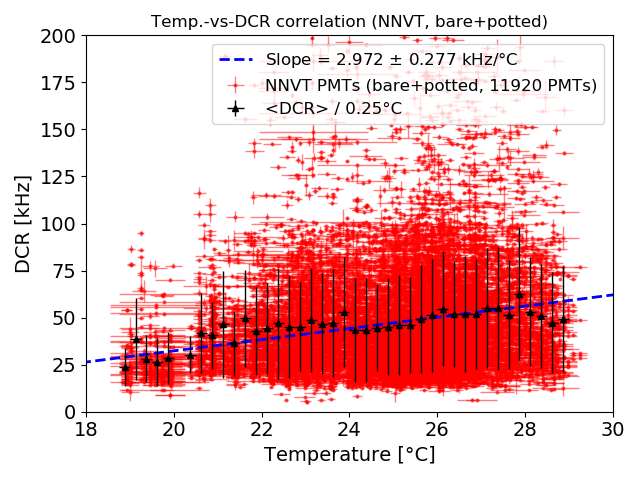}
		\caption{NNVT PMTs (bare and potted)}
		\label{fig:PMT:DCRtrend.b}
	\end{subfigure}	
	\caption{Observed correlation between air temperature in the container and measured DCR of a large sub-sample of PMTs. The black dots indicate the mean and STD of all DCR measurements within the corresponding $0.25^\circ$\,C slice.}
	\label{fig:PMT:DCRtrend}       
\end{figure}

As discussed in~\cite{HamManual,lowTPMT2,largePMTlei,largePMTtemperature}, the actual DCR of a PMT is a function of temperature. We thereby assume that the photocathode's temperature is in equilibrium with the temperature of the surrounding air, i.e. after a several hours stay in the drawers. To estimate the effect of different air temperatures on the DCR results of the JUNO PMTs during their characterizations, data from the temperature monitoring system within the containers was taken into account to the observed DCR of the PMTs. This was done for a large PMT sub-sample containing more than 1,800 HPK PMTs and more than 11,900 NNVT PMTs. Although every PMT acts uniquely, a slight trend could be observed for each PMT type, indicating a mild temperature dependence of about 0.5\,kHz/$^{\circ}$C for the combined HPK PMT sub-sample, and of about 3.0\,kHz/$^{\circ}$C for the combined NNVT PMTs sub-sample within a temperature range of 19$^{\circ}$C to 29$^{\circ}$C as present during the mass characterization campaign, 
see also Fig.\,\ref{fig:PMT:DCRtrend}.

\begin{figure}[!ht]
	\begin{subfigure}[c]{0.49\textwidth}
		\centering
		 \includegraphics[width=\linewidth]{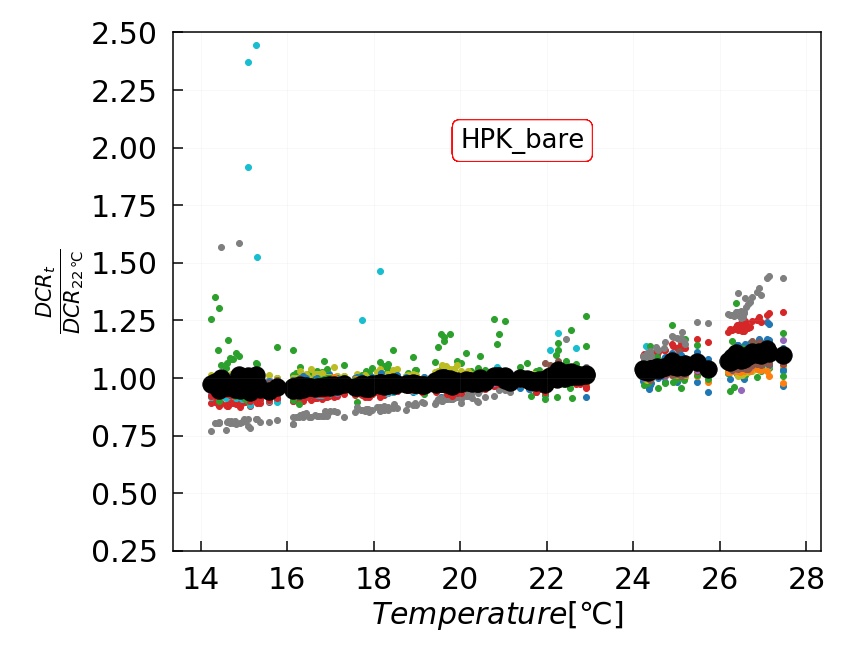}
		\caption{Sample of 26 HPK PMTs}
		\label{fig:dcrvstemperature:HPK}
	\end{subfigure}	
    \begin{subfigure}[c]{0.49\textwidth}
		\centering
		 \includegraphics[width=\linewidth]{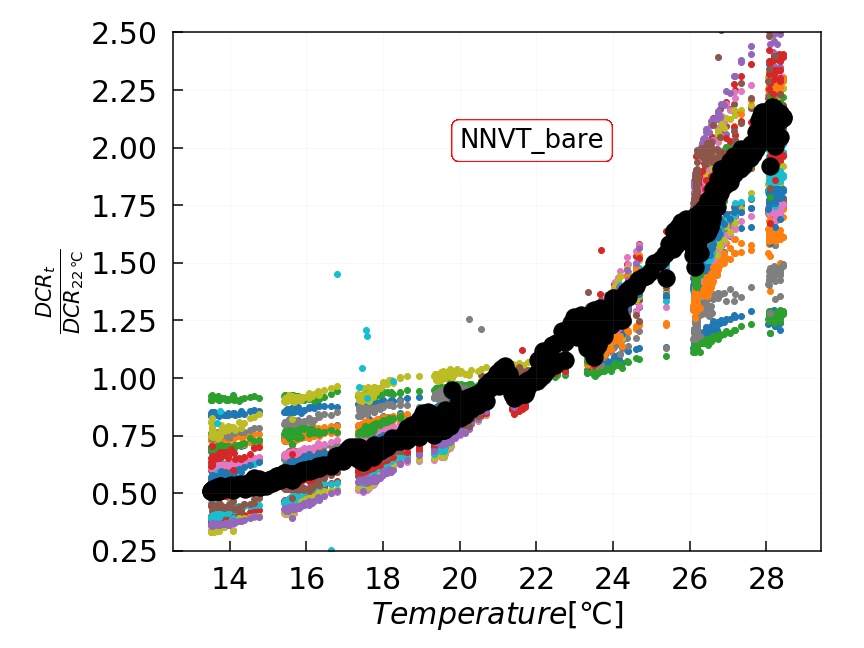}
		\caption{Sample of 26 NNVT PMTs}
		\label{fig:dcrvstemperature:NNVT}
	\end{subfigure}\hfill
	\caption{Measured DCR versus temperature in a specific temperature survey. The black dots represent the averaged trend of all measured PMTs under the same temperature. The discrete points far from the trend are generated by noise during the tests.}
	\label{fig:dcrvstemperature}       
\end{figure}

Additionally, specific temperature surveys were performed using the container system with its HVAC unit to gain a better understanding of the temperature dependence of the measured DCR. In these surveys, the effect of temperature to the measured DCR was investigated by varying the temperature in a range from 14$^{\circ}$C to 28$^{\circ}$C (with a rate of 0.5$^{\circ}$C/h) and monitoring the DCR in parallel. Such a testing cycle containing 26 HPK and 26 NNVT PMTs is shown in Fig.\,\ref{fig:dcrvstemperature}, where the DCR of every measurement was normalized to the value at 22$^\circ$C of each PMT individually. The results of these surveys are confirming a stronger temperature dependence of the NNVT PMTs than for HPK PMTs, which is on average about 0.2\,kHz/$^{\circ}$C ($\sim$2\,\%/$^{\circ}$C relative DCR change) for the combined HPK PMT sample, and about 4.8\,kHz/$^{\circ}$C ($\sim$10\,\%/$^{\circ}$C relative DCR change) for the combined NNVT PMT sample. These values are consistent with the before mentioned estimation between 19$^{\circ}$C and 28$^{\circ}$C in a first order approx.  
The results further indicate, that the observed DCR is significantly increasing for temperatures higher than 22$^\circ$C (particularly for the NNVT PMTs). It should also be noted, although not visible in the plot, that PMTs showing a higher absolute DCR at 22$^\circ$C are more sensitive to temperature changes, compared to PMTs with a lower absolute DCR at 22$^\circ$C.

\subsubsection{Measured dark count rate (DCR)}
\label{sec:3:dcr:distribution}

\begin{figure}[!ht]
	\begin{subfigure}[c]{0.495\textwidth}
		\centering
		\includegraphics[width=\linewidth]{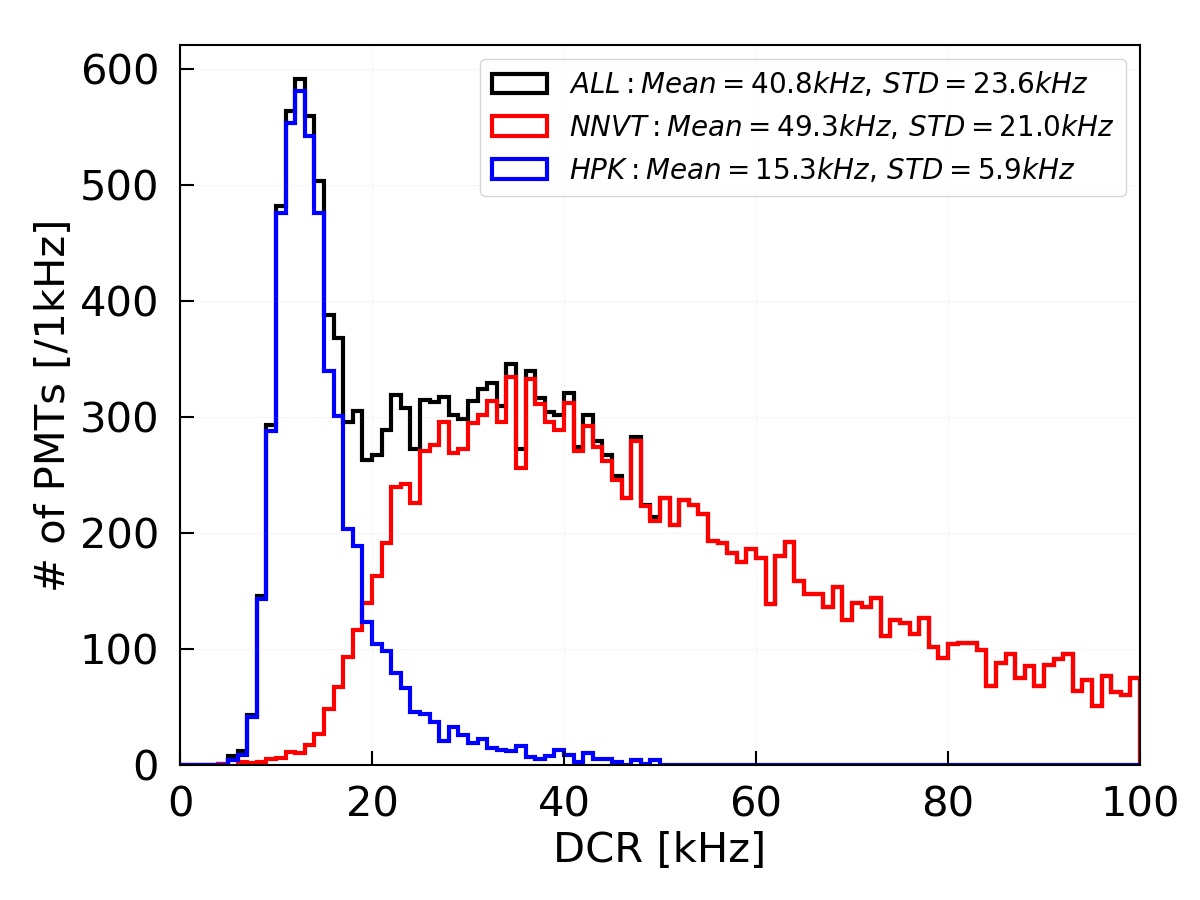}
		\label{fig:DCR:1}
	\end{subfigure}\hfill
	\begin{subfigure}[c]{0.495\textwidth}
		\centering
		\includegraphics[width=\linewidth]{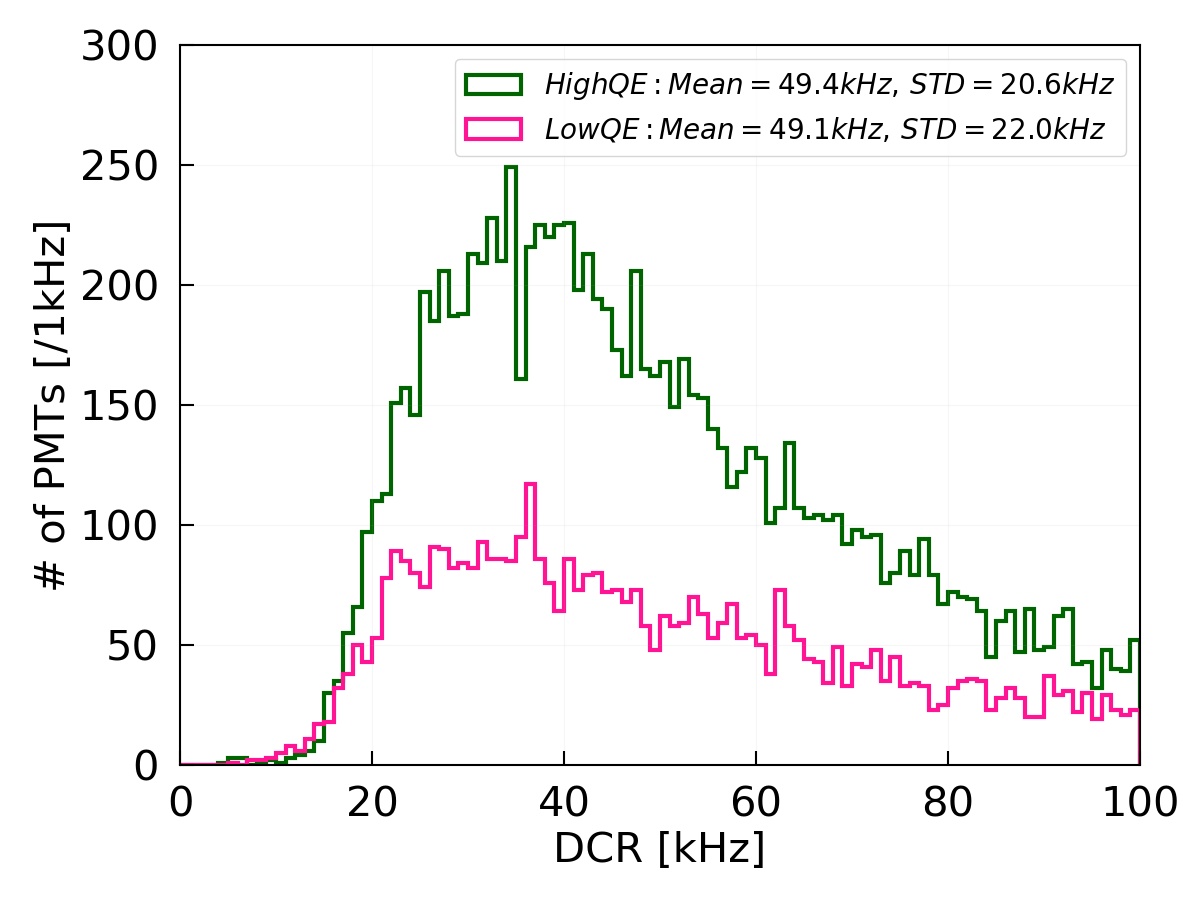}
		\label{fig:DCR:2}
	\end{subfigure}	
	\caption{Measured DCR distribution of qualified PMTs. Left: black solid: all PMTs; red solid: NNVT; blue solid: HPK; Right: dark green solid: High\_QE; deep pink solid: Low\_QE. Note: the DCR of NNVT PMTs is selected with $\leq$ 100\,kHz.}
	\label{fig:DCR}       
\end{figure}

The measured DCR distributions (as shown in Fig.\,\ref{fig:DCR}) of all qualified PMTs were tested by the container systems using the discussed amplitude threshold of 3\,mV (p.e.~threshold of 0.25, refer to Fig.\,\ref{fig:dcrthreshold}), with a cooling time of $\sim 16$ hours, and at an average air temperature of $25 \pm 3^{\circ}$C. The mean value is around 15.3\,kHz for HPK PMTs and 49.3\,kHz for NNVT PMTs. The NNVT DCR distribution shows a long tail which is cut at 100\,kHz according to the requirement of JUNO as listed in Tab.\,\ref{tab:PMT:criteria}. Also, both low- and high-QE MCP-PMTs show a consistent DCR distribution. In the final JUNO detector, an even lower average DCR is expected based on the considerations in sec.\,\ref{sec:3:dcr:temperature}, since the water temperature in the veto pool (where the PMTs will be embedded in) is predicted to be at about $21 \pm 1^{\circ}$C.

\subsection{Transit time spread (TTS)}
\label{sec:2:tts}

\begin{figure}[!ht]
	\begin{subfigure}[c]{0.495\textwidth}
		\centering
		\includegraphics[width=\linewidth]{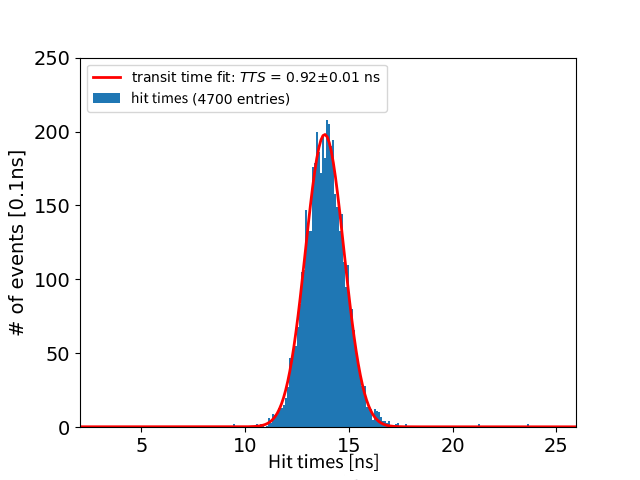}
		\caption{Typical HPK PMT}
		\label{fig:PMT:TTSdist.a}
	\end{subfigure}\hfill
	\begin{subfigure}[c]{0.495\textwidth}
		\centering
		\includegraphics[width=\linewidth]{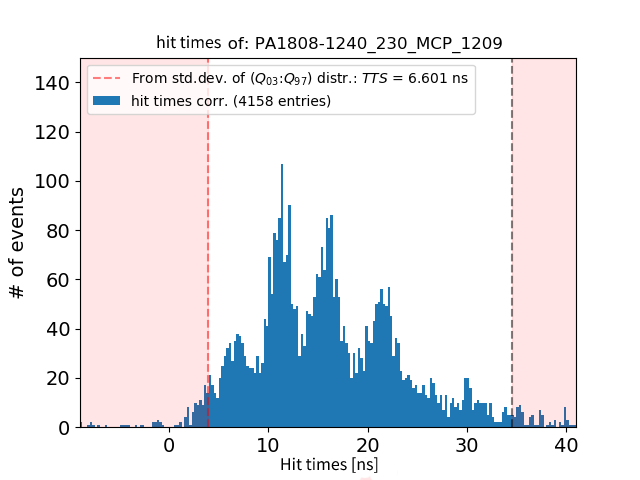}
		\caption{Typical NNVT PMT}
		\label{fig:PMT:TTSdist.b}
	\end{subfigure}	
	\caption{Typical relative transit time (TT) distributions for an HPK (left) and an NNVT PMT (right, indicated ``hit times''). The PMT's TTS is determined using a Gaussian fit in case of the HPK PMTs, and using the std. dev. of a cut distribution (use only quantiles $Q_{03}$ to $Q_{97}$) in case of the NNVT PMTs. Both methods produce consistent results if applied on the data of HPK PMTs.}
	\label{fig:PMT:TTSdist}       
\end{figure}

The transit time spread (TTS) describes the timing resolution of the PMT. It represents the spread of different photo-electron transit times within the PMT bulb, which depend on the interaction point and emission angle of the photo-electron, as well as on the energy of the released photo-electron and the focusing electric field distribution \cite{PMTbook}. For JUNO, the timing resolution of the HPK dynode PMTs is crucial for a precise event reconstruction in the detector. For these PMTs, the distribution of transit times (TT)\footnote{The relative ``transit time'' (TT) calculates in this case from the difference between the trigger time of the laser light pulse and the time when the PMT pulse arises (``hit time'') as shown in Fig.\,\ref{fig:gain:wave}. Its absolute number is arbitrary here, since it depends also on cable lengths and individual response times of the data taking electronics.} follows a Gaussian distribution as shown in Fig.\,\ref{fig:PMT:TTSdist.a}, where the TTS is defined as the $\sigma$ of this distribution, with $\sigma \simeq \mathrm{FWHM} / (2 \sqrt{2 \ln 2}) $~\cite{HamManual}. In case of the NNVT MCP-PMTs, this relative TT distribution is much more complex and contains a varying number of substructures (sub-peaks), see Fig.\,\ref{fig:PMT:TTSdist.b}. These features are related to the PMT design and MCP structures~\cite{MCPPMTTTSsen} and are a consequence of the optimization of the MCP-PMTs for a maximum collection efficiency, which leads to a worse timing resolution~\cite{TTSQianS} (will be discussed in sec.\,\ref{sec:3:uniform:tts}). As a consequence, the TTS can not be described using a Gaussian fit, but rather will be described here as the mean standard deviation of the TT distribution with the edges cut away (only the quantiles $\left[Q_{03}:Q_{97}\right]$ are taken into account) to remove outliers and noise events\footnote{This method has been approved by comparing the TTS results of about 200 HPK PMTs analyzed with both methods (fit and mean std. dev. of cut distribution), showing highly consistent results. $\left[Q_{03}:Q_{97}\right]$ are the quantiles of first 3\% and 97\% of all the entries from left to right.}.

\begin{figure}[!ht]
	\begin{subfigure}[c]{0.49\textwidth}
		\centering
		\includegraphics[width=\linewidth]{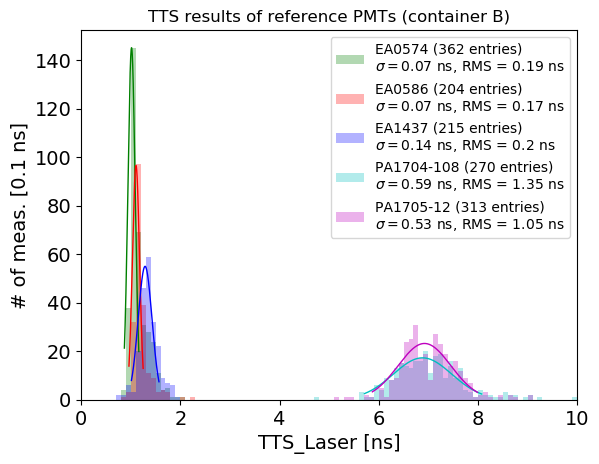}
		\caption{TTS reproducibility of HPK and NNVT monitoring PMTs.}
		\label{fig:PMT:TTSacc}
	\end{subfigure}\hfill
	\begin{subfigure}[c]{0.48\textwidth}
		\centering
		\includegraphics[width=\linewidth]{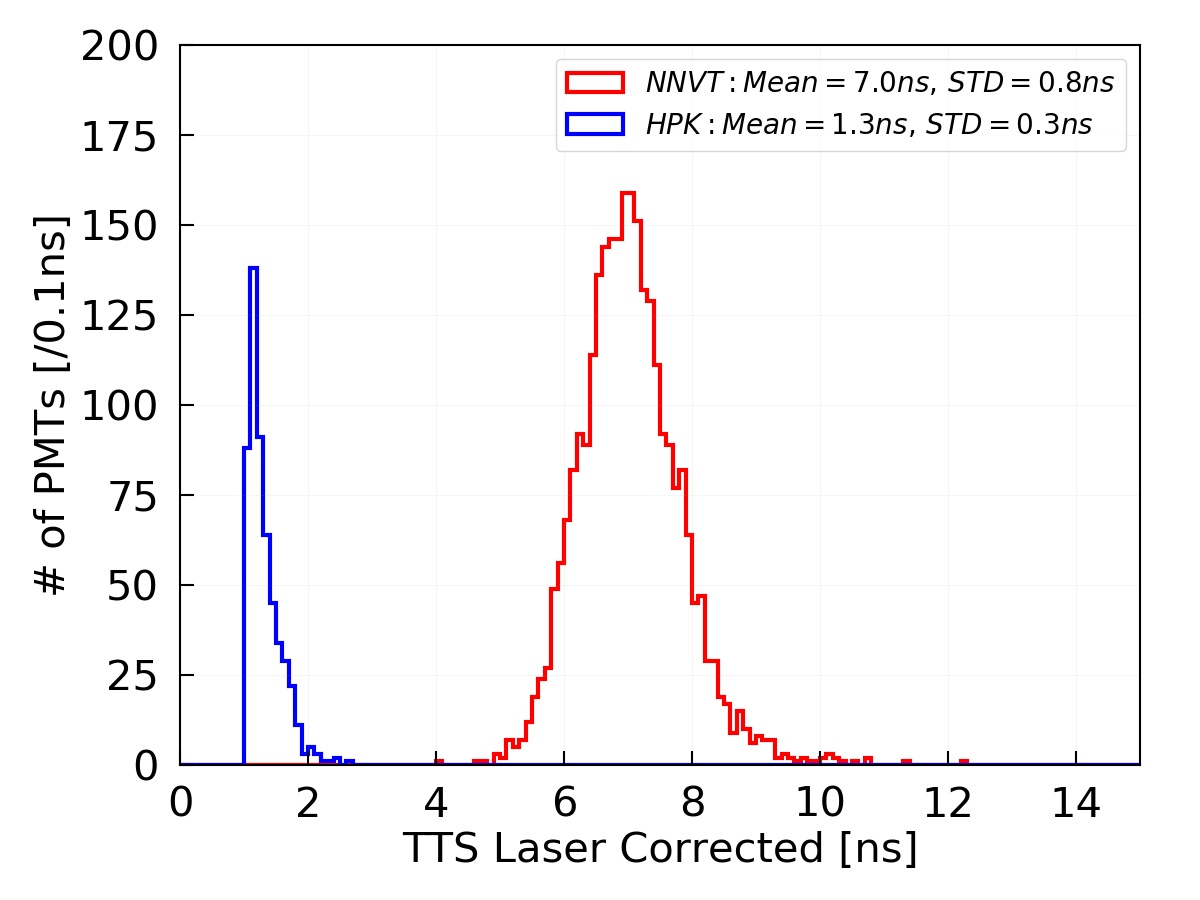}
		\caption{TTS distribution of a sub-sample of qualified bare PMTs.}
		\label{fig:tts}
	\end{subfigure}	
	\caption{Left: TTS results of the monitoring PMTs in container \#A. Combined result distribution of all monitoring PMTs (3 HPK, 2 NNVT, tested in all channels) shows a good reproducibility of the results. Right: TTS distribution of a sub-sample of qualified bare PMTs containing 837 HPK and 3,610 NNVT PMTs, measured with the ps-Laser system (red: NNVT; blue: HPK).}
	\label{fig:PMT:TTS:acc}       
\end{figure}

All valid TTS results discussed in this section are from measurements with the container system using the pico-second laser fiber system\footnote{The container LED system only features light pulses with a pulse width of $\sigma \sim 2$\,ns, which is too large to resolve the TTS of the HPK PMTs, thus it can be used only for a relative comparison, particularly for NNVT PMTs.} \cite{JUNOPMTcontainer,TietzschDiss} and similar settings as used for the SPE features in sec.\,\ref{sec:2:spe} (i.e. a light intensity of \textmu$\simeq0.1-1$\,p.e.), but with larger statistics of up to 50,000 trigger events. The triggers for the data acquisition and light sources are provided by a Keysight 33512B arbitrary waveform generator featuring a jitter of only $\sim 50$\,ps~\cite{containerGen}.
Unfortunately only container $\#$B has a sufficient timing resolution, with a minimum measurable TTS of $\sim 0.8$\,ns (that's the reason why there is a left cut around 0.8\,ns in Fig.\ref{fig:tts}) -- due to a mechanical incident, the fiber system of container $\#$A can not provide a timing resolution of $<2$\,ns anymore. The final accuracy for the TTS measurements achieved with container $\#$B's laser is less than 0.2\,ns (RMS 0.2\,ns) for HPK PMTs and less than 0.6\,ns (RMS 1.4\,ns) for NNVT PMTs \footnote{Different measurement accuracies are due to the different analysis methods related to the PMT type. Values describe repeatability/accuracy for multiple measurements of the same PMT.} (also see Fig.\,\ref{fig:PMT:TTSacc}). Systematic effects introduced by the data taking electronics and the fiber system were measured to be only about 0.6\,ns on average for all channels and were corrected in order to achieve a fair estimation of the PMTs' TTS.

Since only one of containers $\#$A and $\#$B provides a sufficient timing resolution, the TTS was evaluated only for a large sub-sample of all tested PMTs. The results of this sub-sample are representative for the full sample of qualified PMTs for JUNO as presented in the other sections.
Results are available for a total of 837 HPK and 3,610 NNVT PMT of qualified bare PMTs tested with the laser system in container $\#$B, covering about 16.7\,\% of the full qualified HPK PMT sample, and about 24.0\,\% of the full qualified NNVT PMT sample\footnote{In total, TTS data of 1198 qualified HPK and 4042 qualified NNVT PMTs tested in container $\#$B was analyzed, including monitoring PMTs and potted PMT, while the distribution of the full analyzed sample is highly consistent with Fig.\,\ref{fig:tts}.}. The results' distribution is depicted in Fig.\,\ref{fig:tts}. The mean TTS is about 1.3\,ns in $\sigma$ as required for HPK dynode PMTs, with only about $1\,\%$ of the examined PMTs clearly failing the timing performance requirements for JUNO, which is at an acceptable level. Results are also consistent with another study performed earlier, see~\cite{JUNOTTS2} for details. The mean TTS of NNVT MCP-PMTs is about 7.0\,ns, see also Fig.\,\ref{fig:tts}.

\begin{figure}[!ht]
	\begin{subfigure}[c]{0.495\textwidth}
		\centering
		\includegraphics[width=\linewidth]{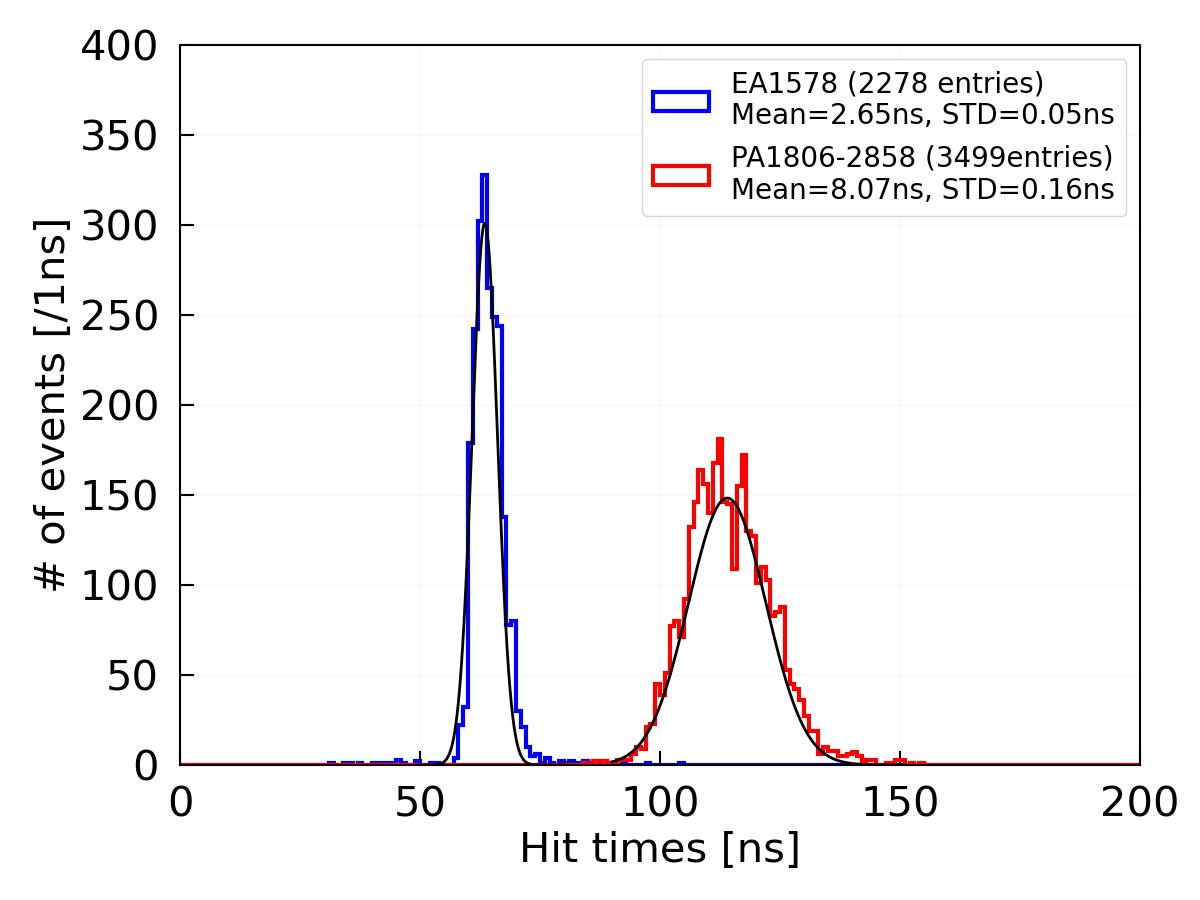}
		\caption{Examples of relative TT, measured with the LED for single HPK or NNVT PMT.}
		\label{fig:tts_led:hittime}
	\end{subfigure}\hfill
	\begin{subfigure}[c]{0.495\textwidth}
		\centering
		\includegraphics[width=\linewidth]{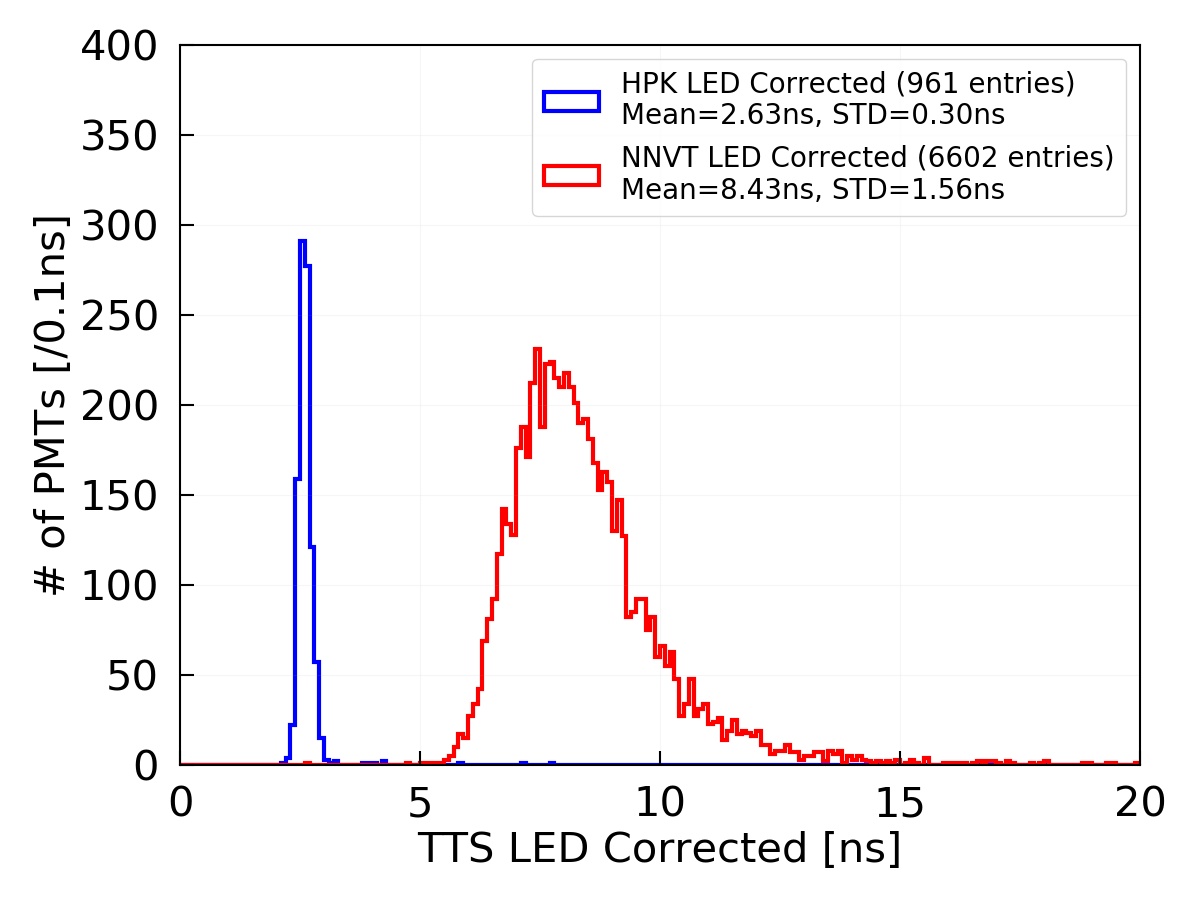}
		\caption{Relative TTS, measured with the LED of a sub-sample of qualified bare PMTs}
		\label{fig:tts_led}
	\end{subfigure}\hfill
	\caption{Relative TT distribution from LED measurements of single PMT and the relative TTS distribution of sub-sample of all qualified bare PMTs measured with LED in the container system. Dark-green, NNVT PMT; blue, HPK PMT.}
	\label{fig:tts_led_full}       
\end{figure}

The relative transit times' distribution can be calculated also from the data shown in secs.\,\ref{sec:3:amplitude} and \ref{sec:3:risefall} using the LED system of the containers (SPE light intensity, 20,000 waveforms)\footnote{The relative TT calculation using the LEDs is analogue to the one using the ps-laser system.}. This is shown in Fig.\,\ref{fig:tts_led:hittime} for an example, single HPK and NNVT PMT. Comparing to the results shown in Fig.\,\ref{fig:PMT:TTSdist.b} using the laser system, the sub-peaks in the relative TT distribution of NNVT MCP-PMTs are smeared out by the container LEDs -- this will be further discussed in sec.\,\ref{sec:3:uniform:tts} in the scope of the scanning station (which also uses LEDs).
When calculating the relative TTS of each PMT measured with the LED system of the containers\footnote{In case of the LED measurement, the `relative TTS' describes the combined timing resolution of PMT and light system, in contrast to the laser measurement where the TTS described purely the PMT performance. The timing resolution of the LED itself can refer to \cite{HVSYS-LED}.}, the relative TTS can be described by the sigma of a directly Gaussian fit on the relative TT distribution, as shown in Fig.\,\ref{fig:tts_led:hittime}. Using the monitoring PMTs for checking the repeatability of these measurements, the final accuracy for the relative TTS measurements using the LEDs achieved with containers $\#$A and $\#$B is around $\sigma \sim 0.2$\,ns (RMS $1.9$\,ns) for HPK PMTs and around $\sigma \sim 0.7$\,ns (RMS $1.6$\,ns) for NNVT PMTs, which is consistent to the laser measurements. The typical values of the relative TTS from the LED measurements for all qualified PMTs is $2.6$\,ns for HPK and $8.4$\,ns for NNVT PMTs, as shown also in Fig.\,\ref{fig:tts_led}. The results for both PMT types are larger than the values acquired from the laser measurements in container $\#$B as expected from the smearing due to the LED pulses and differences of the photon distribution on the photocathode between laser and LED system. Therefore, the results acquired using the LED system will be used only for a relative comparison and systematic control, and for comparison with the results of the scanning stations (will be presented in sec.\,\ref{sec:3:uniform:tts}).

\subsection{Pre-pulse and after-pulse ratio}
\label{sec:2:afterpulse}

\begin{figure}[!ht]
	\centering
	\includegraphics[width=0.75\linewidth]{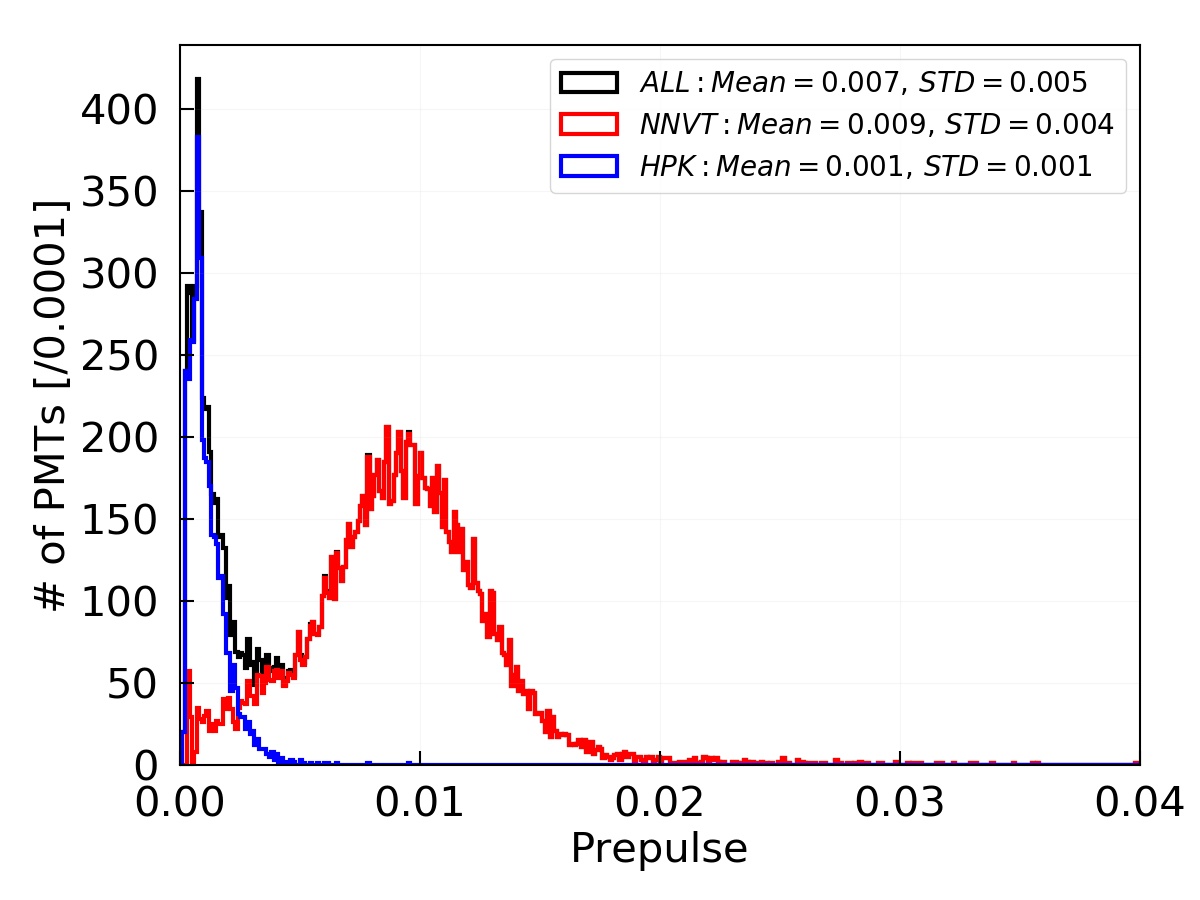}
		\caption{Pre-pulse ratio of all qualified PMTs. Black: all PMTs; red: NNVT PMTs; blue: HPK PMTs.}
	\label{fig:PMT:ppap}       
\end{figure}

Pre-pulses originate mainly from photons which hit the first dynode or MCP directly rather than the photocathode. Hence, they have a smaller pulse amplitude than the main pulses and show up before the main pulse by a few to tens of nanoseconds. With the recorded waveform from the container systems, a charge ratio is checked between the pre-pulse and main pulse in a window of [-80,-10]~ns before the peak of the main pulse. The general results are shown in Fig.\,\ref{fig:PMT:ppap}. The pre-pulse ratio is \textless1\% for HPK PMTs and around 1\% for NNVT PMTs, while there is a larger uncertainty for NNVT MCP-PMTs due to their larger TTS smearing.

\begin{figure}[!ht]
  \includegraphics[width=1.0\textwidth]{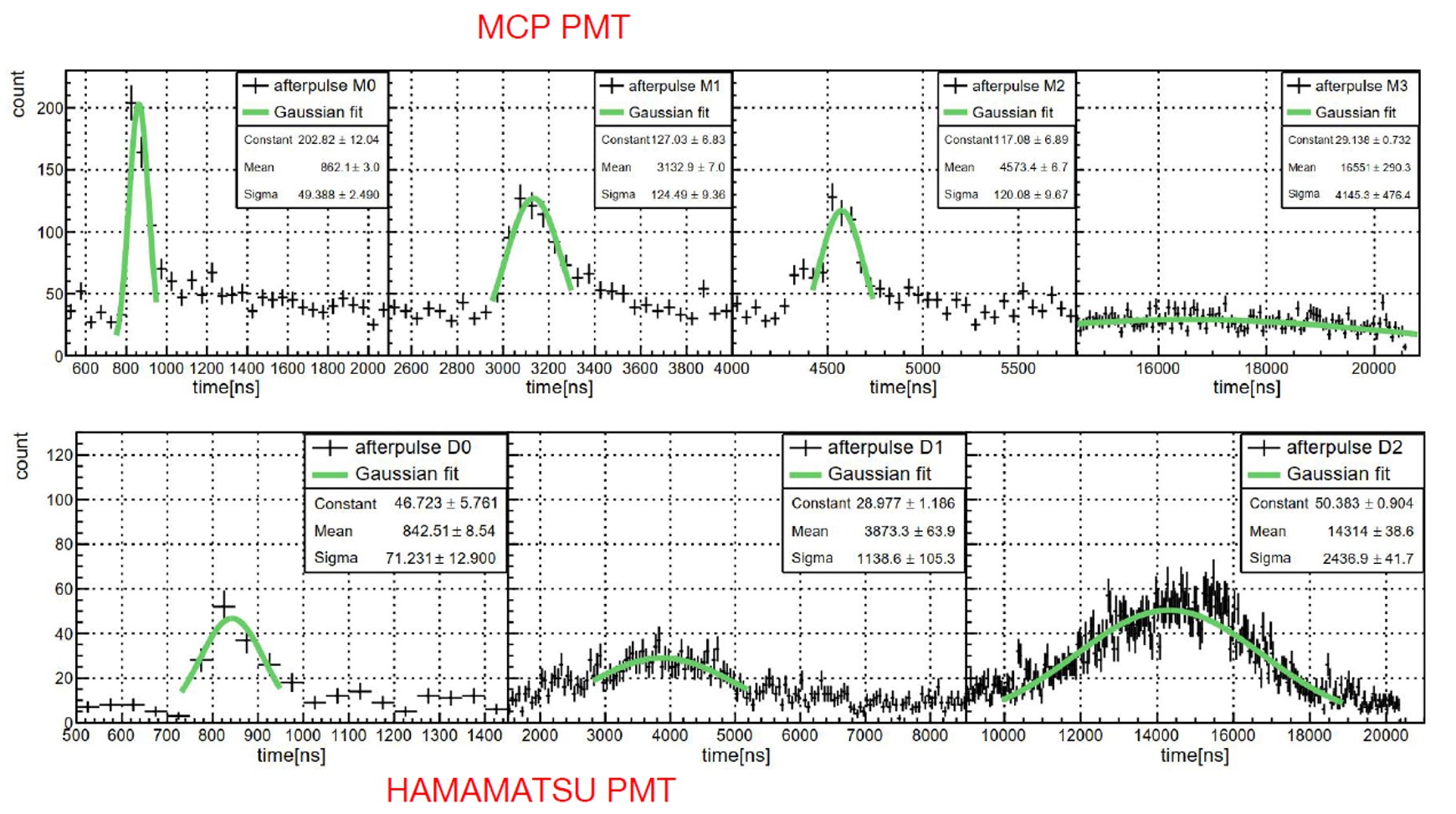}
\caption{Typical time features of after-pulses for two typical PMTs. Top: NNVT; bottom, HPK (Hamamatsu).}
\label{fig:PMT:ap1D}       
\end{figure}

\begin{figure}[!ht]
  \includegraphics[width=1.0\textwidth]{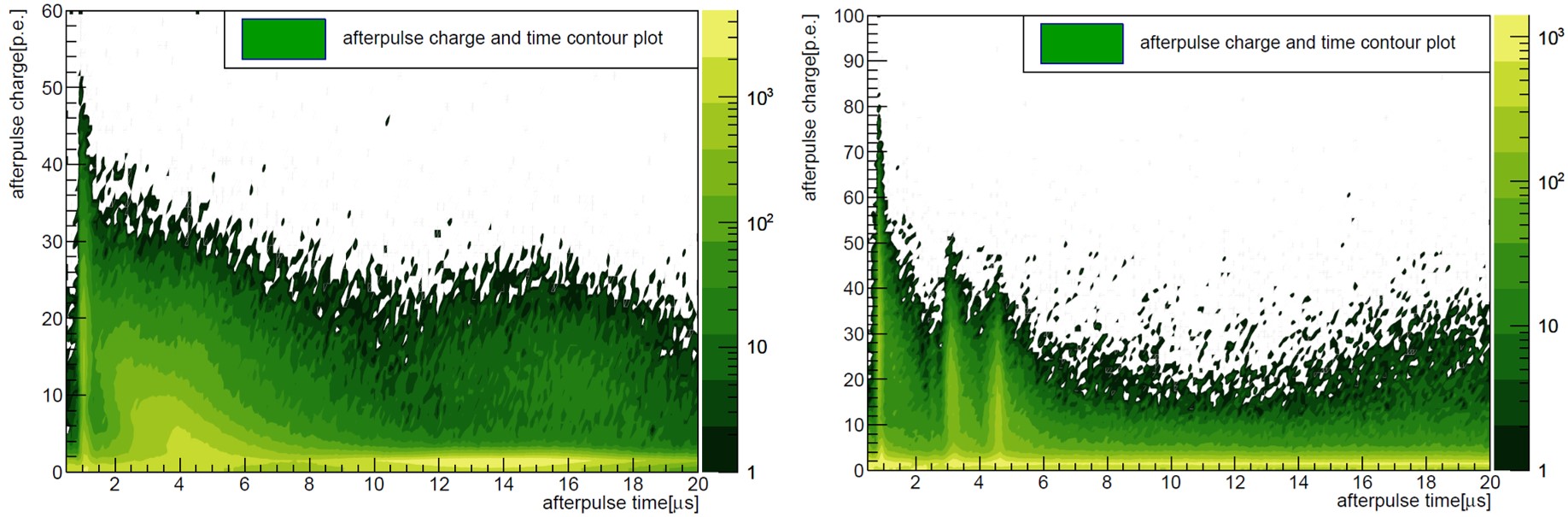}
\caption{After-pulse intensity in p.e. versus time after initial pulse. Left, HPK; Right, NNVT.}
\label{fig:PMT:ap2D}       
\end{figure}

Another possible issue of PMTs is the contamination of the vacuum bulb with gases. Molecules inside the PMT glass envelope can be ionized when the photo-electrons are passing through. These ions will travel back and hit the photocathode, ejecting more electrons. Such events will cause an after-pulse, which features a charge ratio that is proportional to the initial pulse while the after-pulse will be delayed in time by hundreds of nanoseconds to tens of microseconds, depending on the gas molecule, the dimension of the glass bulb and the strength of the applied electric field.
The after-pulse ratio of 20-inch PMTs is calculated in a window from 500\,ns to 20,000\,ns after an initial pulse of about 100\,p.e. with a special LED configuration. This ratio is expected to be less than 15\% of the initial pulse charge. \\
There are already distinct studies about after-pulses~\cite{HamManual,largePMTxia,YBJ8inch,APU2008,APKorea2011}. This study only measured a sample of several $\sim$\,100s PMTs using the scanning station as shown in~\cite{JUNOPMTapYu}. 
The mean after-pulse charge ratio is calculated from 150 NNVT MCP-PMTs and 7 HPK dynode PMTs respectively. This sample leads to the result of 6.7\,$\%$ for NNVT MCP-PMTs and 12.0\,$\%$ for HPK dynode PMTs. At the same time, it was found that the HPK dynode PMTs and NNVT MCP-PMTs have very different after-pulse features, as indicated for an example PMT shown in Fig.\,\ref{fig:PMT:ap1D}: the selected MCP-PMT from NNVT was measured with a $\sim$7.0\% after-pulse charge ratio, showing four typical after-pulse peaks in time around 0.8\,\textmu s (ratio 3.1\,$\%$), 3.2\,\textmu s (ratio 1.4$\%$), 4.6\,\textmu s (ratio 1.7$\%$) and 17.0\,\textmu s(ratio 0.8$\%$, only a hint), while the selected HPK dynode PMT was measured with an after-pulse charge ratio of 13.8\% and typical peaks at 0.8\,\textmu s (ratio 1.1$\%$), 3.9\,\textmu s (ratio 8.7\,$\%$) and 14.2\,\textmu s (ratio 4.0\,$\%$). The intensity in charge of a single after-pulse is mainly at the SPE level, while it can reach up to several O(10)\,p.e.s as shown in Fig.\,\ref{fig:PMT:ap2D}: so in the case of this particular HPK PMT, the charge ratio of the after-pulses at 3.9\,\textmu s is larger than the pulse at 14.2\,\textmu s despite the count rates are vice versa (compare Fig.\,\ref{fig:PMT:ap1D} again).

\subsection{Non-linearity}
\label{sec:2:nonlinaerity}

\begin{figure}[!ht]
  \includegraphics[width=1\textwidth]{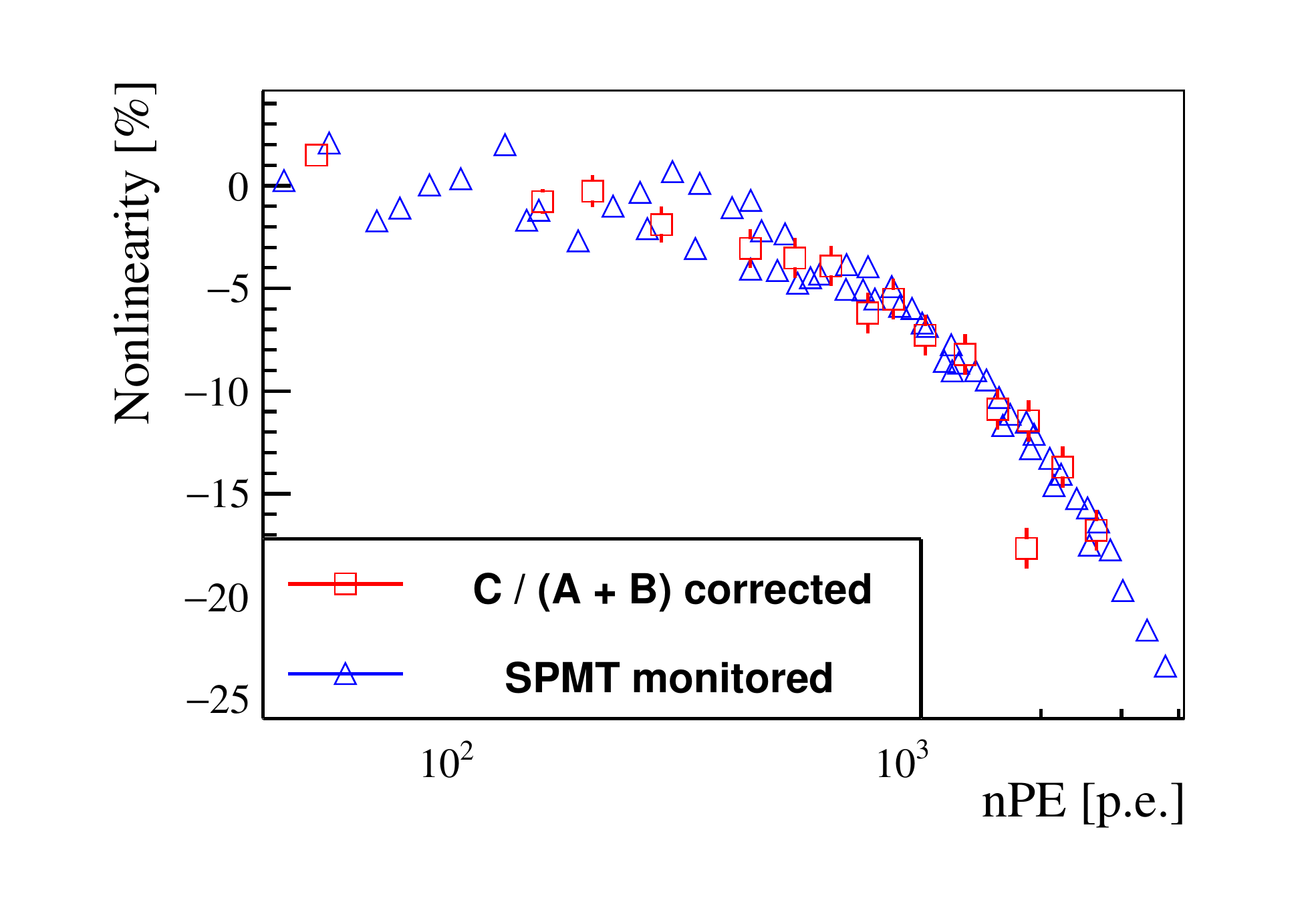}
\caption{Measured linearity of a 20-inch PMT by double-LED method (C/(A+B)) (red rectangular) and cross checking with 3-inch PMT (SPMT monitored) (blue triangle).}
\label{fig:PMT:linearity}       
\end{figure}

Neutrino events detected in JUNO typically correspond to a signal strength of up to few hundreds of p.e.s in each of the 20-inch PMTs. Furthermore, there are huge signals expected from physics or background events with much higher energy such as cosmic muons. A dynamic range in linearity of the PMT response with the designed HV divider is required to reach 1,000\,p.e.s with less than 10$\%$ distortion. The HV divider design for the 20-inch PMTs of JUNO described in~\cite{JUNOPMTsignalover,JUNOPMTsignalopt,JUNOPMTflasher,JUNOPMTinstr} was designed to provide a DC of $\sim$100\,\textmu A for HPK PMTs and $\sim$180\,\textmu A for NNVT PMTs at a working gain of $1\times10^7$. A few samples of both PMT types were measured by the double-LED (LED A, LED B, C is when LED A and B flashing at the same time) method (C/(A+B))~\cite{JUNOPMTlinearity} in a pulsed mode, and a cross-checking method between the 20-inch PMT and the 3-inch PMT. One resulting curve of a NNVT MCP-PMT is shown in Fig.\,\ref{fig:PMT:linearity}, where the PMT response with the designed HV divider can satisfy the requirements, and the results of both methods are consistent. The current design can satisfy the requirement on the linearity response. Another standalone paper will be prepared on this topic.

\subsection{Earth magnetic field (EMF) effect}
\label{sec:2:emf}

\begin{figure}[!ht]
  \centering
    \includegraphics[width=0.75\textwidth]{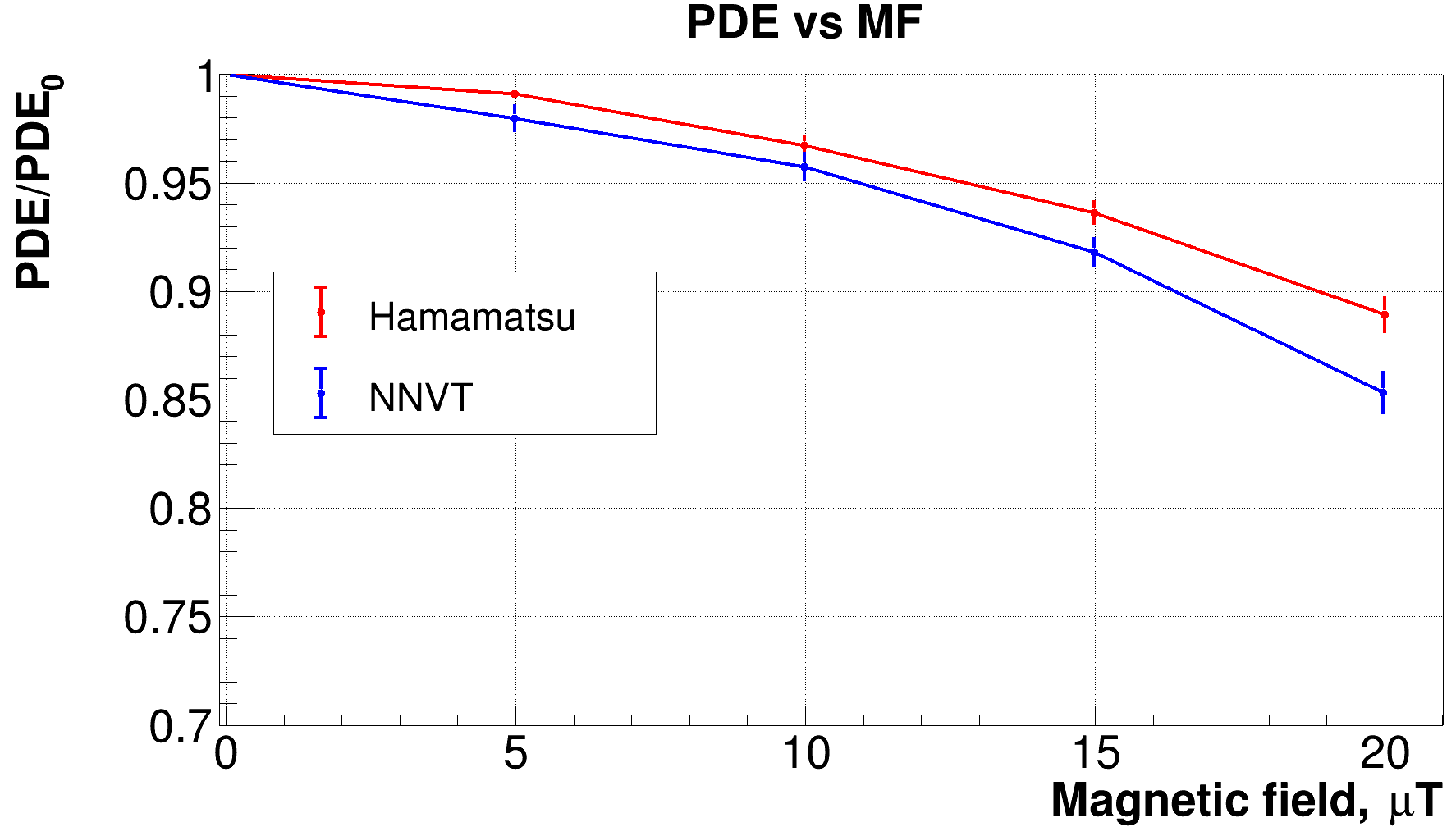}
  \caption{Averaged PMT PDE versus remaining magnetic field (MF) strength tested with 9 HPK and 15 NNVT PMTs.}
  \label{fig:PMT:emf}
\end{figure}

PMTs with large vacuum bulbs are significantly affected by magnetic fields when the photoelectron drifts to the collection dynode or MCP, in fact, their performance is even sensitive to the Earth magnetic field (EMF, $\sim$50\,\textmu T at the JUNO's experimental site). This requires a more thorough characterization on the magnetic field effect to PMTs. Using the scanning station, the EMF's strength around PMT's location can be surveyed in the range -50\textmu T to +50\textmu T by changing the DC configuration of the Helmholtz coils in the housing dark rooms. As shown in Fig.\,\ref{fig:PMT:emf}, the variation of the performance of PMTs is negligible (\textless 1\%$\sim$3\%) when the residual field is less than 10$\%$ of EMF. The direction of the magnetic field is perpendicular to the central axis of the PMT, which is considered as the worst condition, and followed by the direction of the collection box of dynode PMTs or the direction of micro-channel of MCP-PMT. The container system and the scanning stations are operating with the magnetic field suppressed to a level of $<5$\,\textmu T. Following these measurements, the requirement on the residual magnetic field in the JUNO detector is specified to be $\lesssim$5\textmu T. As one can see, NNVT PMTs are affected more strongly than HPK PMTs, which could be explained by differences in the photoelectron focusing process. For more details, see~\cite{JUNOstationIlya}.

\subsection{Uniformity}
\label{sec:2:uniform}

\begin{figure}[!ht]
  \centering\includegraphics[width=0.7\textwidth]{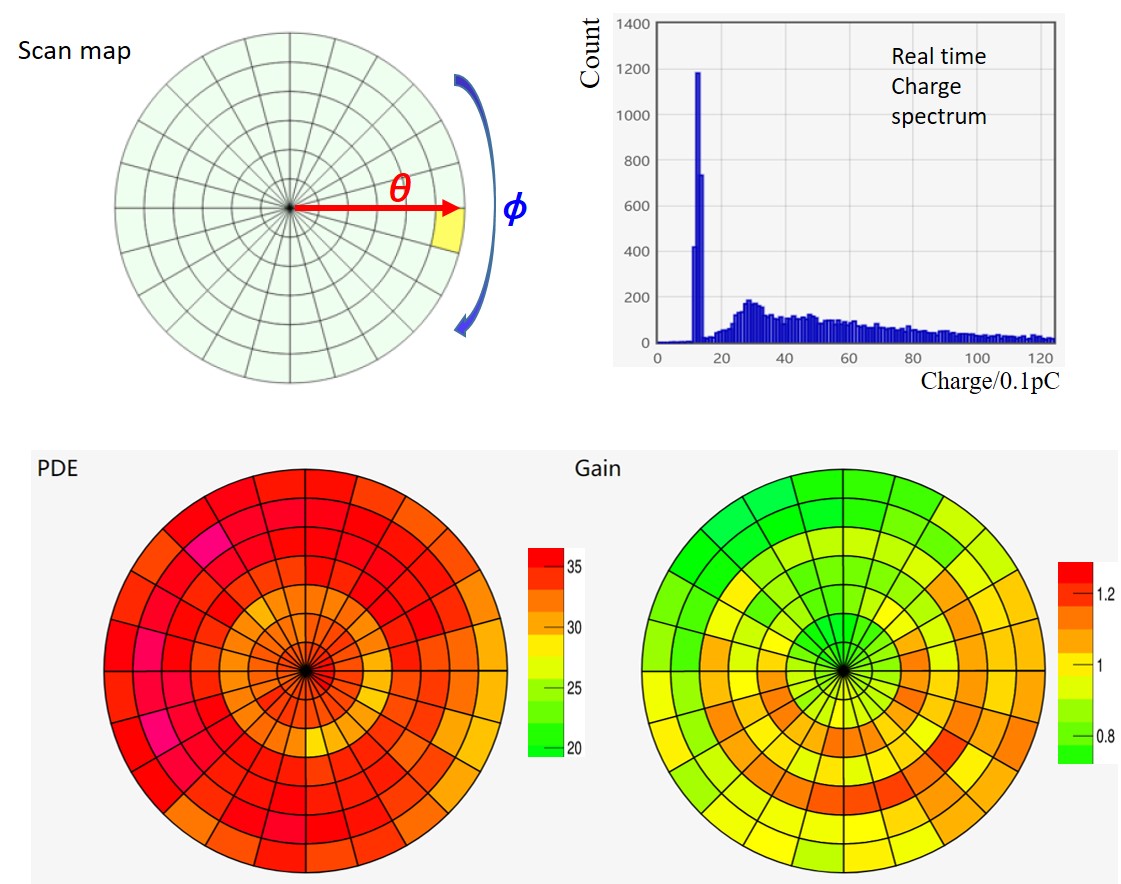}
\caption{Exemplary data of a PMT scan with the scanning station. Top left: scanning process indication; top right: integrated charge spectrum of selected area element; bottom: measured maps of an NNVT MCP-PMT (PA2004-1007) on PDE (bottom left) and gain (bottom right). }
\label{fig:PMT:stationExample}       
\end{figure}

For a 20-inch PMT, the anode uniformities along the full photocathode are significant for a better understanding of the PMT itself and the detector's response. An exemplary PMT surface scan performed by the scanning station is shown in Fig.\,\ref{fig:PMT:stationExample} with 15$^{\circ}$ steps along the azimuthal angle ($\phi$) by the seven LEDs distributed in zenith angle ($\theta$), the online integrated charge spectrum of one spot, and the scanned map on PDE and gain. Thanks to the powerful capability of the scanning stations, detailed studies were made using a few thousand PMTs (585 HPK PMTs, 2,658 NNVT PMTs, including 939 high-QE PMTs, and 1,719 low-QE PMTs) to characterize the uniformity, which will be discussed in detail as follows.

\subsubsection{Uniformity of gain and gain excess normal distribution factor (gENF) }
\label{sec:3:uniform:gain}

\begin{figure}[!ht]
  \begin{minipage}[!ht]{0.495\linewidth}
  \centering
          {\includegraphics[width=\linewidth]{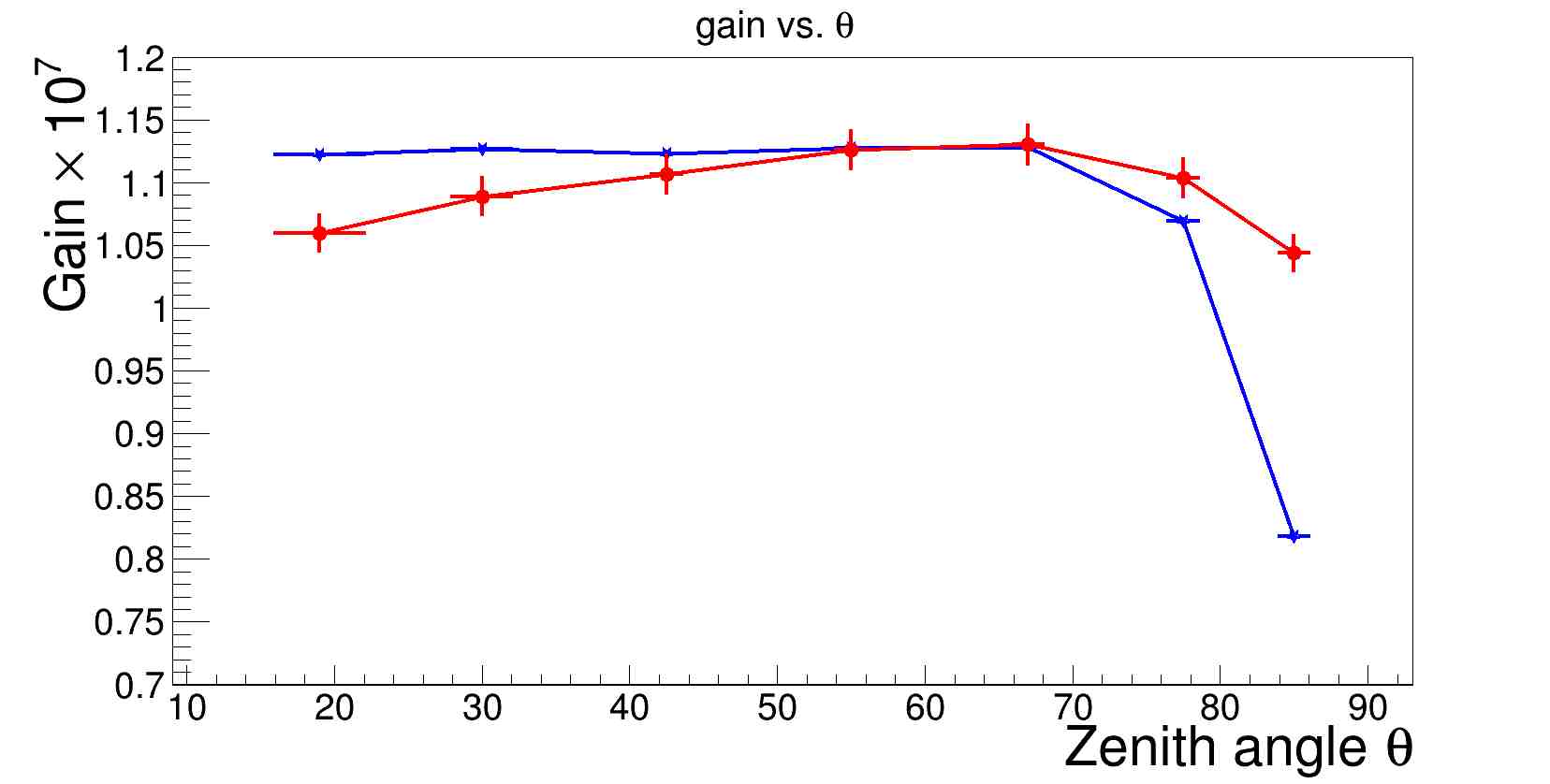}}
  \end{minipage}
  \begin{minipage}[!ht]{0.495\linewidth}
  \centering
    {\includegraphics[width=\linewidth]{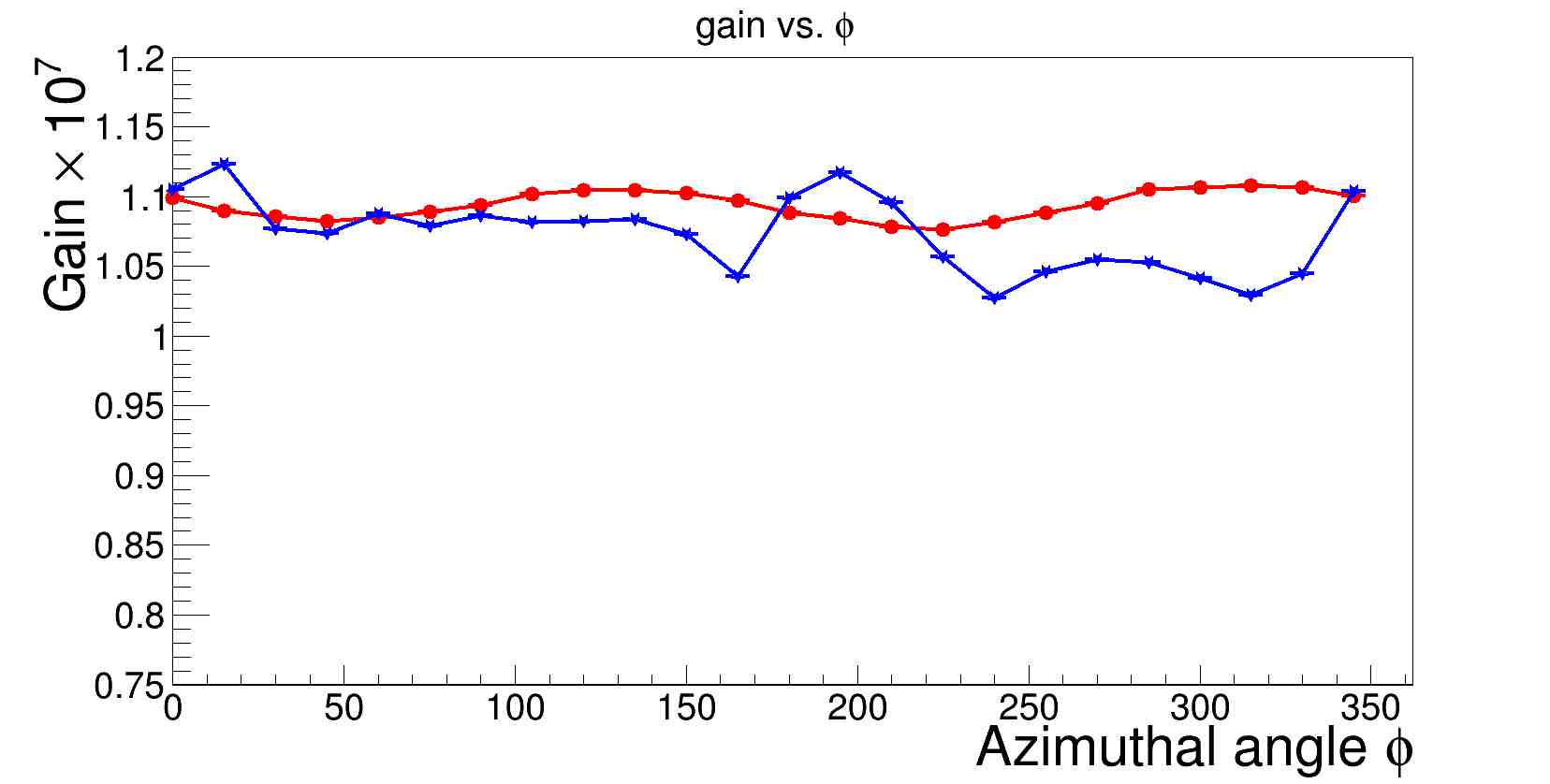}}
  \end{minipage}
\caption{Uniformity of gain along the zenith angle ($\theta$, left) and azimuthal angle (($\phi$, right). HPK: blue; NNVT: red.}
\label{fig:PMT:gainU}       
\end{figure}

The charge response along the whole photocathode surface can be described firstly by its gain. As known, there will be a large response deformation around the equator ($\theta\sim$90$^\circ$) of the glass bulb, compared to the top pole ($\theta\sim$0$^\circ$) of the photocathode. Following the gain definition in sec.\,\ref{sec:2:gain} and applying the determined HV for a gain of $1\times10^{7}$, the gain is determined for each LED spot measurement individually. The uniformity along the whole photocathode in the zenith ($\theta$) and azimuthal ($\phi$) angles is shown in Fig.\,\ref{fig:PMT:gainU}: each point is the average of all the measurements of the full, tested PMT sample with the same zenith angle or azimuthal angle individually, and the uncertainty bar in the vertical axis is calculated from the error of the average. The gain of HPK PMTs is more uniform for $\theta$\textless70$^\circ$ along the zenith angle, but it decreases sharply by $\sim$30\% towards the equator. Along the azimuthal angle, the gain of HPK PMTs shows systematically smaller values ($\sim$5\%) and some sub-structures (peak-to-peak variation $\sim$10\%) around its collection box in a rectangle shape, including the peaks before ($\phi\sim$200$^\circ$), around $\phi\sim$270$^\circ$, and $\phi\sim$0$^\circ$ in particular. On the other hand,
the gain of NNVT PMTs is continuously increasing as the zenith angle increases up to $\theta\sim$70$^\circ$ before drops at the equator with a total variation range of $\sim$10\%.
Along the azimuthal angle ($\phi$), the NNVT PMTs are more uniform than the HPK PMTs and one can observe a two-cycles-oscillation with the valleys located at around $\phi\sim$50$^\circ$ and $\phi\sim$230$^\circ$ with a peak-to-peak variation of $\sim$3\%, which should be related to the direction of the MCP channels.
The averaged gain of the full photocathode of all PMTs tested by the scanning
station is 1.07$\times10^{7}$ for HPK PMTs and 1.09$\times10^{7}$ for NNVT PMTs, and they are systematically higher than the gain measured by the container ($1.00\times10^7$ for HPK PMTs and $1.03\times10^7$ for NNVT PMTs, respectively). The difference is assumed from the amplification factor of the stations, which is not exactly equal to the claimed $\times$10. 

\begin{figure}[!ht]
  \begin{minipage}[!ht]{0.495\linewidth}
  \centering
          {\includegraphics[width=\linewidth]{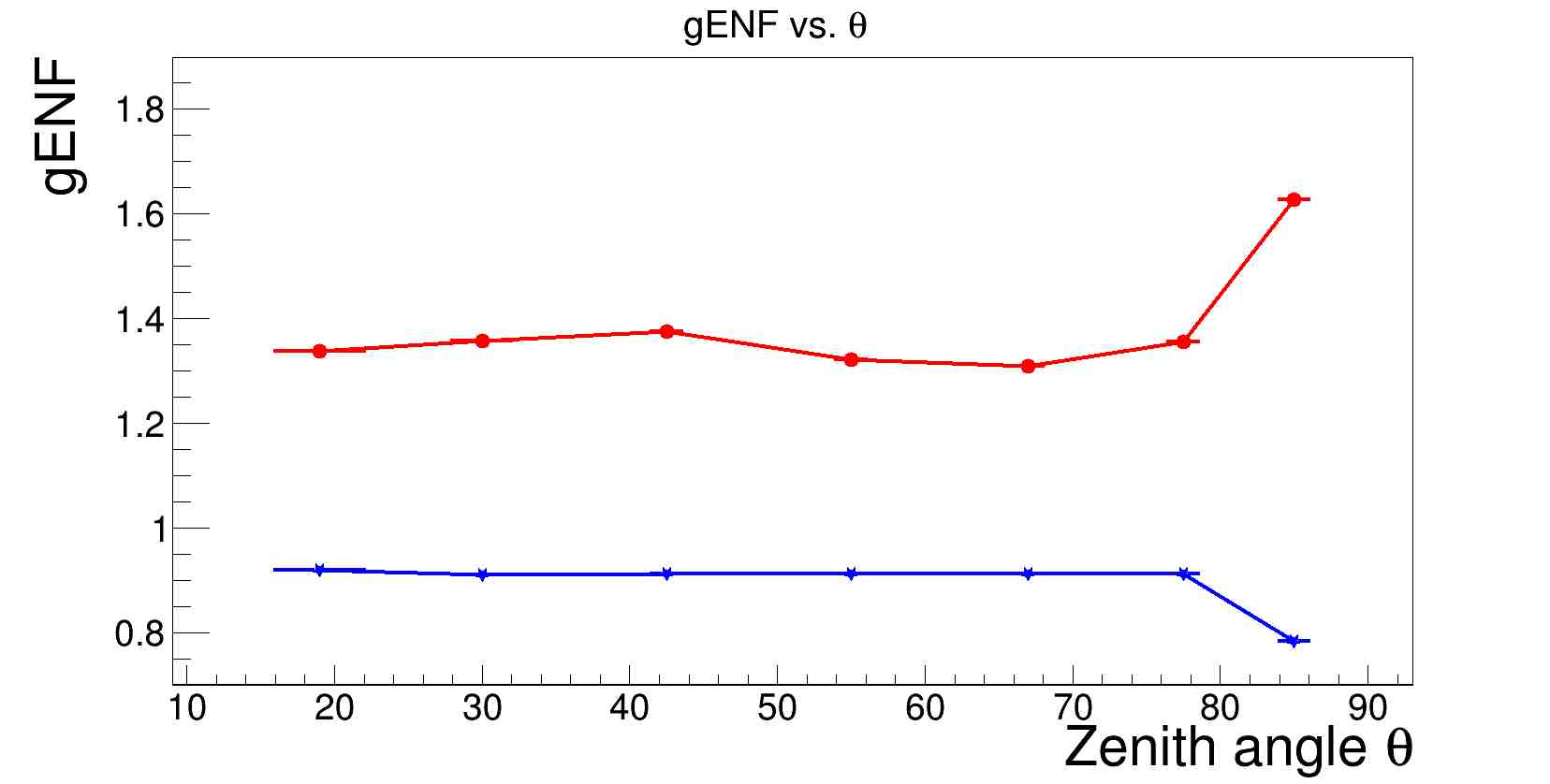}}
  \end{minipage}
  \begin{minipage}[!ht]{0.495\linewidth}
  \centering
    {\includegraphics[width=\linewidth]{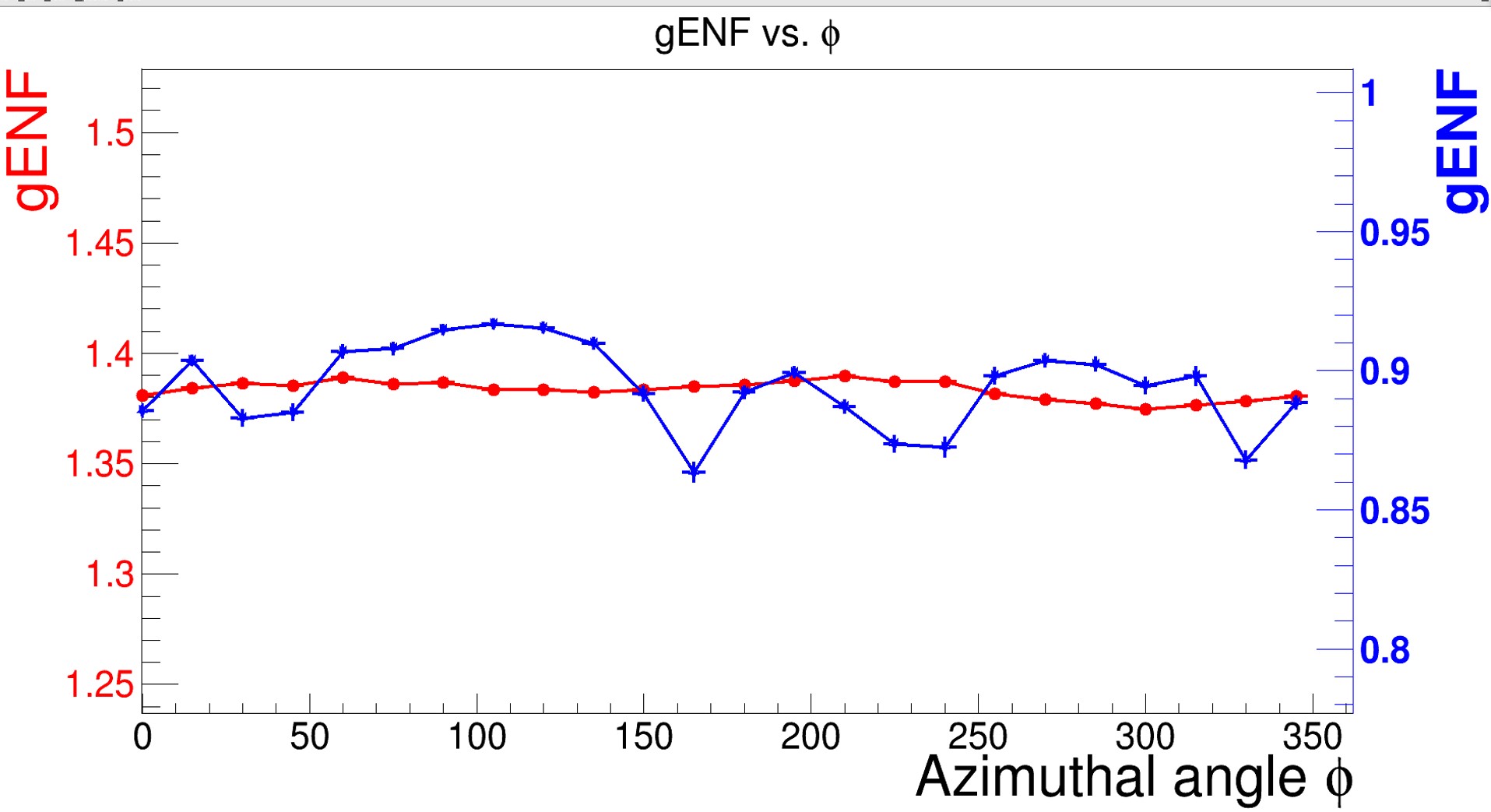}}
  \end{minipage}
\caption{Uniformity of gENF along the zenith angle ($\theta$, left) and azimuthal angle (($\phi$, right). HPK: blue; NNVT: red.}
\label{fig:PMT:stationgENF}       
\end{figure}

As a correlated factor of the gain, the uniformity of gENF (defined in sec.\,\ref{sec:2:genf}) is also checked with the same data set as the gain uniformity, which is shown in Fig.\,\ref{fig:PMT:stationgENF} in the zenith ($\theta$) and azimuthal angles ($\phi$). Firstly, the absolute value difference between HPK and NNVT PMTs is related to the PMT itself as discussed in sec.\,\ref{sec:2:genf}. The gENF on average of the whole photocathode of the full, scanned sample is 0.89 for HPK PMTs and 1.38 for NNVT PMTs. Comparing to the values acquired by the containers, the discrepancy is mainly assumed from the averaging scheme and the contribution of the gENF at larger $\theta$. Except the absolute value difference, the gENF of both PMT types is more uniform along the zenith angle (when $\theta$\textless80$^\circ$) and azimuthal angle. At the same time, there are a few sub-structures worth mentioning:
\begin{itemize}
	\item[(1)] the gENF value shows a larger change for $\theta$\textgreater80$^\circ$.
	\item[(2)] NNVT PMTs show a variation ($\sim$6\%) sub-structure for $\theta$\textless80$^\circ$ (peak at $\theta\sim$50$^\circ$ and valley at $\theta\sim$70$^\circ$).
	\item[(3)] the gENF factor of HPK PMTs along the azimuthal angle shows a small structure ($\sim$4\%), which could be related to that in Fig.\,\ref{fig:PMT:gainU}.
	\item[(4)]the gENF factor of NNVT PMTs along the azimuthal angle shows a two-cycles-oscillation (variation $\sim$1.5\%) at its gain but with opposite phases (peaks at $\phi\sim$60$^\circ$ and $\phi\sim$210$^\circ$).
\end{itemize}

\subsubsection{Uniformity of resolution and excess noise factor (ENF)}
\label{sec:3:uniform:res}

\begin{figure}[!ht]
  \begin{minipage}[!ht]{0.495\linewidth}
  \centering
          {\includegraphics[width=\linewidth]{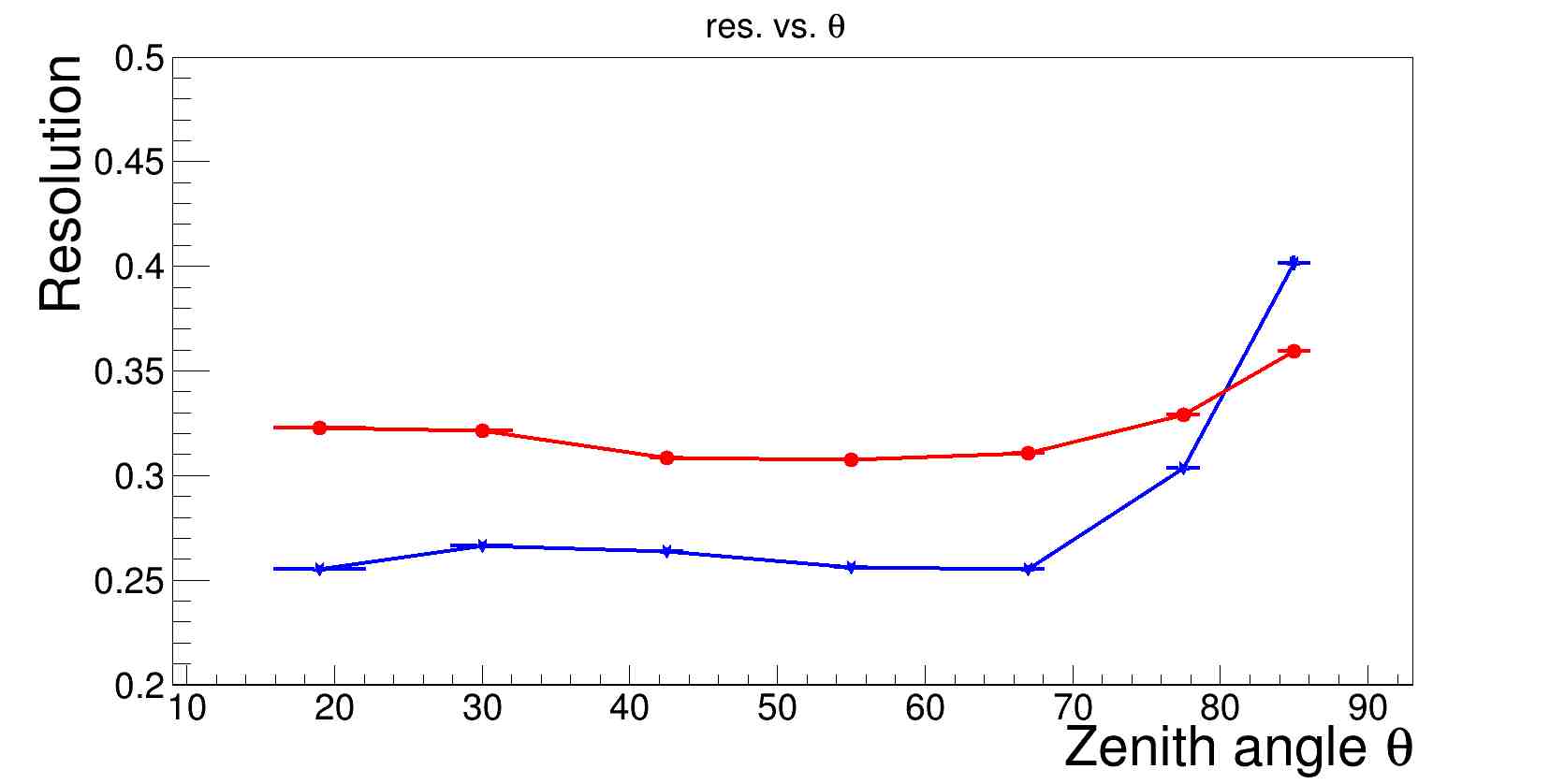}}
  \end{minipage}
  \begin{minipage}[!ht]{0.495\linewidth}
  \centering
    {\includegraphics[width=\linewidth]{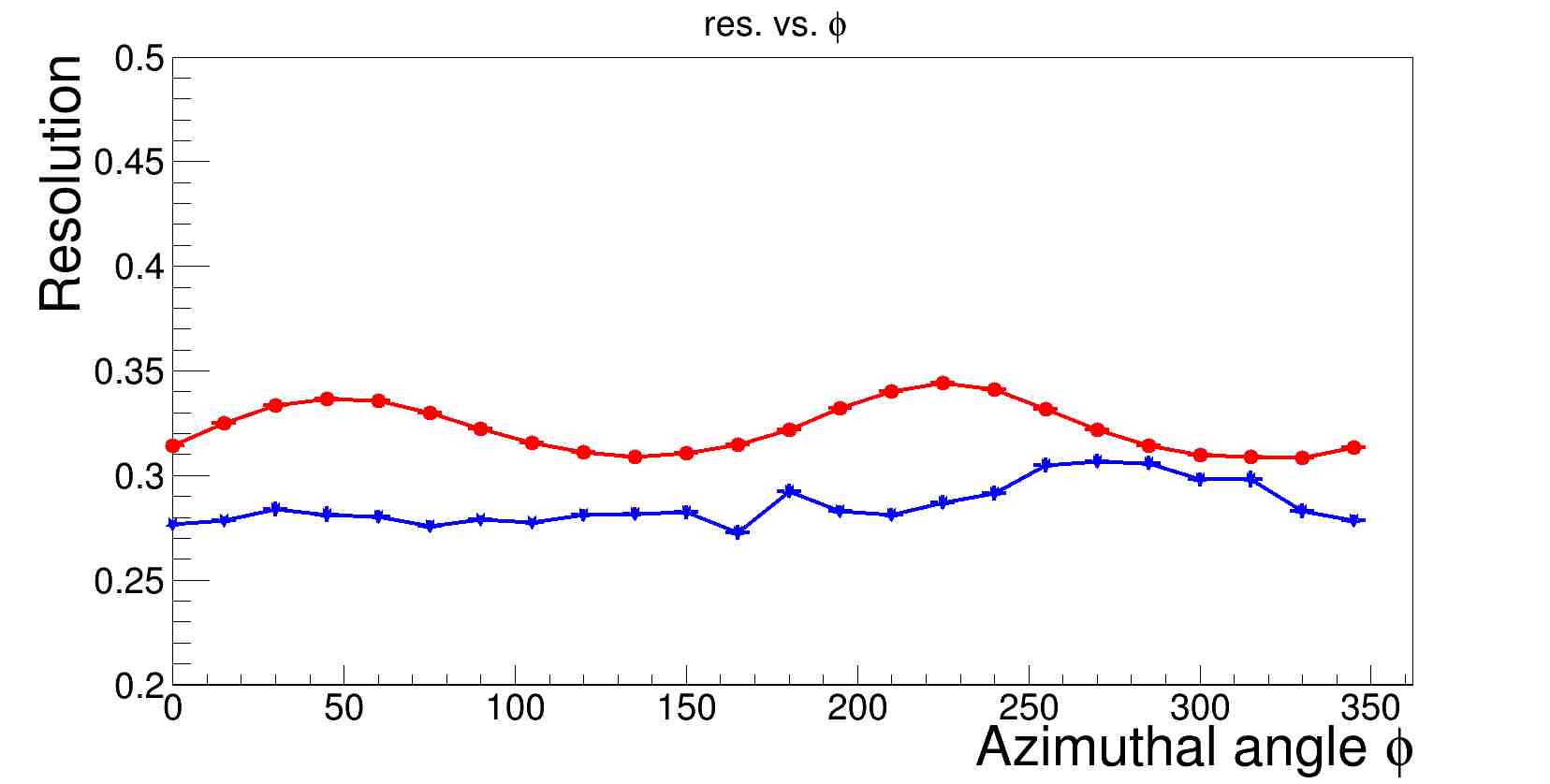}}
  \end{minipage}
\caption{Uniformity of the SPE charge resolution along the zenith angle ($\theta$, left) and azimuthal angle ($\phi$, right). HPK: blue; NNVT: red.}
\label{fig:PMT:res1D}       
\end{figure}

Following the SPE charge resolution definition in sec.\,\ref{sec:3:resolution}, the uniformity of the SPE charge resolution on the whole photocathode is shown in Fig.\,\ref{fig:PMT:res1D} in the zenith ($\theta$) and azimuthal ($\phi$) angles.
Each point is directly averaged from all PMTs tested by the scanning
station with the same zenith angle or azimuthal angle individually. The total mean charge resolution of the whole photocathode and for all the scanned PMTs is 0.28 for HPK and 0.32 for NNVT. Comparing the uniformity results of HPK and NNVT PMTs, the charge resolution is almost constant when $\theta$\textless70$^\circ$ but getting larger towards to the equator for both of them (but in particular for HPK PMTs). Along the azimuthal angle, the charge resolution of both HPK and NNVT PMTs shows a similar trend but in opposite phase at the gain as shown on the right of Fig.\,\ref{fig:PMT:gainU}: it is systematically higher than average around $\phi\sim$270$^\circ$ and small fluctuations can be observed (variation $\sim$14\%) for HPK PMTs and a two-cycles-oscillation (variation $\sim$10\%) for the NNVT PMTs with peaks located at around $\phi\sim$50$^\circ$ and $\phi\sim$230$^\circ$.

\begin{figure}[!ht]
  \begin{minipage}[!ht]{0.495\linewidth}
  \centering
          {\includegraphics[width=\linewidth]{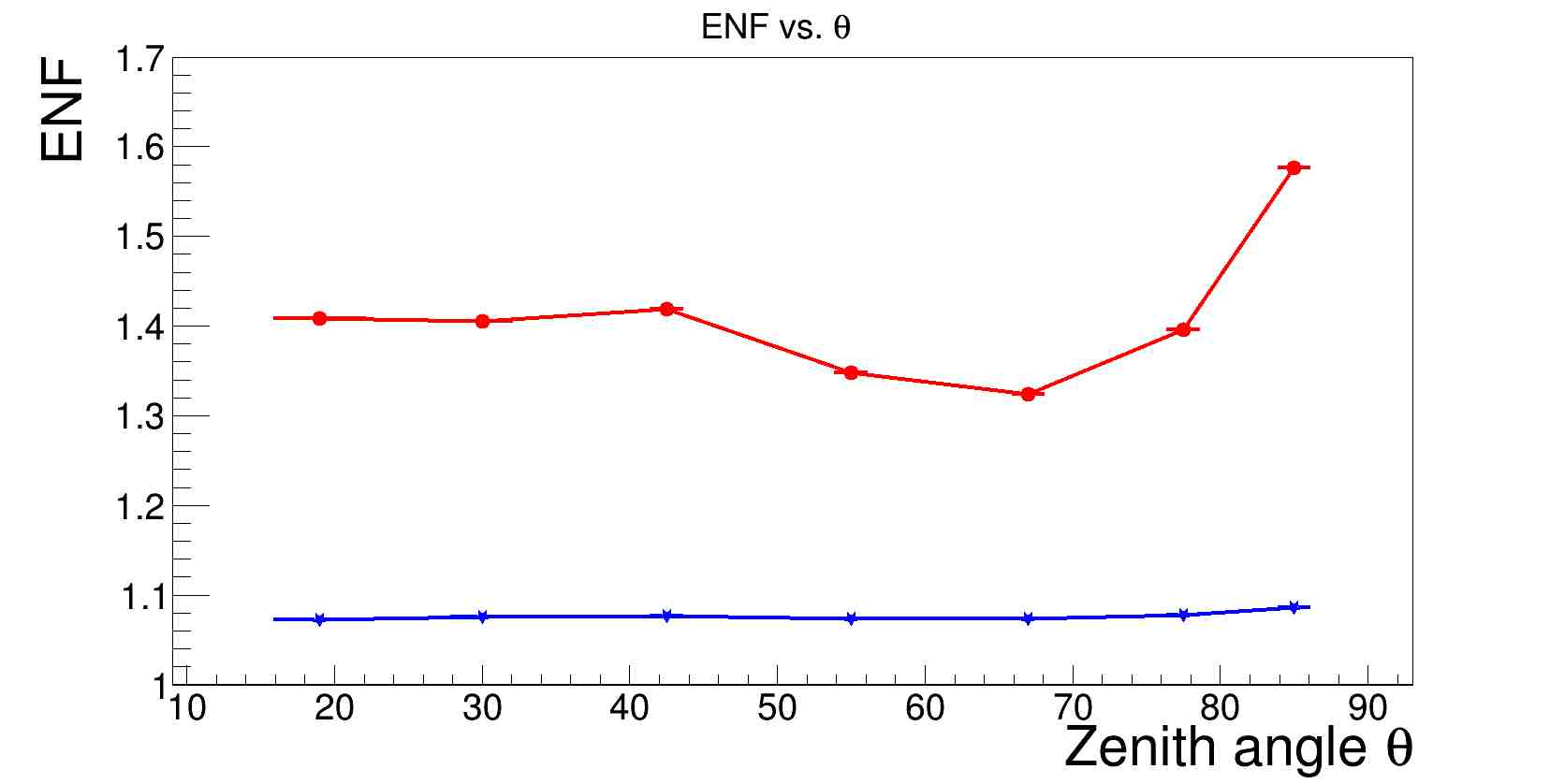}}
  \end{minipage}
  \begin{minipage}[!ht]{0.495\linewidth}
  \centering
    {\includegraphics[width=\linewidth]{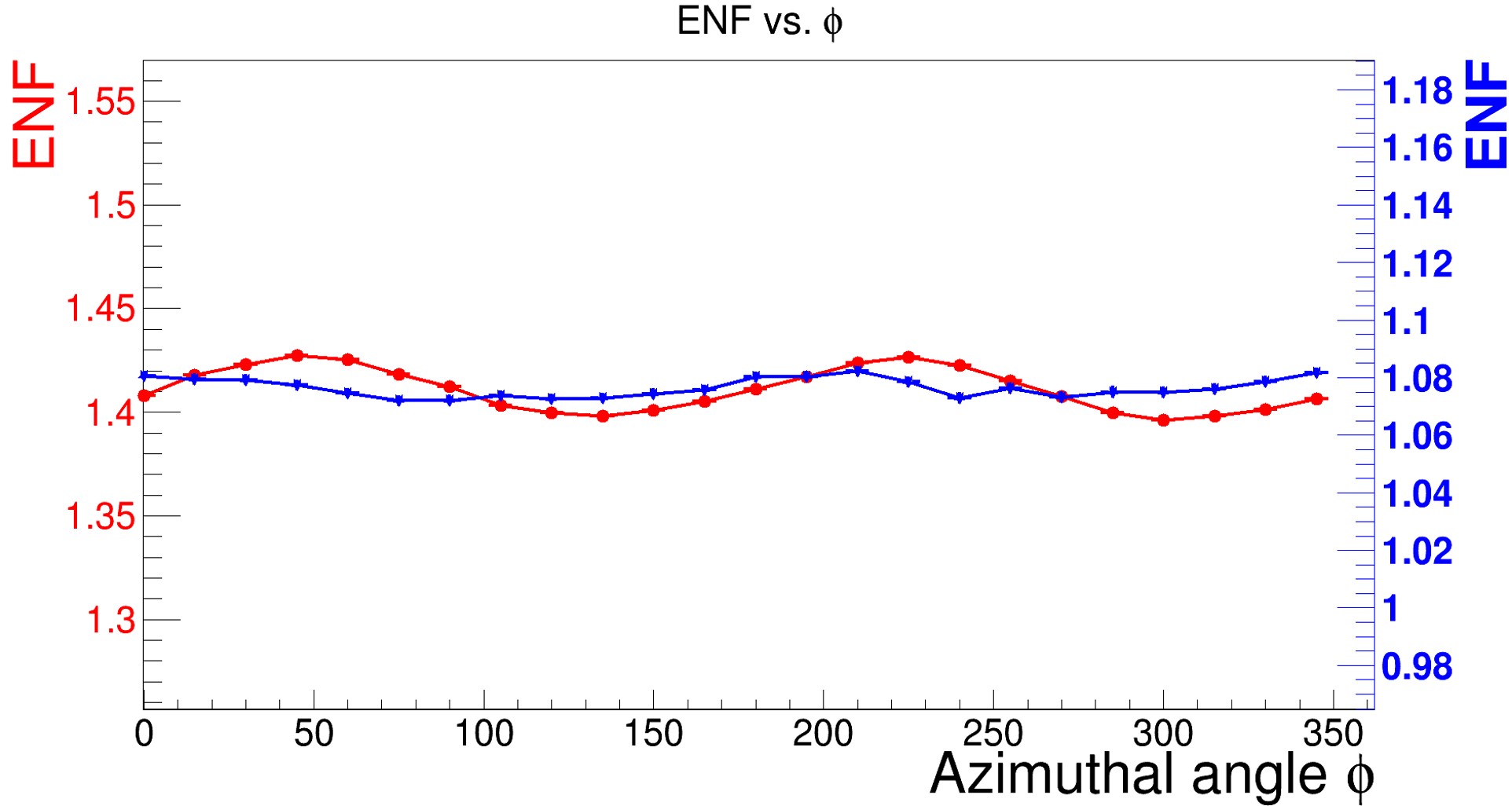}}
  \end{minipage}
\caption{Uniformity of ENF along the zenith angle ($\theta$, left) and azimuthal angle ($\phi$, right). HPK: blue; NNVT: red.}
\label{fig:PMT:stationENF}       
\end{figure}

As a correlated factor to the SPE charge resolution, the uniformity of ENF (defined in sec.\,\ref{sec:3:resolution}) along the whole photocathode is determined for the same data set as done for the charge resolution analysis, see Fig.\,\ref{fig:PMT:stationENF}. The ENF on average over the whole photocathode of all scanned PMTs is 1.08 for HPK PMTs and 1.41 for NNVT PMTs. The HPK PMTs show a more uniform performance along both zenith angle and azimuthal angles. The NNVT PMTs reach $\sim$1.5\% as the zenith angle increases and also show a two-cycles-oscillation with a variation of $\sim$1.5\% along the azimuthal angle with the peaks located at around $\phi\sim$50$^\circ$ and $\phi\sim$230$^\circ$. The NNVT PMTs show a similar trend as in case of the gENF factor and charge resolution along the zenith angle to a max. change of $\sim$18\%, while a clear dependence to azimuthal angles also with a two-cycles-oscillation with peak-to-peak variation of $\sim$2\% peaks located at around $\phi\sim$50$^\circ$ and $\phi\sim$230$^\circ$.

\subsubsection{Uniformity of photon detection efficiency (PDE)}
\label{sec:3:uniform:pde}

\begin{figure}[!ht]
  \begin{minipage}[!ht]{0.495\linewidth}
  \centering
    {\includegraphics[width=\linewidth]{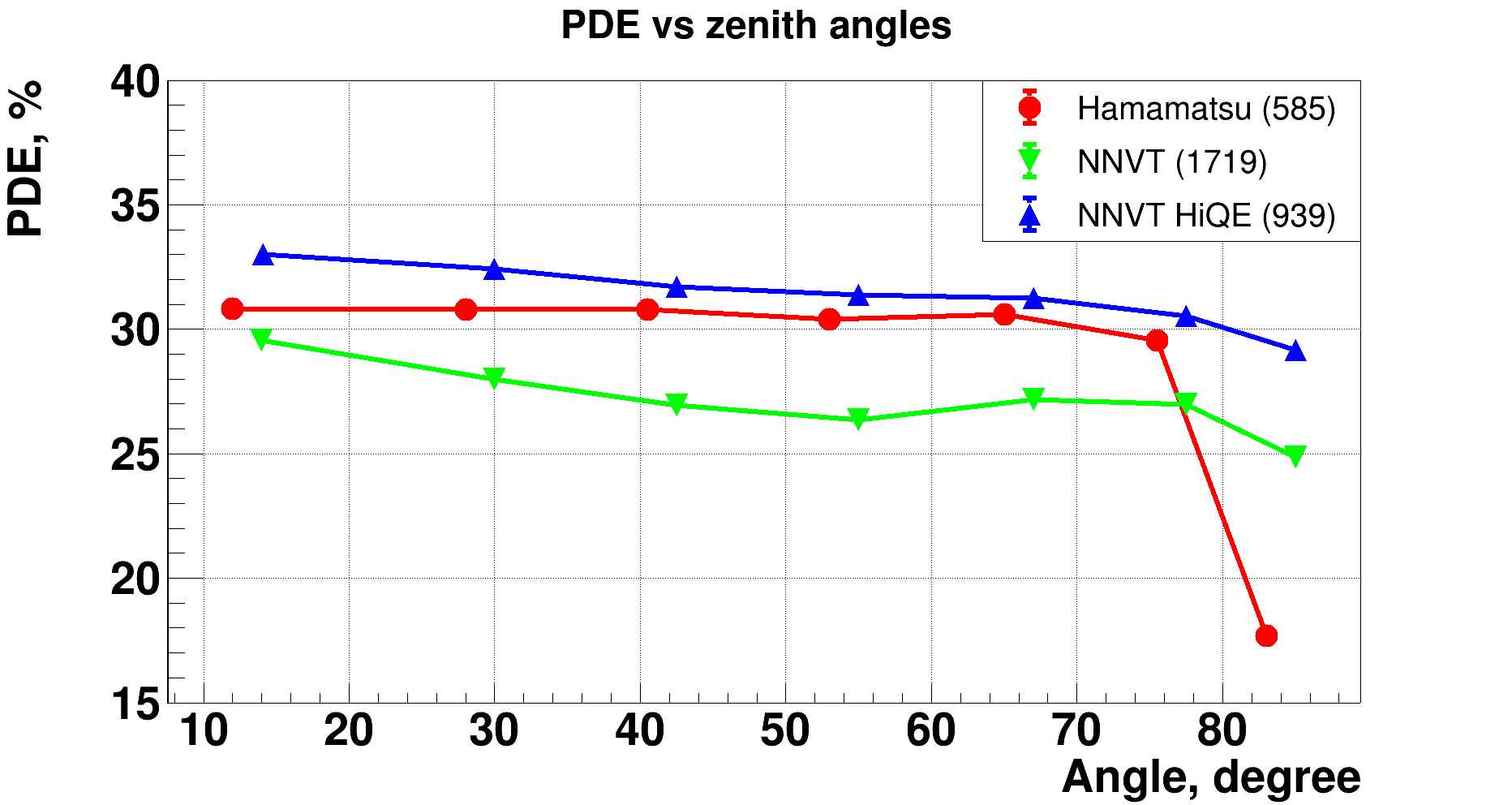}}
  \end{minipage}
  \begin{minipage}[!ht]{0.495\linewidth}
  \centering
    {\includegraphics[width=\linewidth]{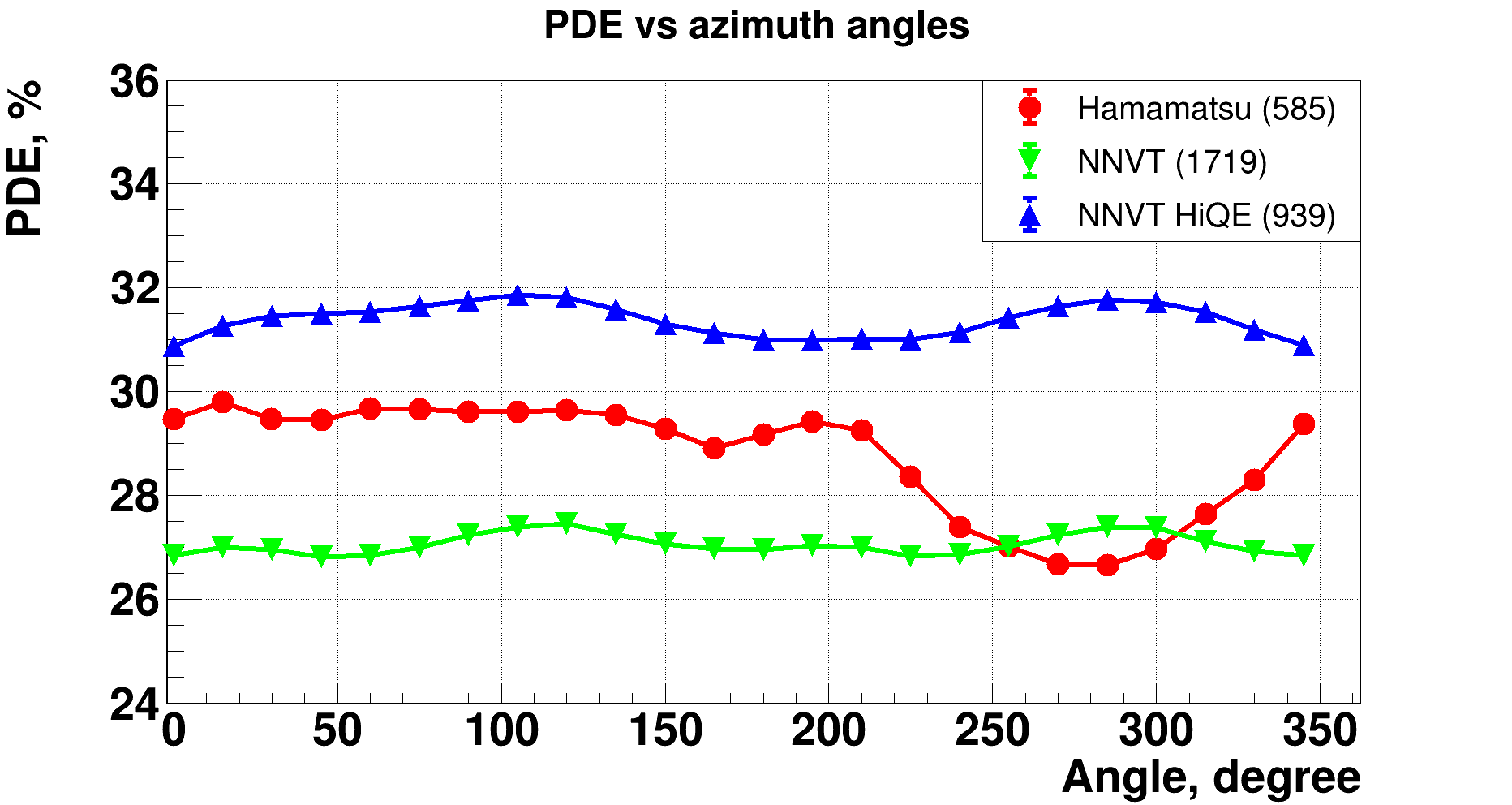}}
  \end{minipage}
\caption{Uniformity of PDE along the zenith angle ($\theta$, left) and azimuthal angle ($\phi$, right). Blue,: HPK PMTs; red: NNVT low-QE PMTs; green: NNVT high-QE PMTs.}
\label{fig:PMT:pde1D}       
\end{figure}

The measurement of the PDE uniformity along the whole photocathode of all scanned PMTs is shown in Fig.\,\ref{fig:PMT:pde1D} and divided in three data sets: HPK (Hamamatsu), High-QE NNVT (HiQE) and Low-QE NNVT (NNVT). The PDE along the zenith ($\theta$) and azimuthal ($\phi$) angle is the averaged value of all the measured spots with the same zenith and azimuthal angle of all the scanned PMTs. The high-QE NNVT PMTs show a better performance than HPK PMTs in both directions: a better uniformity towards the equator and a larger absolute value in both zenith and azimuthal direction. The low-QE NNVT PMTs show a smaller PDE which has slightly better uniformity than for the high-QE PMTs along $\theta$. The HPK PMTs' PDE sharply drops from $\theta\sim$80$^\circ$ to $\theta\sim$90$^\circ$ by about 40\%, and a clear asymmetry (max. $\sim$10\%)  is observed along azimuthal angle related to its focusing box direction around $\phi\sim$270$^\circ$. Both the low-QE and high-QE NNVT PMTs show a two-cycles-oscillation along azimuthal angle (peak-to-peak variation is $\sim$3\%) and peaks located at around $\phi\sim$110$^\circ$ and $\phi\sim$300$^\circ$.

\begin{figure}[!ht]
  \begin{minipage}[!ht]{0.495\linewidth}
  \centering
    {\includegraphics[width=\linewidth]{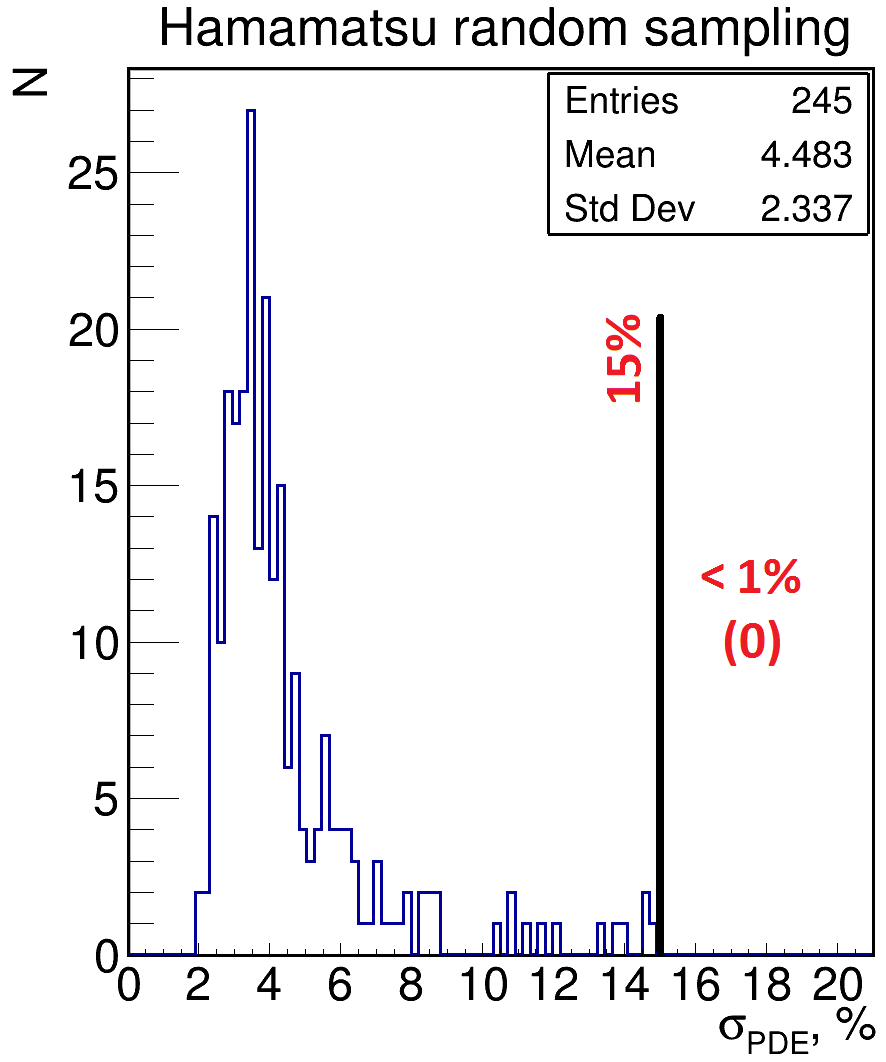}}
  \end{minipage}
    \begin{minipage}[!ht]{0.495\linewidth}
  \centering
          {\includegraphics[width=\linewidth]{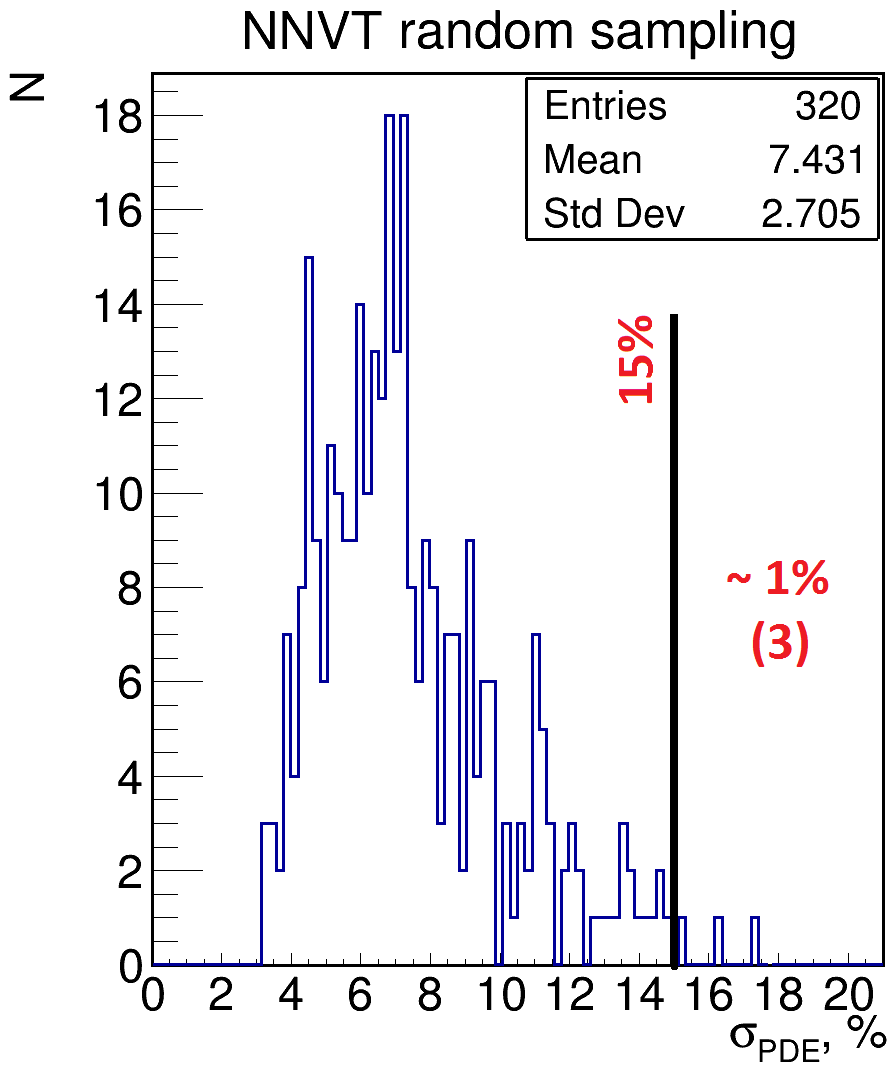}}
  \end{minipage}
  \caption{PDE non-uniformity distributions for randomly sampled PMTs. Black lines show the threshold for PDE non-uniformity (15\%), numbers in brackets indicate failed PMTs. Left: HPK (Hamamatsu) PMTs; right: NNVT PMTs.}
  \label{fig:PMT:sigmaPDE_random}
\end{figure}

Following the requirements, the non-uniformity of PDE (defined as the ratio of the distribution of the PDE of each light spot to their average) should not exceed 15\%. In Fig. \ref{fig:PMT:sigmaPDE_random}, the distributions for 320\,NNVT and 245\,HPK randomly selected PMTs are presented. These samples allow an estimation of the total number of PMTs failing the requirements as (0.5 - 1.7)\%\,@\,C.L.\,=\,68\% for NNVT and (0.0 - 0.5)\%\,@\,C.L.\,=\,68\% for HPK PMTs. Considering such a small number of non-uniform PMTs for the general sample there is no strict necessity to test all the PMTs on the scanning system and no PMT is rejected by the non-uniformity. These distributions will be useful for further detector simulations to evaluate the PMTs' non-uniformity contribution to the JUNO energy resolution.

\subsubsection{Uniformity of relative transition time spread (TTS) and transition time (TT) }
\label{sec:3:uniform:tts}

\begin{figure}[!ht]
  \begin{minipage}[!ht]{0.495\linewidth}
  \centering
          {\includegraphics[width=\linewidth]{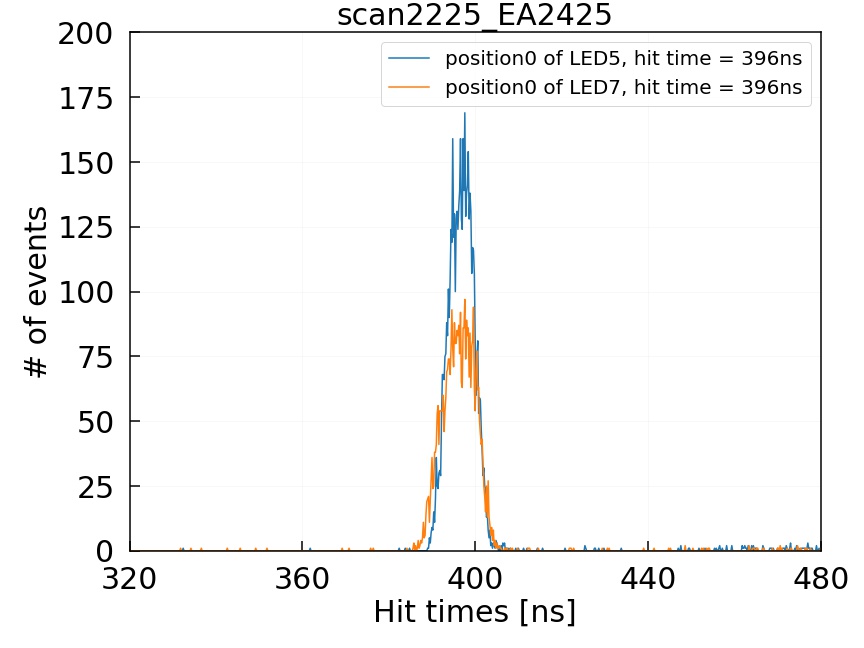}}
\end{minipage}
\begin{minipage}[!ht]{0.495\linewidth}
  \centering
    {\includegraphics[width=\linewidth]{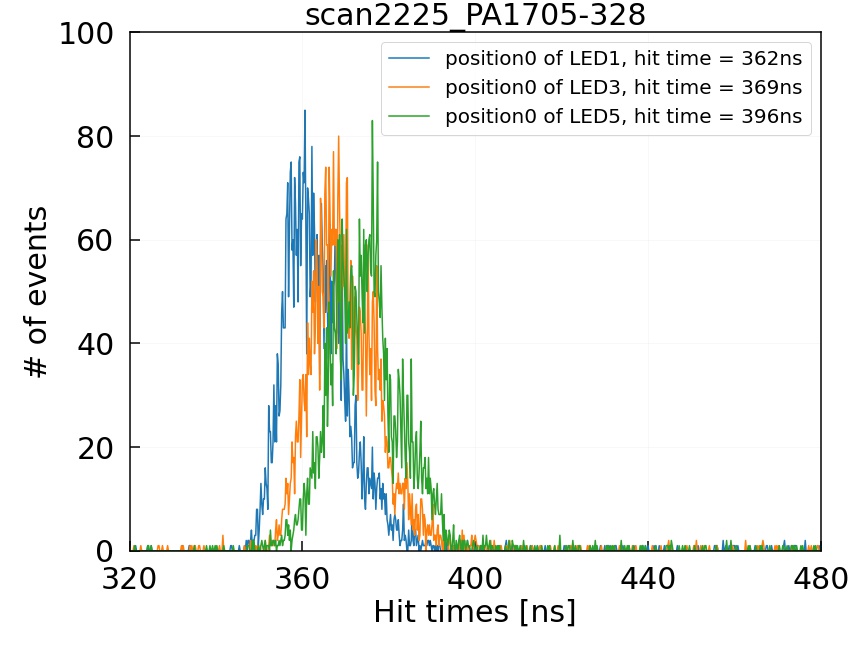}}
\end{minipage}
\caption{PMT hit time (relative TT) distribution of a single NNVT and HPK PMT directly measured in the scanning station with different LEDs (locations) at azimuthal angle $\phi$\,=\,0. Left: HPK (blue: 5th LED; orange: 7th LED.); right: NNVT (blue: 1st LED; orange: 3rd LED; green: 5th LED.).}
\label{fig:PMT:station:hittime}       
\end{figure}

The scanning station only implements the stabilized-LED flashing in pulse mode. The LED flashing intensity detected by PMT is 1$\sim$1.5\,p.e. for the PDE measurement, where the timing resolution can only reach $\sigma\sim$2\,ns to realize a relative TTS comparison along the photocathode\,\cite{HVSYS-LED}. Fig.\,\ref{fig:PMT:station:hittime} shows an example of the hit time (relative TT) distribution of a single HPK and NNVT PMT measurement analyzed with constant fraction discrimination (CFD). The NNVT PMT shows a broader distribution than the HPK PMT. Additionally, a sub-peak structure can be observed, that is similar to the right plot of Fig.\,\ref{fig:PMT:station:hittime}. There is a clear dependence of the typical relative TT value and its distribution spread on the location of the LED spots as shown on the right of Fig.\,\ref{fig:PMT:station:hittime}. We can distinguish some sub-peaks among LED\_1, LED\_3, and particularly LED\_5: the sub-peak structure (as introduced before in Fig.\,\ref{fig:PMT:TTSdist.b}) of the NNVT PMT's relative TT is related to the location of the photon hitting on the photocathode due to its relation on the focusing electric field and MCP itself.

\begin{figure}[!ht]
  \begin{minipage}[!ht]{0.495\linewidth}
  \centering
          {\includegraphics[width=\linewidth]{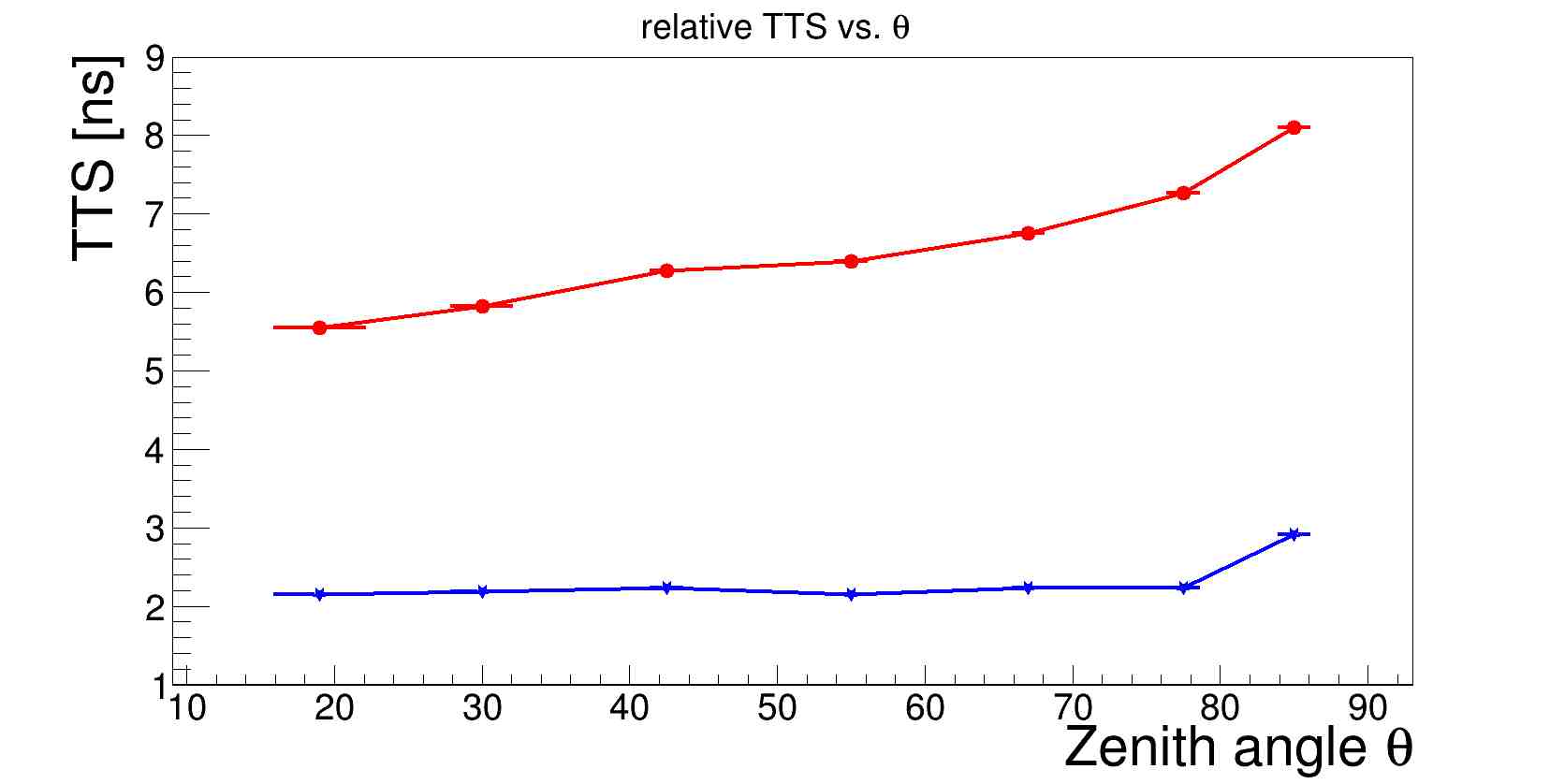}}
\end{minipage}
  \begin{minipage}[!ht]{0.495\linewidth}
  \centering
    {\includegraphics[width=\linewidth]{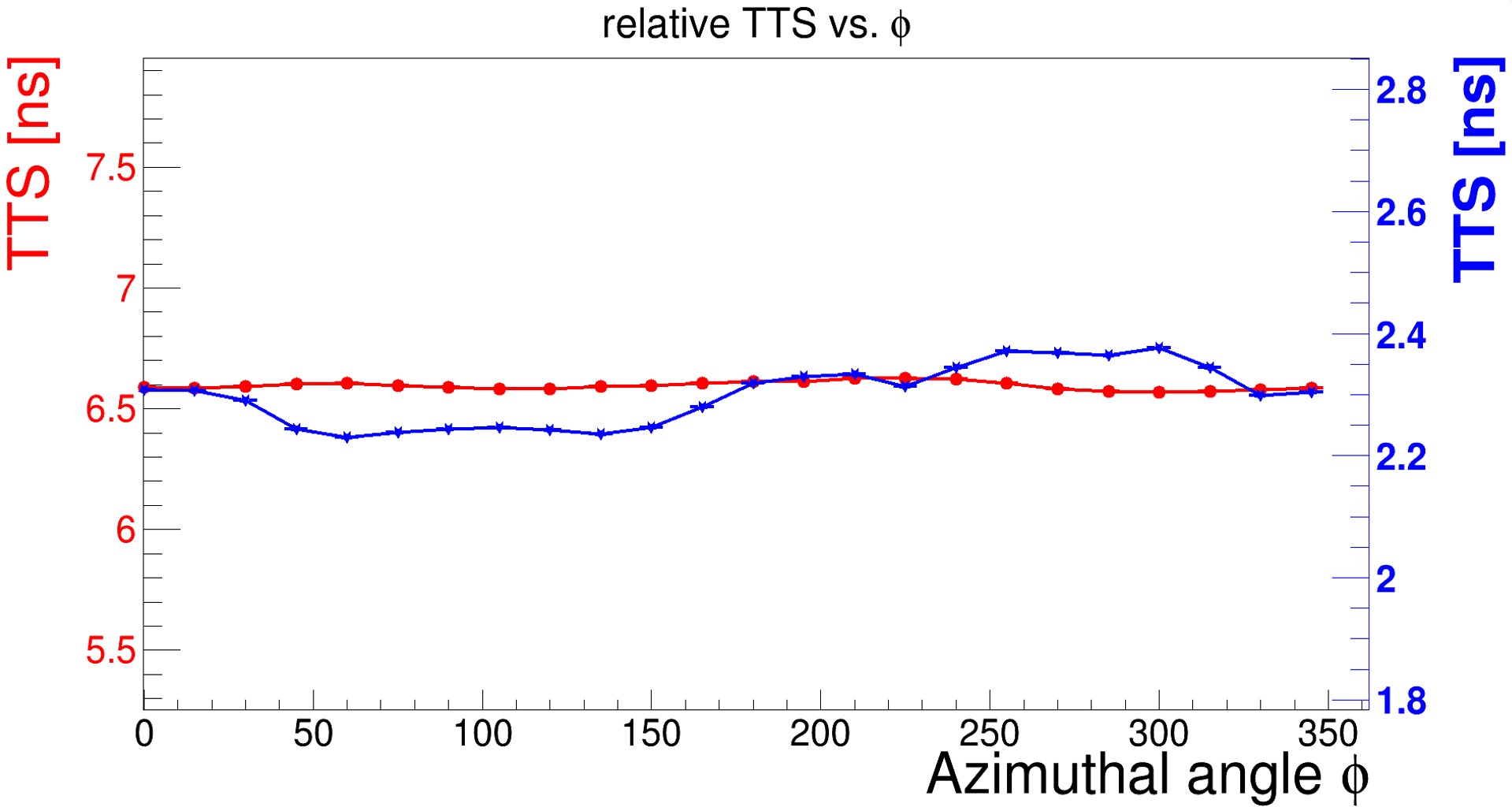}}
  \end{minipage}
\caption{Uniformity of the relative TTS ($\sigma$) along the zenith ($\theta$, left) and azimuthal angle ($\phi$, right) on average of all scanned PMTs with the same zenith or azimuthal angle. Blue: HPK PMTs; red: NNVT PMTs.}
\label{fig:PMT:ttsU}       
\end{figure}

A Gaussian fitting is applied to the relative TT spectrum (Fig.\,\ref{fig:PMT:station:hittime}) around its maximum peak ($\pm$20\,ns) on each light spot of the scanning station measurements, the fitted mean is treated as the relative TT and the $\sigma$ as the relative TTS. Fig.\,\ref{fig:PMT:ttsU} shows the average of the relative TTS of all scanned PMTs along the respective zenith or azimuthal angle. The relative TTS average over the whole photocathode of all scanned PMTs is 2.3\,ns for HPK PMTs and 6.6\,ns for NNVT PMTs, where the discrepancy to the results of the container LEDs is mainly due to the averaging scheme and ignoring the later discussed relative TT difference. HPK dynode PMTs show a better uniformity (max. variation $\sim$0.1\,ns) along the zenith angle for $\theta$\textless80$^{\circ}$. On the other hand, the NNVT MCP-PMTs show a larger TTS than HPK PMTs and a clear zenith angle dependence: the larger $\theta$, the larger the relative TTS (relative TTS increases 0.25\,ns/10$^{\circ}$ when $\theta$\textless80$^{\circ}$ on average with a linear assumption). The HPK PMTs also show some tiny structures along the azimuthal angle in the range of 0.15\,ns and a a single-cycle-oscillation related to its focusing box direction at $\phi\sim$270$^\circ$.

\begin{figure}[!ht]
  \begin{minipage}[!ht]{0.495\linewidth}
  \centering
          {\includegraphics[width=\linewidth]{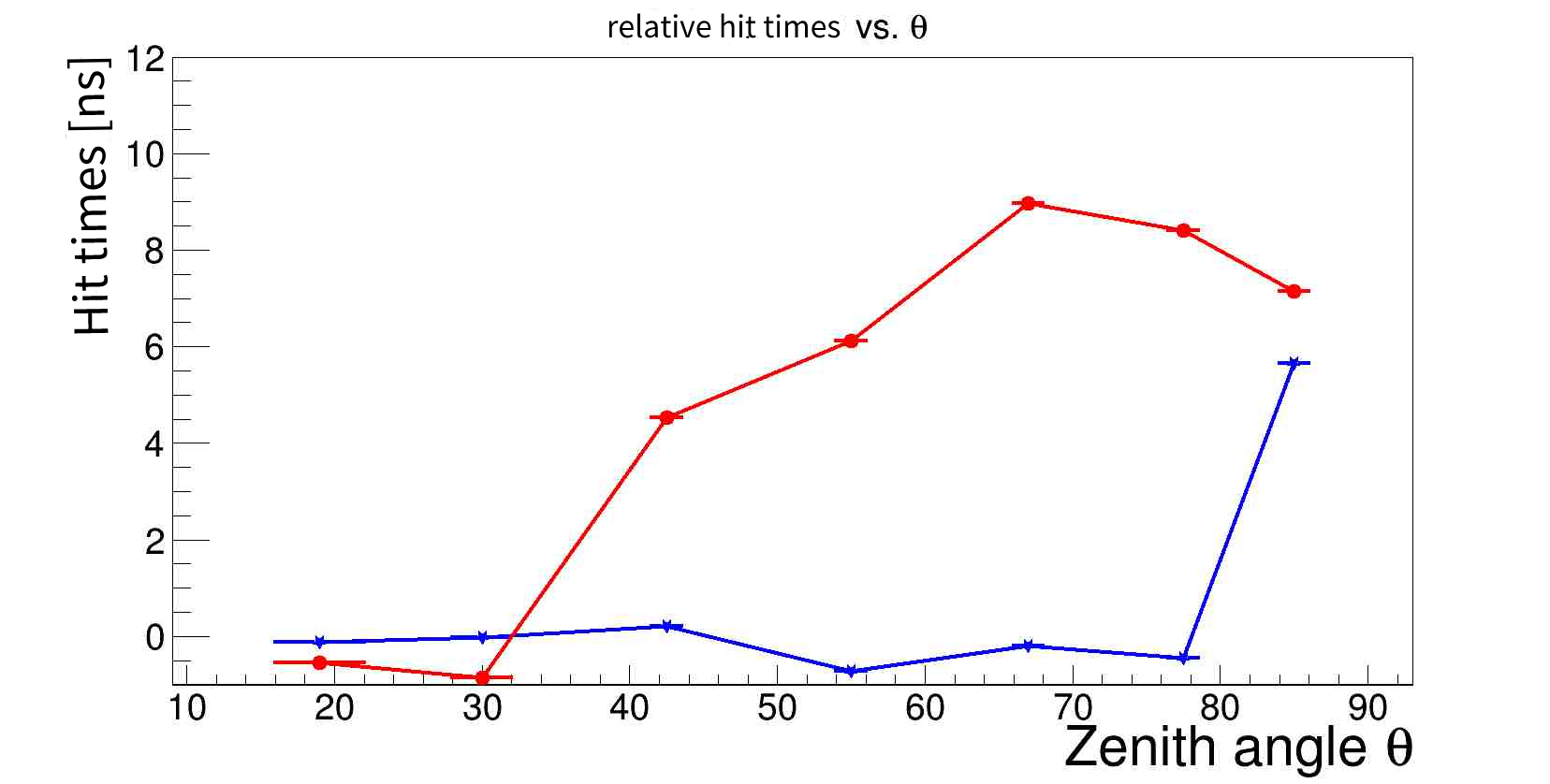}}
  \end{minipage}
  \begin{minipage}[!ht]{0.495\linewidth}
  \centering
    {\includegraphics[width=\linewidth]{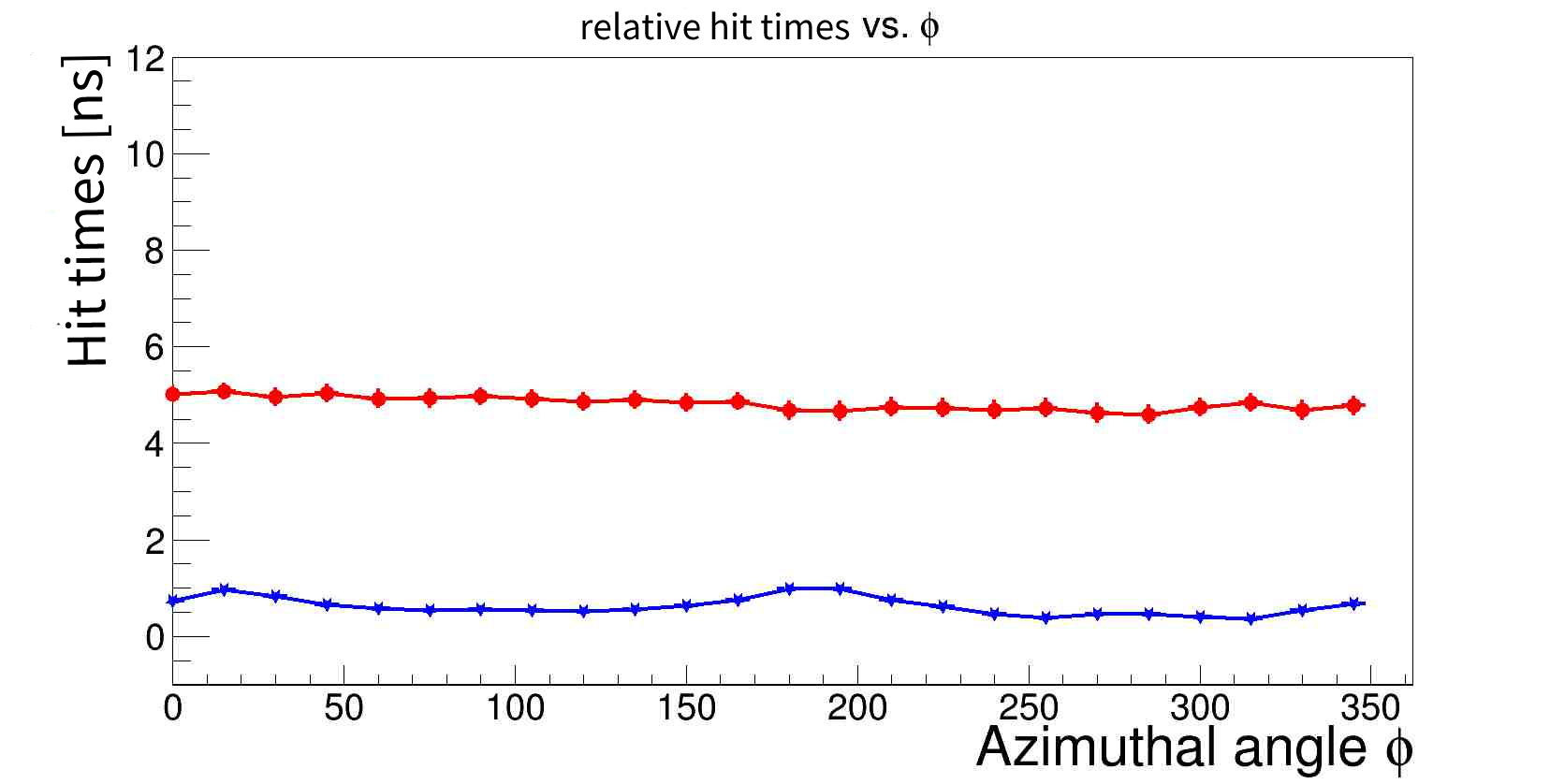}}
  \end{minipage}
\caption{Uniformity of relative TT (normalized to the LED\_0 angle\_0 of each scanned PMT itself) along the zenith angle ($\theta$, left) and azimuthal angle ($\phi$, right) on average of all measurements with the same zenith angle or azimuthal angle of all the scanned PMTs. Blue: HPK PMTs; red: NNVT PMTs.}
\label{fig:PMT:ttU}       
\end{figure}

At the same time, the uniformity of the transit time (TT) is also determined for both HPK and NNVT PMTs, which could also contribute to the TTS, i.e. for NNVT PMTs. Considering the absolute difference on TT between HPK and NNVT PMTs (as shown on the right of Fig.\,\ref{fig:PMT:station:hittime}), as well as variations among the PMTs, the relative TT (fitted mean of the PMT hit time spectrum as previously discussed) normalized to LED\_1 angle\_0 ($\phi$\,=\,0) of each scanned PMT itself is averaged and determined for all the scanned PMTs. The relative TT non-uniformity in average of the whole photocathode of all scanned PMTs is 0.6\,ns for HPK PMTs and 4.8\,ns for NNVT PMTs. The NNVT PMTs show a clear dependence on the zenith angle ($\theta$) as shown on the left of Fig.\,\ref{fig:PMT:ttU} (relative TT increases by a factor of $\sim$5 between $\theta\sim$30$^\circ$ to $\theta\sim$66$^\circ$), while for the HPK PMTs the relative TT is almost constant and shows a sharp rise for  $\theta$\textgreater80$^\circ$. Along the azimuthal angle, the HPK PMTs show a two-cycles-oscillation (peak-to-peak amplitude 0.6\,ns) and peaks located at around $\phi\sim$15$^\circ$ and $\phi\sim$190$^\circ$), while the NNVT PMTs remain almost flat across the range of azimuthal angles.

\subsubsection{Uniformity of PDE vs. Earth magnetic field (EMF)}
\label{sec:3:uniformemf}

\begin{figure}[!ht]
  \centering
  \includegraphics[width=0.75\textwidth]{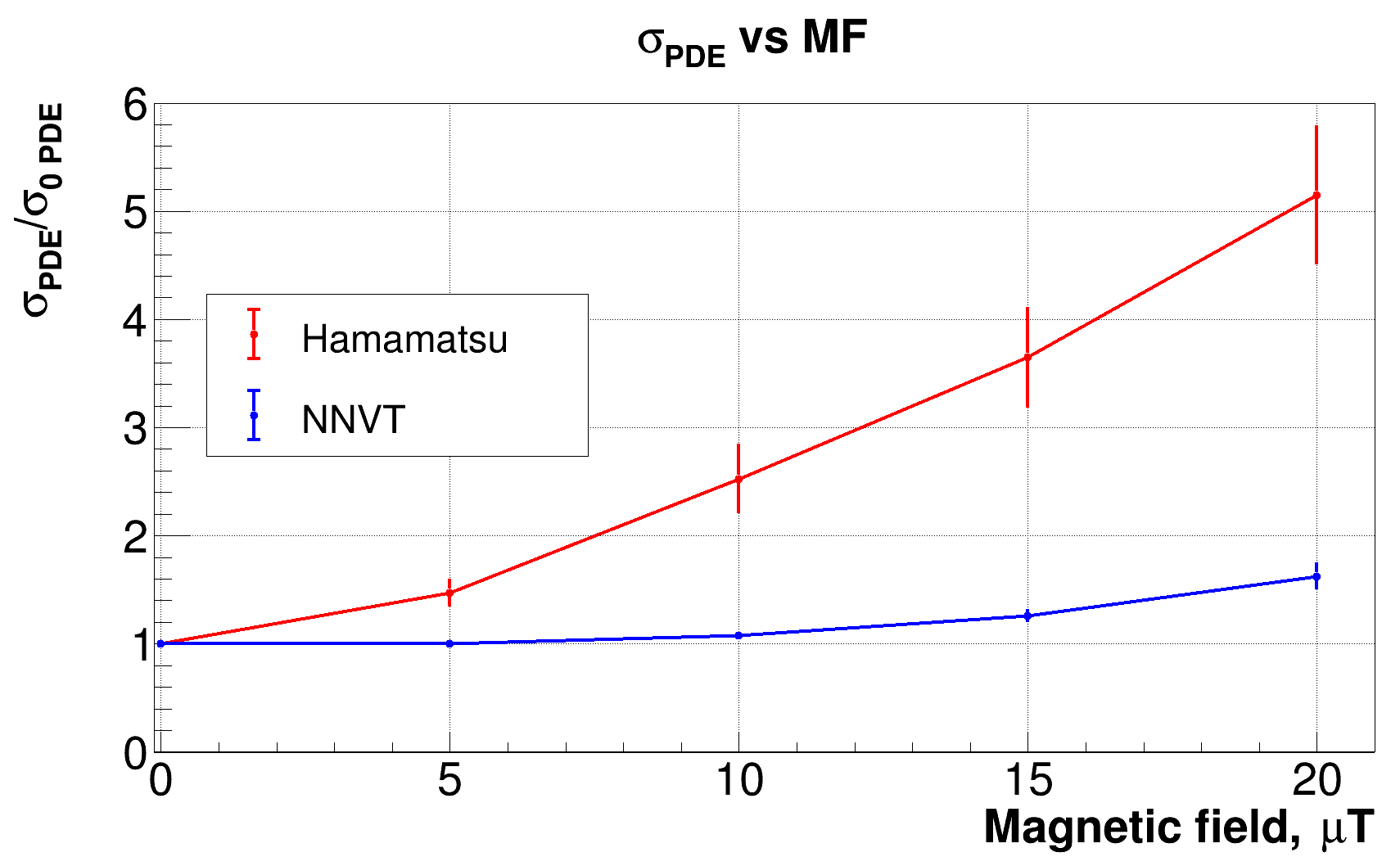}
\caption{PDE uniformity versus magnetic field strength. Indicated data points represent means and standard deviations of the tested samples (9 HPK PMTs and 15 NNVT PMTs).}
\label{fig:PMT:pdevsemf}       
\end{figure}

Another concern of JUNO is raised regarding the correlation of the PDE uniformity versus the magnetic field strength when JUNO's EMF shielding system has an imperfect shielding near the edges and corners of the detector, in particular for the PMTs used in the JUNO VETO system. The PDE uniformity versus magnetic field strength is further determined with the same data set as done for the PDE variation in sec.\,\ref{sec:2:emf}. The sigma ($\sigma_{\textit{PDE}}$) of the 168 scanned PDEs' distribution of a single PMT scanning is used to represent the PDE uniformity. The relative ratio of $\sigma_{\textit{PDE}}$ to that with zero-residual MF is used to represent the uniformity of PDE versus the remaining EMF. The averaged relative ratio of all the tested PMTs (9 HPK and 15 NNVT) is shown in Fig.\,\ref{fig:PMT:pdevsemf}. In contrast to the PDE versus MF, it seems that the uniformity of PDE of NNVT MCP-PMTs has a better tolerance than HPK dynode PMTs. 

\subsection{PMT long term stability}
\label{sec:2:aging}

A slow aging of the PMTs used in JUNO is one of the crucial features required by JUNO for a long-term running over in a 20 to 30\,years period. Long term stability measurements are of interest especially for the newly developed NNVT MCP-PMTs. Lots of studies are performed on the MCP or MCP-PMT lifetime, even with atomic layer deposition (ALD) protections~\cite{lifetimeMCPPMT3,MCPPMTlifeBelleII,lifetimeMCPPMT,lifetimeMCPPMT2}. Concerning this issue, some R\&D work had been done in~\cite{JUNOMCPPMTagingWenwen}, providing an expectation for a use over 20 years under JUNO conditions: maximum anode output charge is about 54.6\,C (assuming a gain of $ 1 \times 10^7$, $\sim$50\,kHz p.e.s from dark noise and signals per PMT, $\sim$4\,Hz muon through JUNO CD LS and averaged $\sim$1,000\,p.e.s per single PMT per muon).
At the Zhongshan Pan-Asia 20-inch PMT testing and potting station, the container $\#$C (as previously discussed) is re-configured to test the aging effect of HPK and NNVT PMTs under different configurations. The first loading completed its run after approx. 5 months and was intended partially for system debugging, the second loading intended for PMT long-term testing is still ongoing for more than 7 months now, where 32 PMTs (4 HPK PMTs + 28 NNVT PMTs) are separated into four groups:
\begin{itemize}
    \item[1.] Constant light to simulate 1\,MHz DCR and additional 10\,Hz pulses to generate 1,000\,p.e.s/pulse (0.140\,C/day).
    \item[2.] 10\,Hz pulses (1,000\,p.e.s/pulse) to simulate accelerated aging (assuming PMT itself with 50\,kHz DCR, 0.008\,C/day).
    \item[3.] Constant light to simulate 1\,MHz DCR (0.138\,C/day).
    \item[4.] 1\,Hz pulsed LEDs (1,000\,p.e.s) to simulate the future JUNO detector (assuming 50\,kHz DCR, 0.007\,C/day).
\end{itemize}

\begin{figure}[!ht]
 \includegraphics[width=0.33\textwidth]{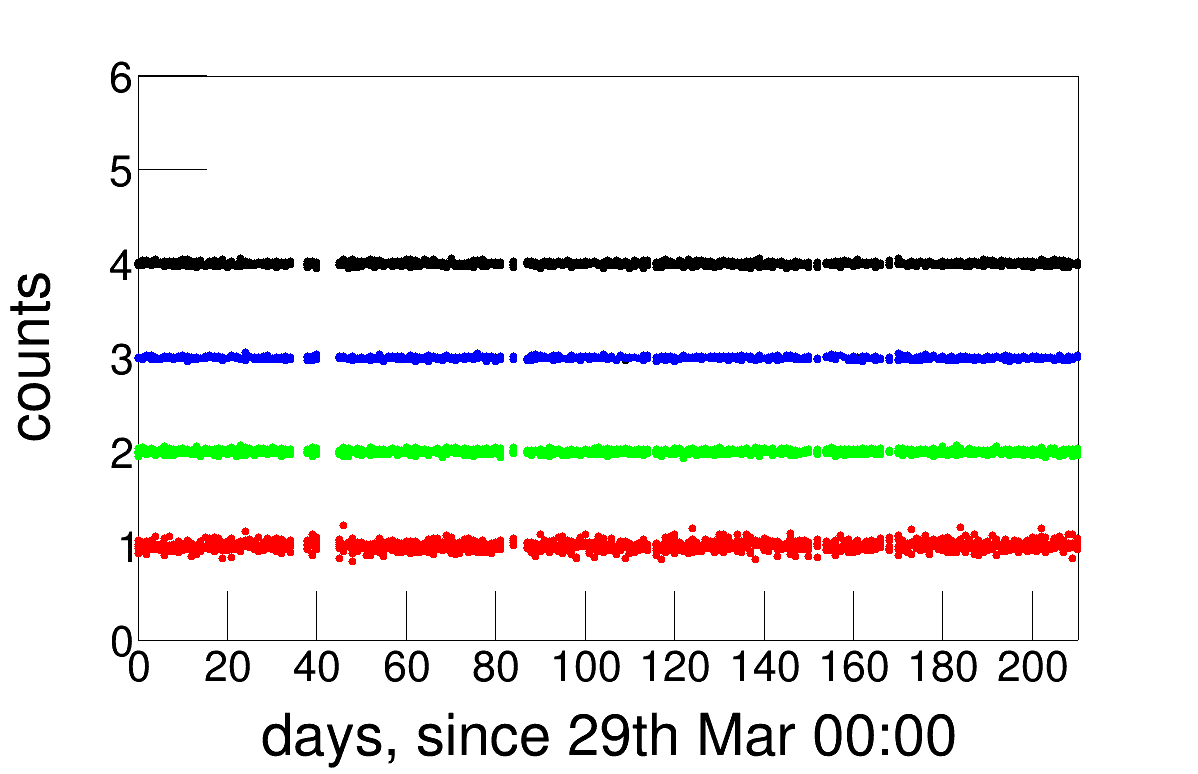}
\includegraphics[width=0.32\textwidth]{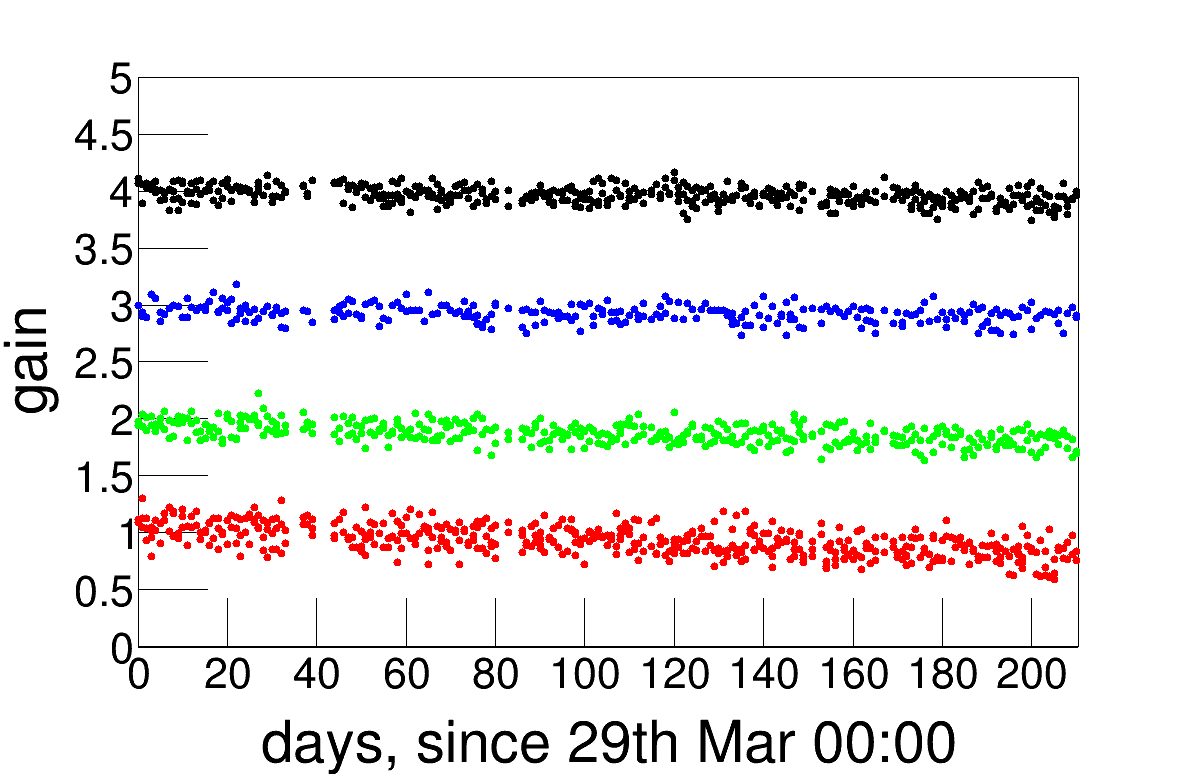}
 \includegraphics[width=0.33\textwidth]{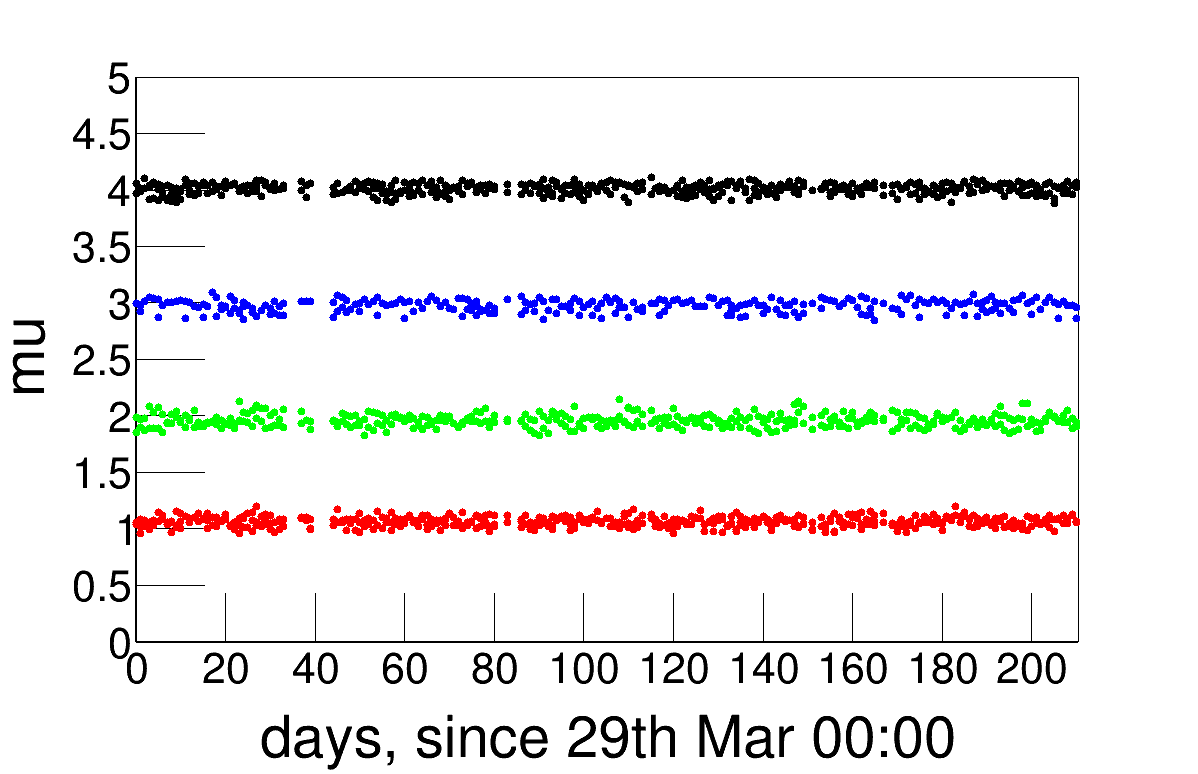}
\caption{Monitored DCR (left), gain (middle) and light intensity (measured number of p.e.s in \textmu, proportional to PDE, configured with constant light intensity, right) versus operating time under different configurations. There was an artificial shift introduced between each testing group on the plots to improve differentiation of the trend of each configuration: the first curve from top in black is group 4, the second curve from top in blue is group 3, the third curve from top in green is group 2, and the fourth curve from top in red is group 1, The y-axis is normalized to the individual starting points.}
\label{fig:PMT:aging}       
\end{figure}

Fig.\,\ref{fig:PMT:aging} shows the daily monitored DCR, gain, and light intensity (proportional to the PDE, configured with constant light source intensity). The DCR here is derived daily from the random coincidence of dark counts to the acquired waveform time window, which is not sensitive to instantaneous variations of the DCR.
A linear fit versus the testing day is applied to each monitored curve. No obvious degradation of the DCR and measured light intensity are identified so far ($\leq$0.0002/day, within the monitoring uncertainty), while the monitored gain of group 1 (output charge up to max. 19.6\,C in total) shows a clear decreasing trend of 0.001/day (monitoring uncertainty 0.0002/day), which is as expected from the aging of the MCP plates\,\cite{MCPPMTlifeBelle,MCPPMTlifeBelleII,MCPPMTlifeDIRC,JUNOMCPPMTagingWenwen,lifetimeMCPPMT,lifetimeMCPPMT3}. On the other hand, the monitored gain of other groups hints at a decreasing trend of $\sim$0.0002/day (monitoring uncertainty 0.0002/day), which still needs further monitoring. Even if the gain shows an obvious degradation in case of the NNVT MCP-PMTs of group 1, the requirement is still satisfied for a runtime of 20\,years running (\textless 50\% in 20\,years) and deviations can be compensated by slightly increasing the working voltage of PMTs.

\section{Waterproof potted PMT}
\label{sec:1:exppotted}

All JUNO PMTs, regardless of their placement in the CD or veto, will be operated in pure water. Therefore, the JUNO PMTs must be potted (waterproof encapsulated with the PMT base firmly soldered to the PMT) before their installation~\cite{JUNOdetector,JUNOPMTinstr}. The potted PMTs not only need to be tested once more to guarantee the functionality after the potting process, bet also need to be crosschecked for any change in characteristics.

There are several differences between the testing configurations of bare PMTs and potted ones:
\begin{itemize}
	\item[(1)] The HV divider will be soldered to PMT pins rather than using a pluggable HV divider, even though their designs are the same.
	\item[(2)] The HV-signal decoupler (specified for a positive HV) will be realized in another separate circuit, housed in a small aluminum box (as used for the functionality tests with the potted PMTs in containers \#A and \#B), and integrated as part of the 1F3 underwater electronics (container \#D, as used in the JUNO detector) rather than being integrated into the pluggable HV divider.
	\item[(3)] Another 2\,m (CD PMT) or 4\,m (veto PMT) extension of the HV/signal cable (including an additional SHV connector -- the bellow length of the potting structure is 0.5\,m shorter than the individual cables; another 2\,m extension cable is used during all the testings for easy connection) is necessary, in contrast to the bare PMTs' testing.
\end{itemize}

The potted PMTs were tested with the same containers as the bare PMTs (containers $\#$A and $\#$B), using again the commercial data acquisition electronics (tested 1,121 HPK and 3,708 NNVT PMTs in total). Some of them were tested also together with the 1F3 electronics (in container $\#$D, tested 732 HPK and 1112 NNVT PMTs in total), representing the same configuration as in the JUNO detector later on~\cite{JUNOdetector,JUNOEmbedded}.  
Since most of the investigated parameters are consistent with the results for the bare and potted measurements, only parameters only parameters which have been affected after potting will be discussed further in the following subsections. 

\subsection{HV of potted PMTs}
\label{sec:2:potted:HV}

\begin{figure}[!htb]
	\begin{subfigure}[c]{0.495\textwidth}
		\centering
		\includegraphics[width=\linewidth]{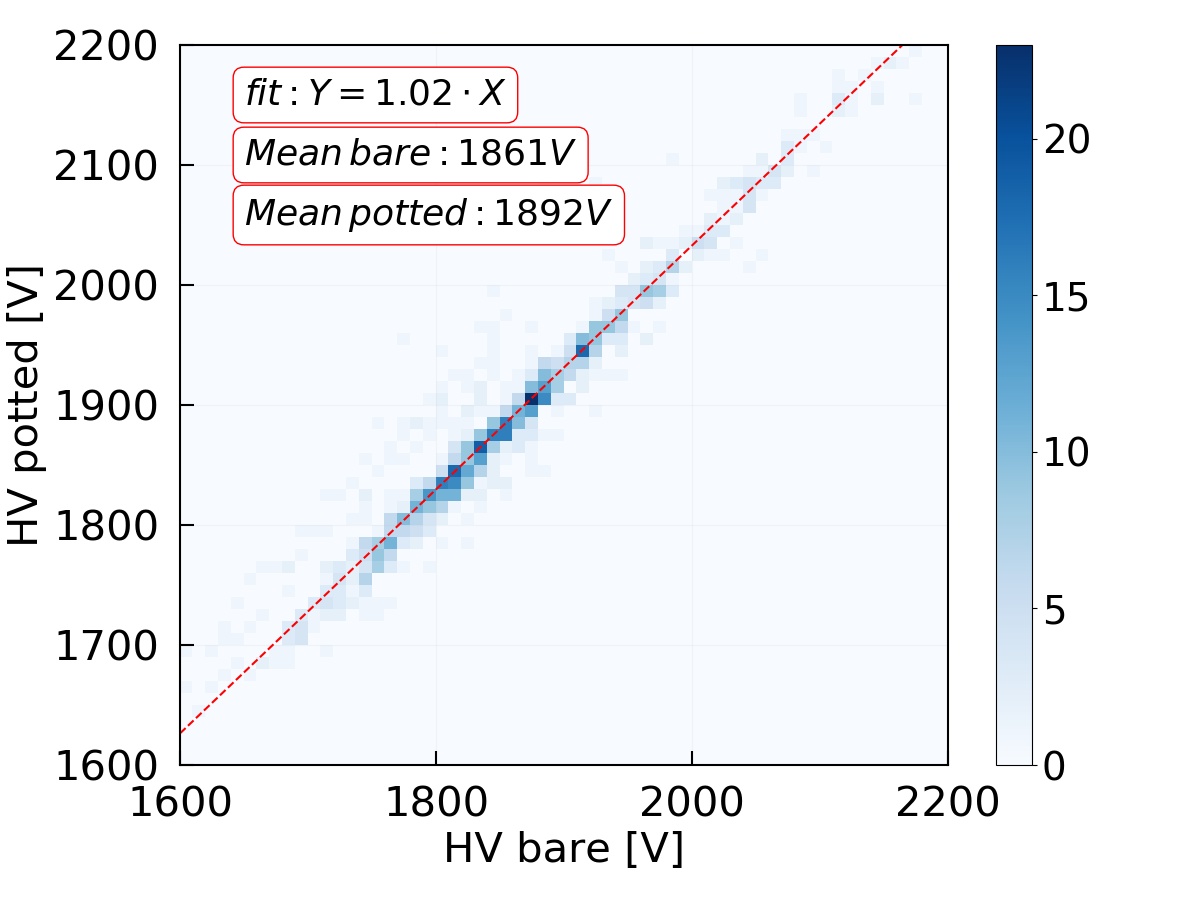}
		\label{fig:hvpotab3}
	\end{subfigure}\hfill
	\begin{subfigure}[c]{0.495\textwidth}
		\centering
		\includegraphics[width=\linewidth]{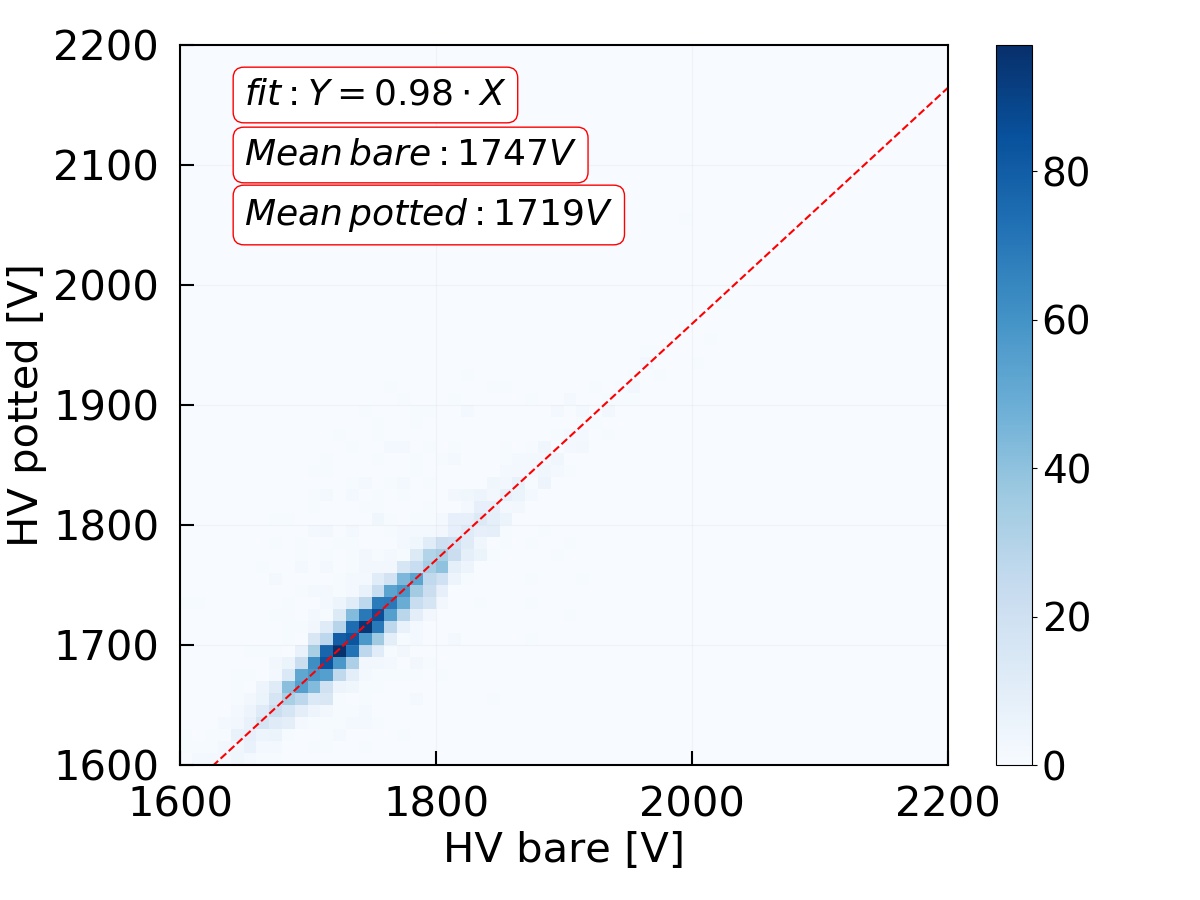}
		\label{fig:hvpotab2}
	\end{subfigure}\hfill
	\caption{ Working HV of potted PMTs for a gain of 1$\times$10$^7$. Left: HV of bare vs. potted HPK PMTs; right: HV of bare vs. potted NNVT PMTs.}
	\label{fig:hvpotab}       
\end{figure}

After the potting, the working HV of the PMTs to apply a gain of $1\times10^7$ is re-calibrated following the procedure presented in sec.\,\ref{sec:2:hv}. A comparison of the new HV working point with the measured values of the bare PMTs show a good correlation (see Fig.\,\ref{fig:hvpotab}). The HV increases by 30\,V for HPK PMTs on average, and decreases by 30\,V for NNVT PMTs on average based on the testing results from containers \#A and \#B. The determined HV values after potting are closer to the suggested values by the manufacturers. The variation is assumed to arise from the resistance change during production compared to the pluggable HV dividers for the HPK PMTs, and from shorter wires between HV divider and PMT pins in case of the NNVT PMTs.

\subsection{Dark count rate (DCR) of potted PMTs}
\label{sec:2:potted:dcr}

\begin{figure}[!htb]
	\centering
	\includegraphics[width=0.75\linewidth]{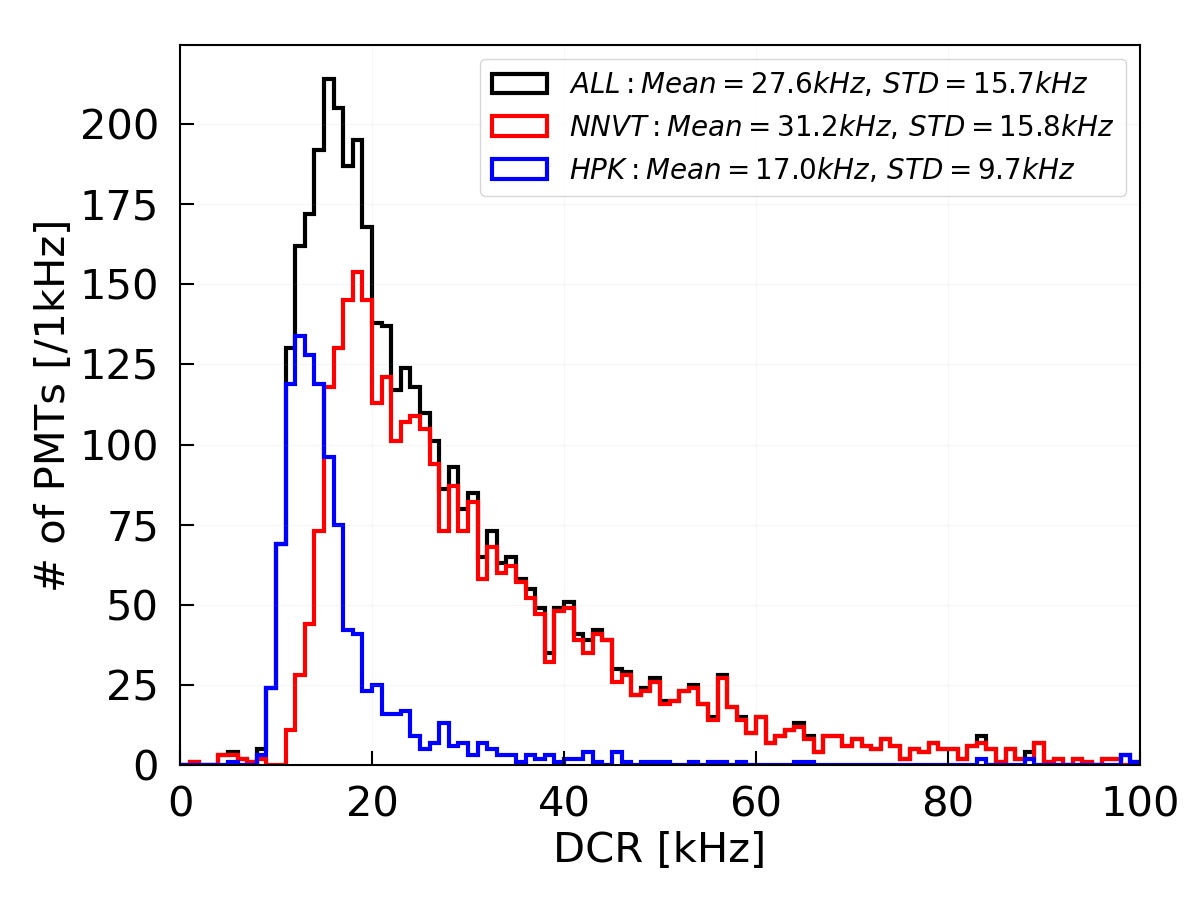}
	\caption{DCR of potted PMTs at a gain of 1$\times$10$^7$, measured in containers $\#$A and $\#$B. Black: all PMTs; blue: HPK; and red: NNVT.}
	\label{fig:dcrpotab}       
\end{figure}

It was found that after the waterproof potting of the PMTs, the measured DCR was decreased by a factor of about 0.59 on average for NNVT PMTs (Fig.\,\ref{fig:dcrpotab}), while HPK PMTs keep their mean value stable. This effect was verified in detail for the noise level, temperature and other conditions. The mean DCR for NNVT PMTs decreases to 31\,kHz tested with containers \#A and \#B. We conclude that this may be caused by the connection difference of the final soldered dividers and the used pluggable HV dividers, which could generate some additional noise for bare NNVT PMTs and increase the HV working point.

\subsection{Rise-time, fall-time and FWHM of potted PMTs}
\label{sec:2:potted:risefall}

\begin{figure}[!htb]
	\begin{subfigure}[c]{0.32\textwidth}
		\centering
		\includegraphics[width=\linewidth]{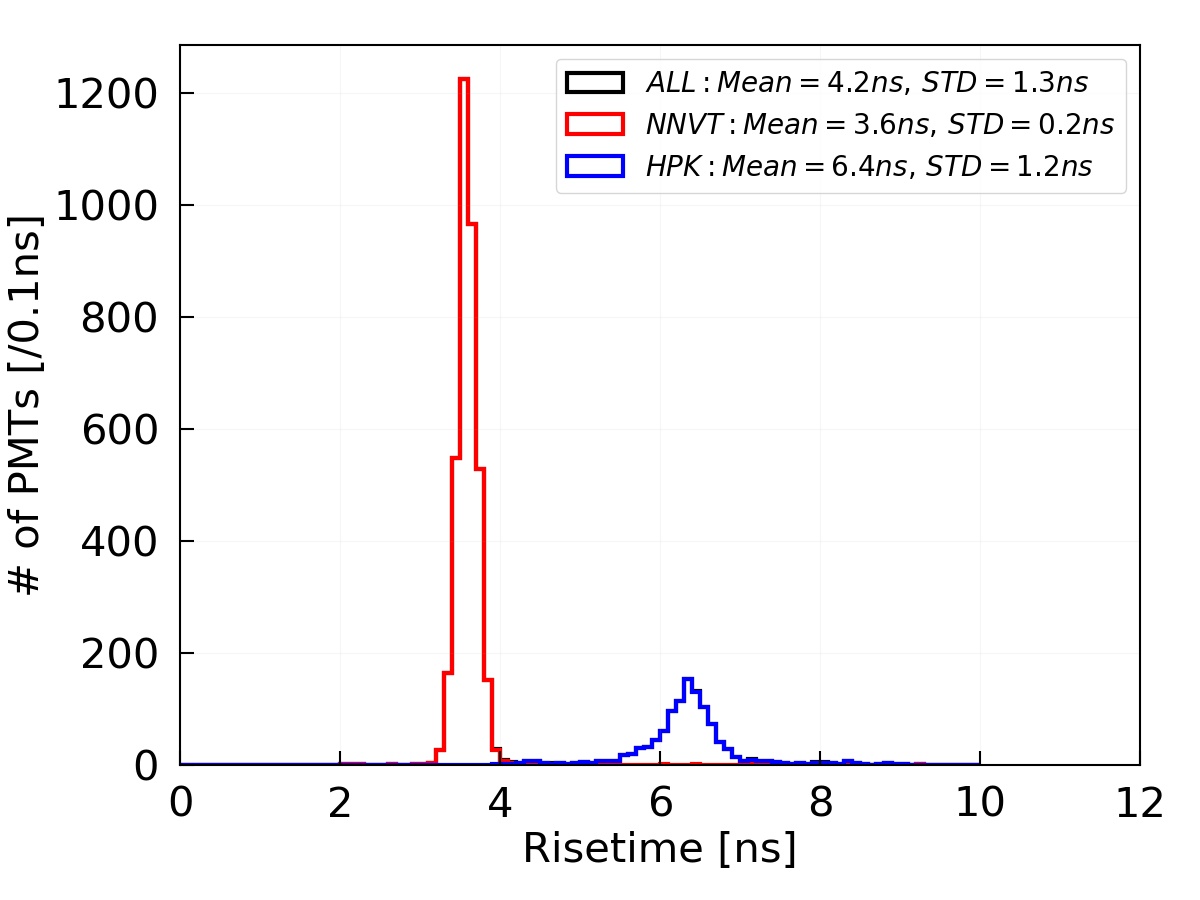}
	\end{subfigure}	
	\begin{subfigure}[c]{0.32\textwidth}
		\centering
		\includegraphics[width=\linewidth]{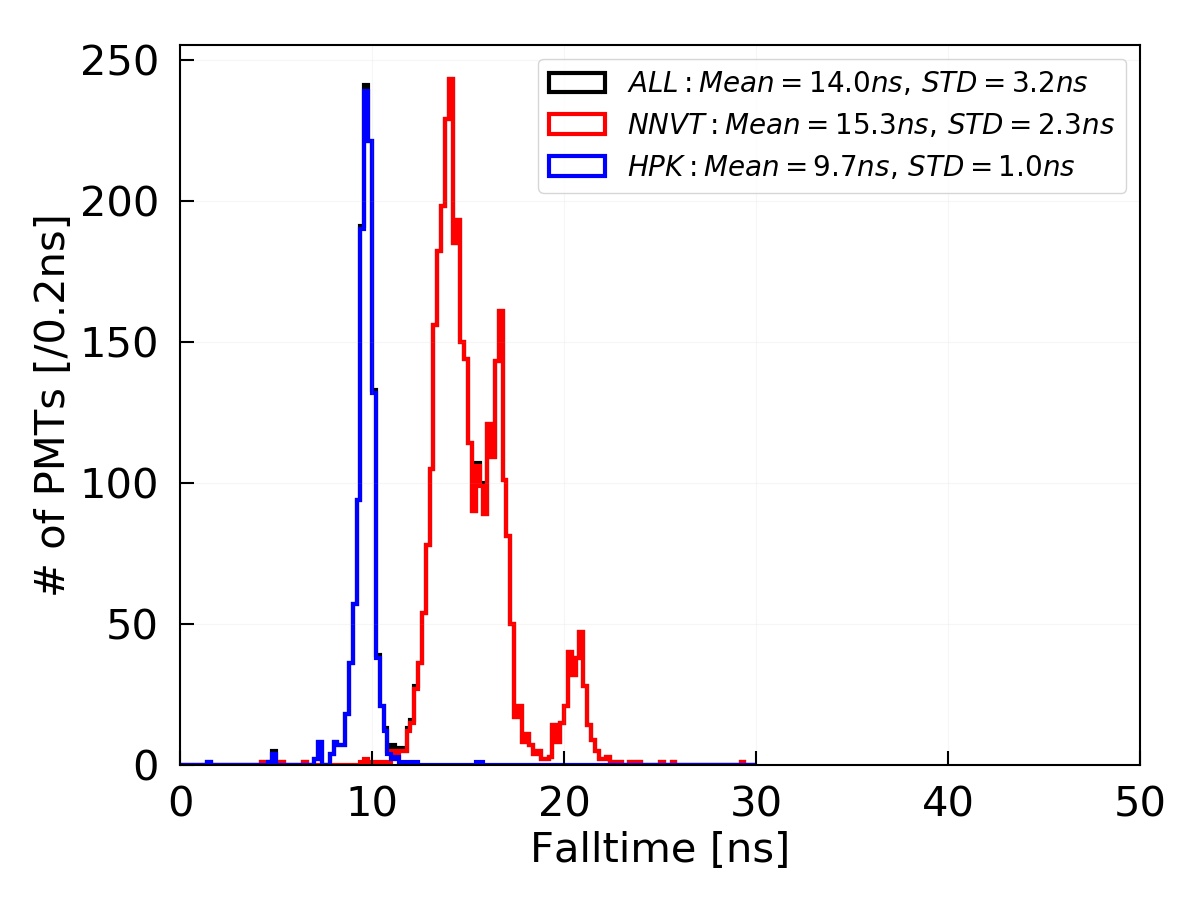}
	\end{subfigure}	
	\begin{subfigure}[c]{0.32\textwidth}
		\centering
		\includegraphics[width=\linewidth]{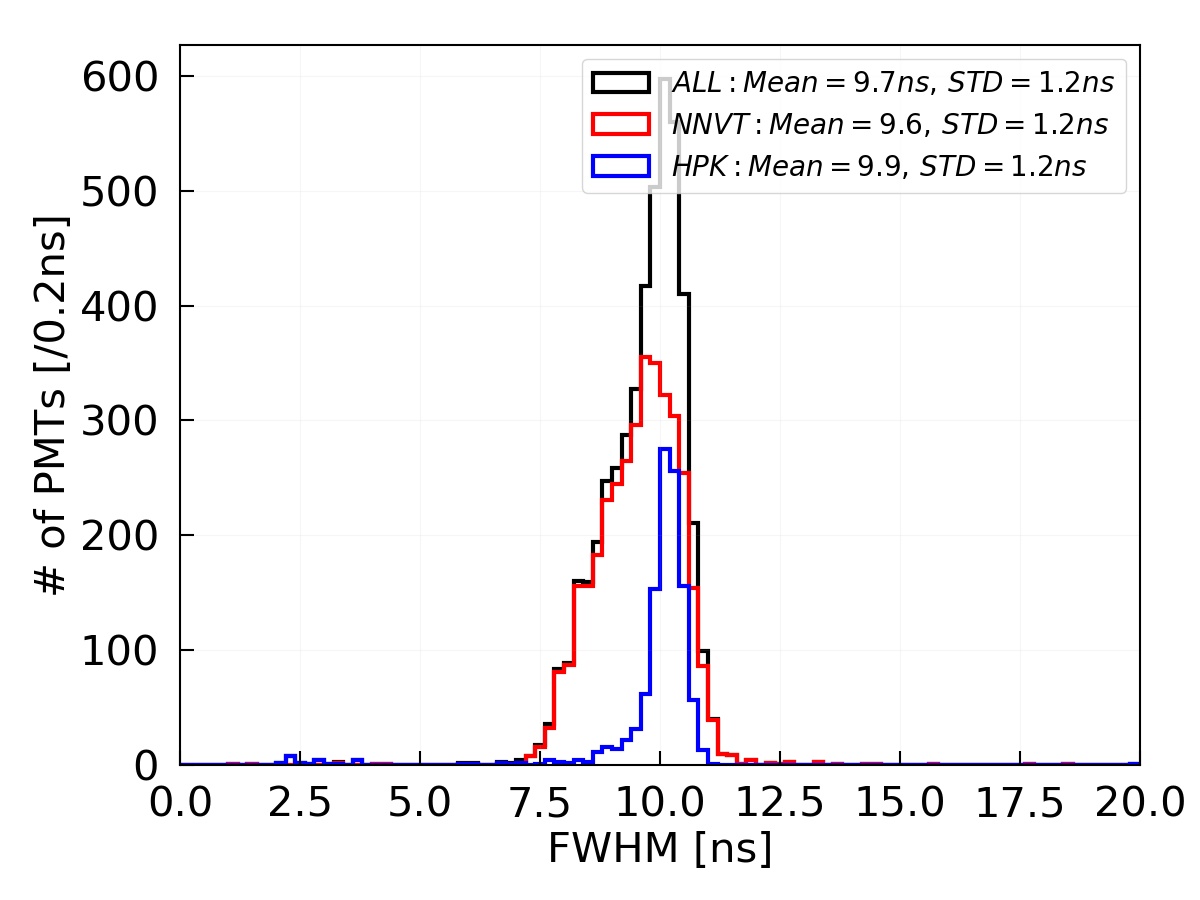}
	\end{subfigure}
		\caption{Measured rise-time (left), fall-time (middle) and FWHM (right) of potted PMTs at a gain of 1$\times$10$^7$ from containers \#A and \#B. Black: all PMTs; blue: HPK; red: NNVT.}
	\label{fig:time:potted}       
\end{figure}

SPE pulses are extracted from the waveforms of each potted PMT as mentioned
in sec.\,\ref{sec:2:spe} and illustrated in Fig.\,\ref{fig:gain:wave}. As known, the rise- and fall-time are more related to the PMT itself, while they also can be regulated by the HV divider design, for example, towards a slower behaviour. Fig.\,\ref{fig:time:potted} shows the rise-time, fall-time and FWHM of SPE pulses for all tested potted PMTs, measured with the container system. The average values for HPK PMTs and NNVT PMTs individually are 6.2\,ns and 3.6\,ns for rise-time, 9.6\,ns and 15.3\,ns for fall-time, and 9.6\,ns and 9.9\,ns for FWHM. The multi-peak structure (around 17\,ns and 21\,ns) of the fall-time distribution of NNVT PMTs is the joint contribution from the impedance mismatch among some of the PMTs, HV dividers, cables, SHV connectors and electronics. On the other hand, the HPK PMTs show a more uniform distribution of their waveforms among the tested PMTs. 

\section{Summary}
\label{sec:1:summary}

The performance of the JUNO 20-inch PMT system is one of the critical items to reach the designed 3\% energy resolution at 1\,MeV. In this article, we have described the facilities, the testing procedure and the characterization of more than 20,000 20-inch PMTs qualified for use in the JUNO detector, including both PMT types: HPK dynode PMTs and NNVT MCP-PMTs.

The testing systems, the container system and the scanning stations, have been running successfully over the whole testing campaign, achieving a satisfying noise level, measurement uncertainty, consistency between the testing systems, and monitored stability. With these systems, the absolute PDE was studied and measured based on the photon counting method, and most of the related PMT parameters were measured from the 20-inch PMTs, including the HV, gain, S/N, SPE amplitude, DCR, risetime, falltime, FWHM, charge resolution, ENF, gENF, TTS, pre-pulse ratio, after-pulse ratio, and the uniformity along the whole photocathode for various parameters such as the gain, charge resolution, PDE, TTS, and the EMF sensitivity.

\begin{table}[!ht]
\centering
\caption{Main parameters of qualified 20-inch PMTs}
\label{tab:parameter:summary}       
\begin{tabular}{lll}
\hline
 Parameters & PMT type & Bare \\
            &          & (Mean)  \\
\hline \hline
\multirow{2}{*}{Gain ($\times\,10^7$)}  & HPK & 1.00 \\
     & NNVT & 1.03  \\
\hline
\multirow{2}{*}{HV (V)}  & HPK & 1863  \\
  & NNVT & 1748 \\
\hline
\multirow{2}{*}{S/N}  & HPK & 13.0 \\
  & NNVT & 13.4  \\
\hline
\multirow{2}{*}{SPE amplitude (mV)}  & HPK & 6.5   \\
  & NNVT & 7.5  \\
\hline
\multirow{3}{*}{PDE (\%)(corrected)}  & HPK & 28.5 \\
  & NNVT & 30.1 \\
  & (High-QE/Low-QE) & (31.3/27.3) \\
\hline
\multirow{3}{*}{DCR (kHz)}  & HPK & 15.3 \\
  & NNVT & 49.3 \\
  & (potted) & (31.2) \\
  \hline
\multirow{2}{*}{P/V}  & HPK & 3.8 \\
  & NNVT & 3.9\\
  \hline
\multirow{2}{*}{Risetime (ns)}  & HPK & 6.9  \\
  & NNVT & 4.9 \\
  \hline
\multirow{2}{*}{Falltime (ns)}  & HPK & 10.2  \\
  & NNVT & 17.3 \\
  \hline
\multirow{2}{*}{FWHM (ns)}  & HPK & 11.6  \\
  & NNVT & 7.9  \\
 \hline
\multirow{2}{*}{TTS ($\sigma$,ns) (container laser)}  & HPK & 1.3   \\
  & NNVT & 7.0  \\
\hline
\end{tabular}
\end{table}

All received 20-inch PMTs from HPK and NNVT have completed their qualification checks during 2017-2021.
All tested PMTs have shown excellent performance, also the newly developed MCP-PMTs, and most of the PMTs reached the JUNO requirements. As a result, about 2,400 PMTs (9.3\%), out of the tested 22,400 PMTs have been rejected due to failing the specifications required by JUNO for visual inspection and/or electrical parameters. Except the visual defects, most of them are failures due to a low PDE (i.e. for the early version of low-QE MCP-PMTs), or due to a high DCR.
The photocathode uniformity was also checked in sub-samples of PMTs with a performance around the requirements boundary, as well as for a randomly selected sample totally covering about 12\% of all HPK PMTs and about 18\% of all NNVT PMTs. Finally, a total of 20,062 PMTs were successfully qualified for use in JUNO (typical parameters shown in Tab.\,\ref{tab:parameter:summary}), with a mean PDE of 28.5\% for HPK dynode PMTs and 30.1\% for NNVT MCP-PMTs (low-QE PMTs 27.3\%, and high-QE PMTs 31.3\%), and the expected mean PDE of the JUNO CD (17,612 PMTs, the PMTs will be randomly distributed by type (Dynode / MCP) as well as by performance in the CD.) and veto detector (2,400 PMTs) is 30.1\% and 26.1\% with the proposed schema, respectively. All the measured parameters of the qualified PMTs are stored in the JUNO PMT database, so that they can be accessed and used by the collaboration for detector simulations, during installation and commissioning, and eventually for the data analysis.

Furthermore, about 5,000 of the waterproof potted PMTs (encapsulated with the PMT base firmly soldered to the PMT) are tested again after the potting in the testing facilities. Additionally, dedicated tests including JUNO final electronics have been performed as well to study the long-term behaviour of the JUNO PMTs under normal conditions and higher strain to simulate and investigate possible aging effects.

\section*{Acknowledgments}
\label{sec:1:acknow}

We are grateful for the ongoing cooperation from the China General Nuclear Power Group.
This work was supported by
the Chinese Academy of Sciences,
the National Key R\&D Program of China,
the CAS Center for Excellence in Particle Physics,
Wuyi University,
and the Tsung-Dao Lee Institute of Shanghai Jiao Tong University in China,
the Institut National de Physique Nucl\'eaire et de Physique de Particules (IN2P3) in France,
the Istituto Nazionale di Fisica Nucleare (INFN) in Italy,
the Italian-Chinese collaborative research program MAECI-NSFC,
the Fond de la Recherche Scientifique (F.R.S-FNRS) and FWO under the ``Excellence of Science – EOS” in Belgium,
the Conselho Nacional de Desenvolvimento Cient\'ifico e Tecnol\`ogico in Brazil,
the Agencia Nacional de Investigacion y Desarrollo and ANID - Millennium Science Initiative Program - ICN2019\_044 in Chile,
the Charles University Research Centre and the Ministry of Education, Youth, and Sports in Czech Republic,
the Deutsche Forschungsgemeinschaft (DFG), the Helmholtz Association, and the Cluster of Excellence PRISMA$^+$ in Germany,
the Joint Institute of Nuclear Research (JINR) and Lomonosov Moscow State University in Russia,
the joint Russian Science Foundation (RSF) and National Natural Science Foundation of China (NSFC) research program,
the MOST and MOE in Taiwan,
the Chulalongkorn University and Suranaree University of Technology in Thailand,
University of California at Irvine and the National Science Foundation in USA.





\bibliography{allcites}   

\end{document}